\documentclass[aps,prb,amsmath,twocolumn,superscriptaddress]{revtex4-2}

\usepackage{amsmath}
\usepackage{amssymb}
\usepackage{tikz}

\usepackage{graphicx}
\usepackage{bm} 
\usepackage{url}
\usepackage{verbatim}   

\usepackage{color}

\usepackage[utf8]{inputenc}
\usepackage{makecell}
\usepackage{multirow}

\DeclareMathOperator{\Tr}{tr}

\usepackage[percent]{overpic}
\usepackage[export]{adjustbox}
\usepackage{stackengine}
\usepackage{appendix}

\usepackage{placeins} 
\usepackage{flafter}  

\begin{document}



\title{Multirate characterization of relaxation mechanisms for two nonequivalent nuclear spins 1/2 in a liquid using maximally entangled pseudo-pure quantum states  }




\author{Georgiy Baroncha}
\email{georgiy.baroncha@uni-leipzig.de}
\affiliation{Felix Bloch Institute for Solid State Physics, University of Leipzig,
Linnéstraße 5, 04103 Leipzig, Germany}

\author{Alexander Perepukhov}
\affiliation{Moscow Institute of Physics and Technology, Institutsky per. 9, Dolgoprudny, Moscow  region,  141700, Russia.
}
 
\author{Boris V. Fine}
\email{boris.fine@uni-leipzig.de}
\affiliation{Institute for Theoretical Physics, University of Leipzig,
Br{\"{u}}derstr. 16, 04103 Leipzig, Germany}%

\date{\today}

\begin{abstract}

Multirate characterization of spin responses in nuclear magnetic resonance (NMR) is a promising approach to fingerprinting  complex molecules in the presence of multiple relaxation mechanisms. Here we present experimental and theoretical investigations simultaneously accessing 8 relaxation rates describing the density matrix of two adjacent non-equivalent nuclear spins 1/2 ($^1$H and $^{\ 13}$C) belonging to a molecule in a liquid. The selected nuclear pair is stable with respect to chemical exchange. Some of the rates are obtained from conventional measurements of inversion recovery and nuclear Overhauser effect, while other, less conventional ones, are extracted from the relaxation initialized by the maximally entangled pseudo-pure Bell states (Bell PPSs) of the spin pair. The Bell PPSs are created using a hereby introduced method based on a detuned Hartmann-Hahn double resonance condition. Microscopic theory behind the measured relaxation rates is presented, and its consistency is demonstrated by several parameter-free tests. In particular, it is shown both theoretically and experimentally, that the eigenmodes of the off-diagonal relaxation of the two-spin density matrix can be selectively initialized using Bell PPSs.  Our multirate analysis suggests that the measured off-diagonal relaxation is partly due to an unconventional mechanism arising from very weak $J$-couplings of the spin pair with fluctuating distant nuclear spins. Furthermore, we identify a dimensionless ratio of diagonal relaxation rates, which is determined exclusively by intra-pair magnetic dipolar interaction and hence possesses a universal value for a broad class of nuclear spin pairs. This value is consistent with both our experiments and other experiments reported in the literature.

\end{abstract}

\maketitle

\section{Introduction}

Relaxation of nuclear spins in a liquid measured by nuclear magnetic resonance (NMR) is sensitive to the structure and dynamics of molecules carrying those spins. Detailed characterization of this relaxation can reveal difficult-to-access microscopic parameters. NMR experiments pursuing such an agenda often focus on longitudinal $T_1$-relaxation, transverse $T_2$-relaxation and the nuclear Overhauser effect (NOE)\cite{Abragam-61,Slichter-96,Levitt-2013,Kowalewski-2017}. However, the above kinds of experiments deliver only a fraction of microscopic information for a strongly coupled pair of nuclear spins 1/2. The density matrix of two spins 1/2 is determined by 15 parameters, whose relaxation is characterized in its entirety by a 15$\times$15 matrix of coefficients, which we call ``rates".   

A promising and not yet fully exploited resource for NMR relaxation studies is the use the initial pseudo-pure states (PPSs) of nuclear spin pairs where the pure part of the PPS is a maximally entangled Bell state. Such a state is to be referred to below as ``Bell PPS." As discussed later, Bell PPSs at typical temperatures of NMR experiments are not entangled in the sense of the standard quantum-informational definition of entanglement for mixed quantum states\cite{Braunstein-1999,Fine-2005}; yet the NMR experiments are only sensitive to the maximally entangled pure parts of Bell PPSs.   Bell PPSs have been of interest for NMR for decades: they are often used in the context of the quantum computing agenda \cite{CORY199882, Chuang1998}, but their utility in NMR is much broader\cite{doi:10.1021/ja0490931, doi:10.1073/pnas.1010570107, doi:10.1126/science.1167693,  article1, CLAYTOR201481, doi:10.1063/1.1893983,   doi:10.1063/1.2778429, doi:10.1063/1.5074199, soton347308, doi:10.1063/5.0006742, https://doi.org/10.1002/cmr.a.20100, doi:10.1021/ja0647396, doi:10.1126/sciadv.aaz1955,PhysRevLett.112.077601,Samal_2010}.

In this work, we investigate a pair of two non-equivalent nuclear spins 1/2, generate their Bell PPSs using hereby introduced method based on detuned Hartmann-Hahn double resonance condition and then combine conventional relaxation measurements with those initialized by the Bell PPSs to obtain 8 relaxation rates for the same system. 
The measured rates characterize both diagonal and off-diagonal relaxation of the density matrix in the quantum basis imposed by external magnetic field. We show, in particular, both experimentally and theoretically that Bell PPSs can be used to selectively initialize individual eigenmodes of the off-diagonal relaxation. 
 
On the theoretical side, we analyze the measured rates using Redfield's equations\cite{Redfield}, thereby connecting those rates to various relaxation mechanisms. The mechanisms are separated into the one due to the intra-pair magnetic dipolar interaction \mbox{(IPMDI)} and those described as originating from fluctuating local fields that are not affected by the spin state of the selected nuclear pair. This analysis then leads us to identifying significant contributions from the currently underappreciated off-diagonal relaxation mechanism associated with very small $J$-couplings of the selected spin pair to distant nuclear spins. Such a mechanism, besides being a necessary part of multirate analysis\cite{Kowalewski-2017}, is also potentially useful for assessing small $J$-couplings not resolvable directly in NMR spectra, which, in turn, can help to characterize the connectivity and the local geometry of complex molecules. 


Our theoretical description is to be subjected to a number of experimental tests. Particularly notable among them is the parameter-free test validating the  NMR relaxation theory of Bloembergen, Purcell and Pound\cite{BPP} (BPP) and of Solomon\cite{Solomon}: it involves a combination of four diagonal relaxation rates, where all contributions from non-IPMDI mechanisms are canceled.   
IPMDI played the central role in the BPP-Solomon theory, and, despite the theory being well established, the parameter-free tests of it have been somewhat elusive --- consequence of multiple mechanisms involved in the relaxation. In particular, the theory of Solomon\cite{Solomon} initially assumed that the \mbox{IPMDI} completely determines NOE in hydrofluoric acid HF --- in disagreement with experiment, which was later explained\cite{Solomon-Bloembergen-1956,Abragam-61} by the $J$-coupling modulated by chemical exchange. Chemical exchange was also part of the reason\cite{Meiboom-1961,Abragam-61} why the simple equality $T_1 = T_2$ for the longitudinal and the transverse relaxation times  was not observed for water --- contrary to the prediction of the BPP theory based on the IPMDI mechanism only. Given the above examples, it is natural to explore the parameter-free tests of the BPP-Solomon theory for nuclear spin pairs stable towards chemical exchange, such as the one investigated in this article. 

In what follows,  Section~\ref{exp} introduces the system to be investigated, Section~\ref{theoretical} describes the theoretical basis of our analysis, Section~\ref{Experiments} presents the experimental investigations. The experimental results are compared with the theoretical predictions and interpreted in terms of microscopic relaxation mechanisms in Section~\ref{discussion}. Section~\ref{sec:ratio} discusses literature-based parameter-free tests of IPMDI relaxation mechanism. Section~\ref{conclusions} contains summary and conclusions.


\section{System of interest }
 \label{exp}

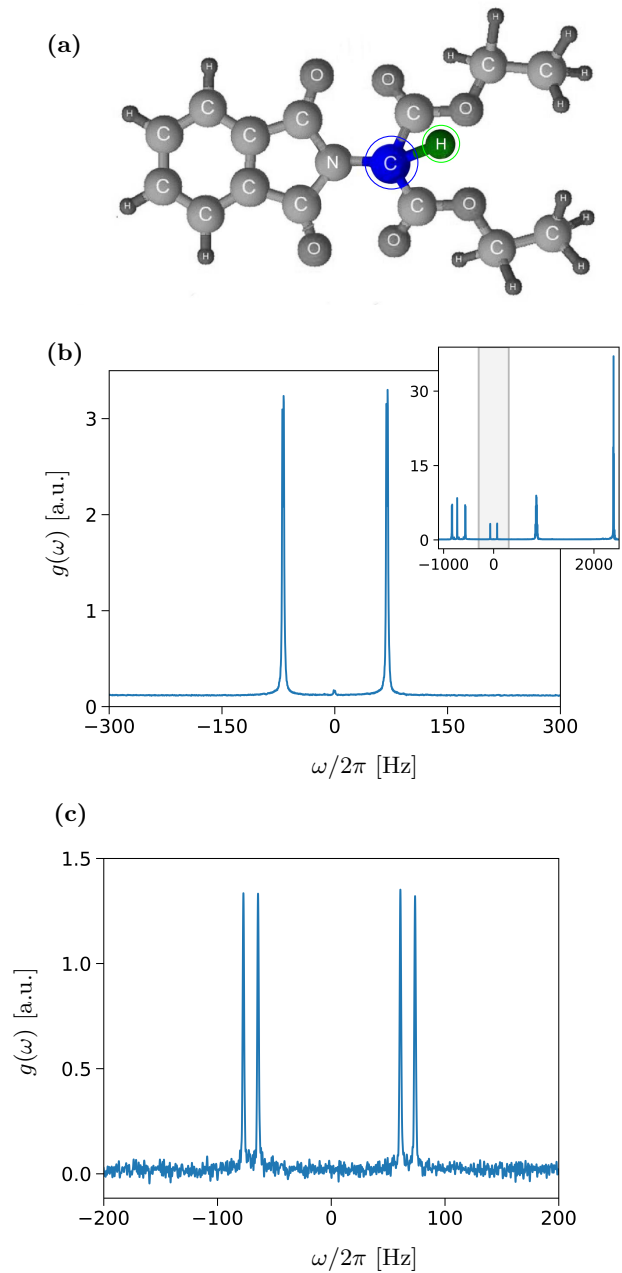
\begin{figure}[h!]
   
    \centering
    \begin{tikzpicture}
    \node[inner sep=0pt] (duck) at (0,0)
    {\includegraphics[width=0.85\columnwidth]{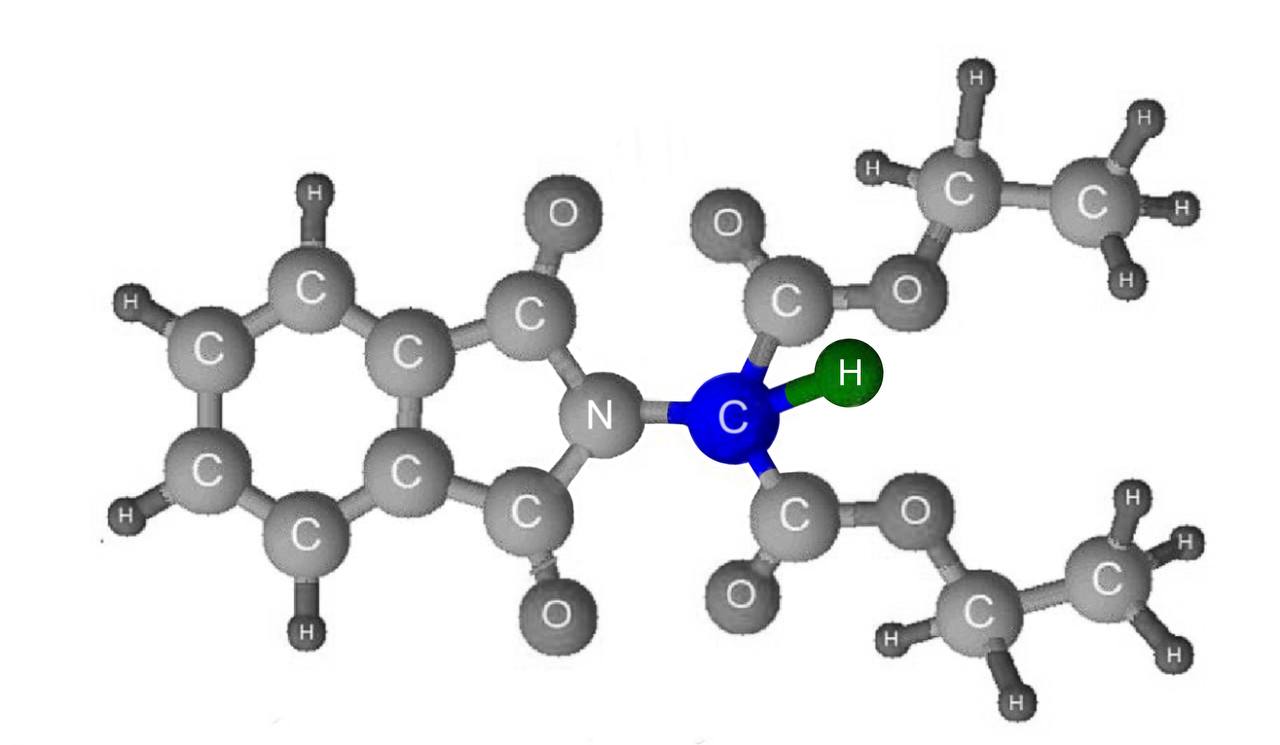}};
    \node[align=center,fill=white] at (-3.8, 1.3) {\textbf{(a)}};
    \draw[blue] (0.55,-0.25) circle (10pt);
     \draw[green] (1.21,-0.00) circle (7pt);
    \end{tikzpicture}

    \vspace{10pt}
    
    \begin{tikzpicture}
    \node[inner sep=0pt] (duck) at (0,0) {\includegraphics[width=0.85\columnwidth]{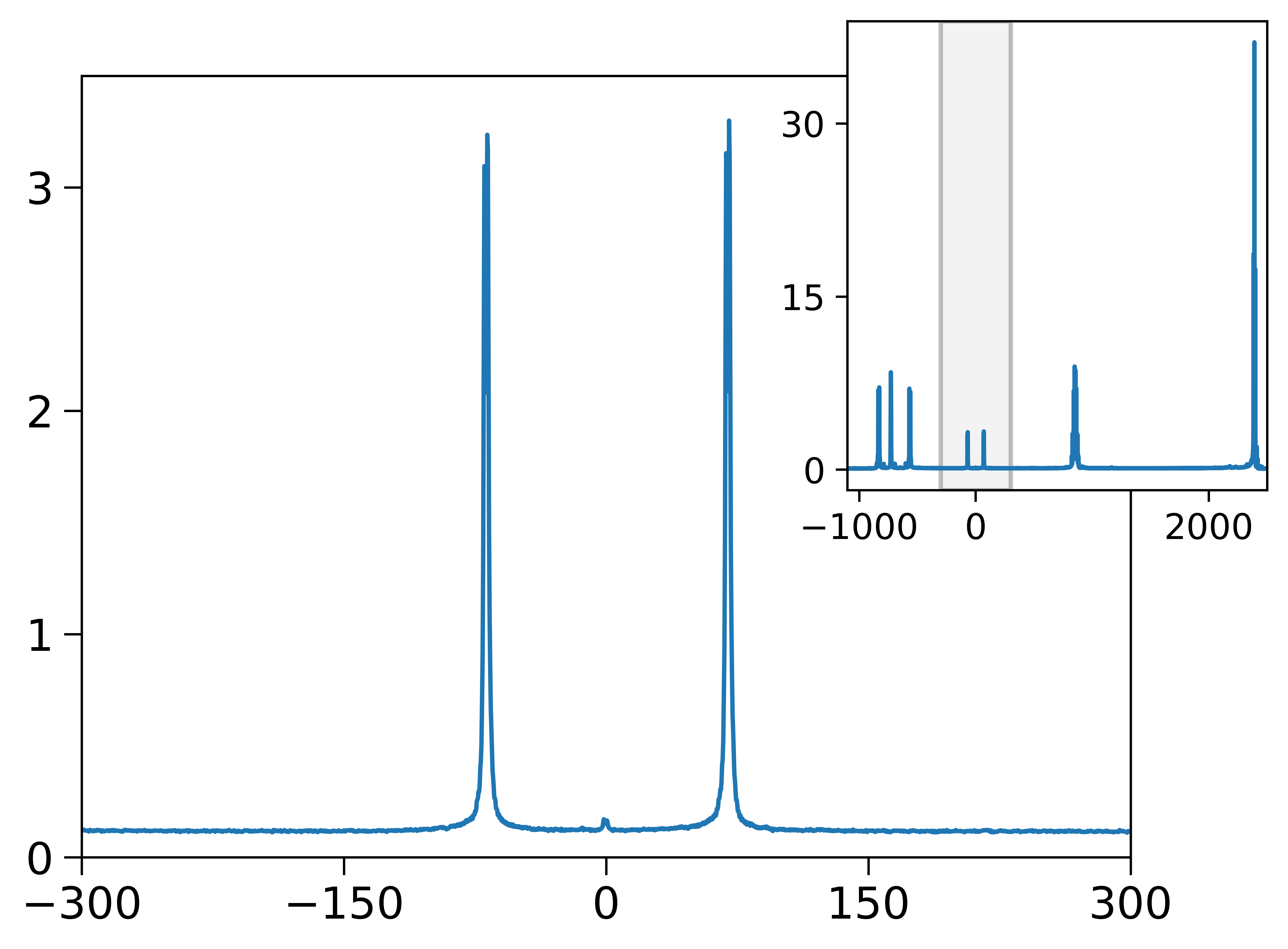}};
    \node[align=center,fill=white] at (-3.8, 2.5) {\textbf{(b)}};
    \node[align=center,fill=white] at (-3.9, 0.3) {{\rotatebox {90}{\text{$g(\omega)$ [a.u.]} }}};
    \end{tikzpicture}
    \text{\qquad \quad $\omega/2\pi$ [Hz] }

    \vspace{5pt}
    \hspace{-31pt}
    \begin{tikzpicture}
    \node[inner sep=0pt] (duck) at (0,0)
    {\includegraphics[width=0.82\columnwidth]{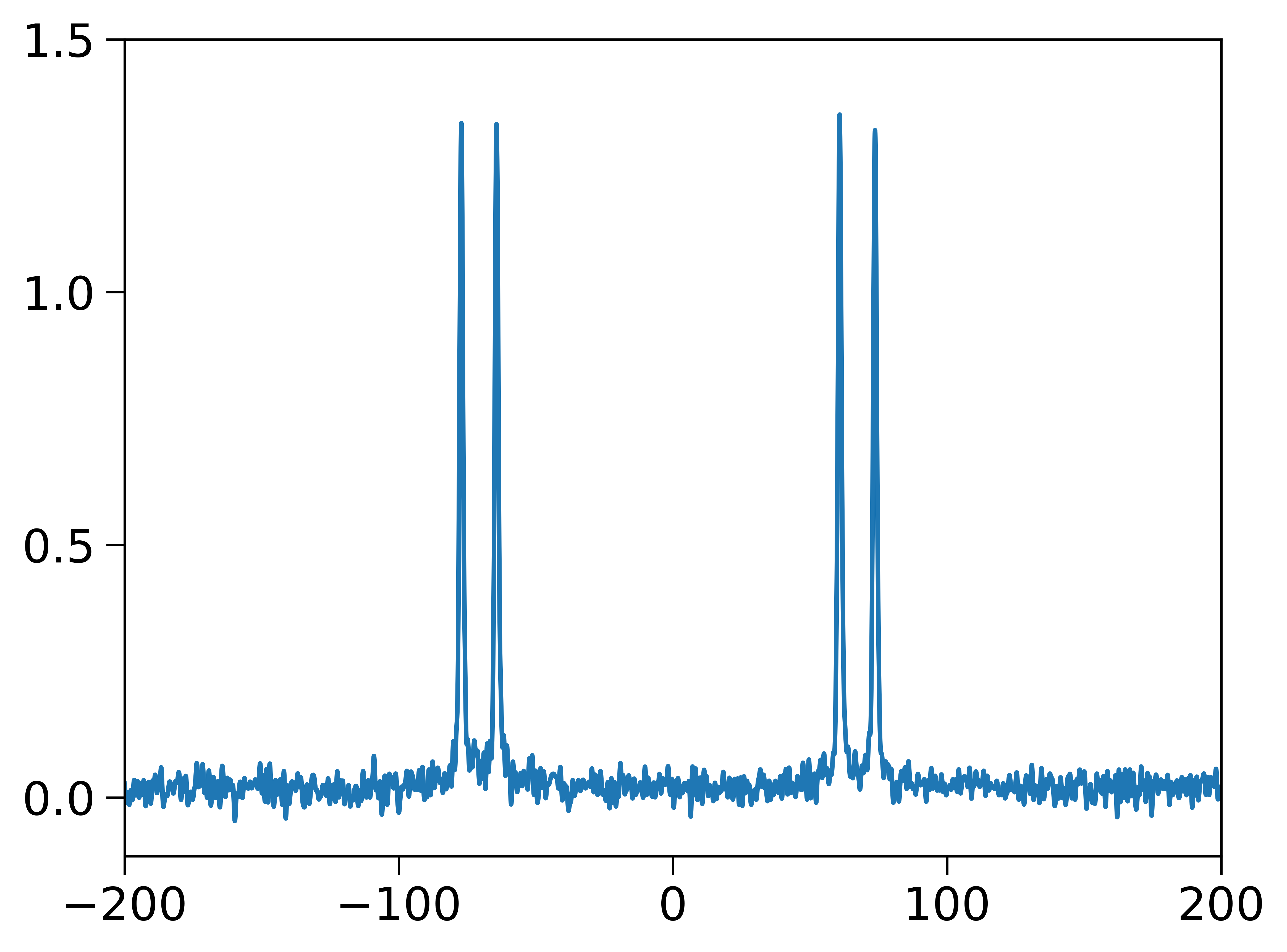}};
    \node[align=center,fill=white] at (-3.3, 3.0) {\textbf{(c)}};
    \node[align=center,fill=white] at (-3.9, 0.3) {{\rotatebox {90}{\text{$g(\omega)$ [a.u.]} }}};
    \end{tikzpicture}
    \text{\qquad \quad $\omega/2\pi$ [Hz] }
    \caption{
\label{fig1}
 (a)  Diethylphthalimidomalonate($-2-^{13}$C,$^{15}$N molecule with the measured pair of nuclear spins $^1$H~--$^{\ 13}$C highlighted by colored circles.  (b) Spectrum $g(\omega)$ of the  highlighted $^1$H, where $\omega$ is the frequency offset. The peak doublet originates from the  $J_{\text{HC}}$-coupling. The inset shows the spectrum of all $^1$H nuclei belonging to the molecule, with the frequency range of the main panel indicated by the gray background.    (c) Spectrum $g(\omega)$  of the highlighted $^{13}$C consisting of the $J_{\text{HC}}$-split doublet with further small splitting due to the coupling to the nearby $^{15}$N spin, whose influence on the measured pair is discussed in  Appendix~\ref{N15}.
}
\end{figure}

We investigate experimentally and theoretically a pair of adjacent nuclear spins 1/2, $^1$H and $^{13}$C, belonging to the molecule diethylphthalimidomalonate($-2-^{13}$C,$^{15}$N) in the solution C$_6$D$_6$. The molecular structure is shown in Fig.~\ref{fig1}(a) with the selected $^1$H~--$^{\ 13}$C pair highlighted. The molecule is 99.5 percent enriched by $^{13}$C at the highlighted position and by $^{15}$N. The $J$-coupling constant for the $^1$H~--$^{\ 13}$C pair is $J/(2 \pi \hbar) \equiv J_{\text{HC}}/(2 \pi \hbar) = 138$~Hz.
The $J$-couplings of  $^{13}$C and $^1$H  to $^{15}$N are much smaller: $J_{\text{CN}} /(2 \pi  \hbar) = 13.1$~Hz and $J_{\text{HN}} /(2 \pi  \hbar) = 1.7$~Hz, respectively. The $J$-coupling of the $^1$H~--$^{\ 13}$C pair to other nuclear spins belonging to the molecule is even weaker, as it was not resolved in the measured spectra.  
The equilibrium NMR spectra of $^1$H and $^{13}$C are shown in Figs.~\ref{fig1}(b,c). All $^1$H spectra in this work originate from a single measurement, while the $^{13}$C spectra are the averages over four measurements.

The NMR experiments were performed using NMR spectrometer Varian Inova 500 with the magnetic field $B  = 11.7 $ T corresponding to the $^1$H Larmor frequency $\Omega_1 = 500 $~MHz. 

As explained in Appendix~\ref{N15}, the influence of $^{15}$N adjacent to the selected $^1$H~--$^{\ 13}$C pair on the relaxation of that pair can be neglected in our experiments.

\section{Theory}
\label{theoretical}
\subsection{General formulation}
We consider two nuclear spins $1/2$,  $\mathbf{S}_1$ and $\mathbf{S}_2$ (corresponding respectively to $^1$H and $^{\ 13}$C)  with different gyromagnetic ratios $\gamma_1$ and $\gamma_2$  belonging to a molecule in a liquid placed in a strong static magnetic field $B$ directed along the $z$-axis. These spins will be referred to as the ``central pair.'' 
Thermal fluctuations in the liquid lead to random reorientations of the molecule on the time scale much faster than the Larmor precession periods of either nucleus.  The secular  part of the time-averaged spin-pair Hamiltonian in the laboratory reference frame is 
\begin{equation}
\label{1}
    \mathcal{H}_{\text{av}} = - \hbar\Omega_1 S_{1z} - \hbar\Omega_2 S_{2z} + J {S}_{1z} {S}_{2z},
\end{equation}
where $\Omega_1 = \gamma_1 B$ and $\Omega_2 = \gamma_2 B$ are the Larmor frequencies, and $J$ is the constant of the $J$-coupling. In the rest of the article, we often  use the ``double-rotating" reference frame, where spin coordinates rotate with respect to the $z$-axis of the laboratory frame with frequencies $-\Omega_1$ or $-\Omega_2$ for spins $\mathbf{S}_1$ and $\mathbf{S}_2$, respectively; we do this  without changing the notations for spin variables ${S}_{nx}$ and ${S}_{ny}$.

The Hamiltonian of the spin pair also includes the fluctuating part, which, while averaging to zero, determines the relaxation of the system: 
\begin{equation}
    \mathcal{H}_{\text{fluct}} = \mathcal{H}^d + \mathcal{H}^{\alpha}
    \label{Hmicro}
\end{equation}
where $\mathcal{H}^d$ is  the fluctuating part of the intra-pair spin-spin interaction,  and $\mathcal{H}^{\alpha}$ the coupling to  the environment of the pair. 
We assume that the intra-pair term $\mathcal{H}^d$ is dominated by IPMDI, which, in the laboratory reference frame has the form
\begin{equation}
    \label{Hd}
    \mathcal{H}^{d}_{\text{lab}} = 
\frac{\gamma_1 \gamma_2 \ \hbar^2}{r^3} \left\{   
    \mathbf{S}_1 \cdot \mathbf{S}_2 
    - 3 \ \frac{(\mathbf{S}_1 \cdot \mathbf{r}) \ (\mathbf{S}_2 \cdot \mathbf{r})}{r^2}
    \right\} ,
\end{equation} 
where $\mathbf{r}$ is the displacement vector between the  spins $\mathbf{S}_1$ and $\mathbf{S}_2$. 
The coupling $\mathcal{H}^{\alpha}$ to the spin-pair environment   is to be represented in the laboratory reference frame as 
\begin{equation}
\mathcal{H}^{\alpha}_{\text{lab}} = \hbar \sum\limits_{n = 1, 2} \alpha_{nx}(t) S_{nx} + \alpha_{ny}(t) S_{ny} + \alpha_{nz}(t) S_{nz}
\label{Halpha}
\end{equation}
where $\boldsymbol{\alpha}_n(t)$ denotes the the total fluctuating local field, classical or quantum,  acting on spin $\mathbf{S}_n$ and defined in units of frequency $\omega$.  The field $\boldsymbol{\alpha}_n(t)$   can be decomposed as follows:
\begin{equation}
\boldsymbol{\alpha}_n =  \boldsymbol{\alpha}_n^{\text{\tiny CSA}} + \boldsymbol{\alpha}_n^{\text{\tiny D}} +
\boldsymbol{\alpha}_n^{\text{\tiny J}} + \boldsymbol{\alpha}_n^{\text{\tiny SR}} + \boldsymbol{\alpha}_n^{\text{\tiny E}} 
\label{alpha-terms}
\end{equation}
where 
$\boldsymbol{\alpha}_n^{\text{\tiny CSA}}$ is associated with fluctuating chemical shift anisotropy (CSA),
\begin{equation}
\boldsymbol{\alpha}_n^{\text{\tiny D}} =  \hbar \gamma_n \sum_{k}^{\text{distant spins}} \gamma_k   
     \ \frac{\mathbf{S}_k \  - \  3 \ (\mathbf{S}_k \cdot \mathbf{r}_{nk}) \  \mathbf{r}_{nk}}{r_{nk}^2}
\label{alpha-D}
\end{equation}
is the magnetic dipole field from distant nuclear spins $\mathbf{S}_k$ -- in particular, other $^1$H on the same molecule, 
\begin{equation}
\boldsymbol{\alpha}_n^{\text{\tiny J}} \ = \mathbf{n}_z \sum_{k}^{\text{distant spins}} J_{nk} S_{kz} 
    \label{alpha-J}
\end{equation} 
is the local field due to $J$-couplings $J_{nk}$ to distant spins, with $\mathbf{n}_z$ being the unit vector along the $z$-axis,
$\boldsymbol{\alpha}_n^{\text{\tiny SR}}$  the field due to the spin-rotation coupling to the molecular angular momentum, and $\boldsymbol{\alpha}_n^{\text{\tiny E}}$ the field from electronic paramagnetic centers in the liquid. The fields $\boldsymbol{\alpha}_n^{\text{\tiny CSA}}$, $\boldsymbol{\alpha}_n^{\text{\tiny D}}$, $\boldsymbol{\alpha}_n^{\text{\tiny SR}}$, $\boldsymbol{\alpha}_n^{\text{\tiny E}}$ are fairly standard\cite{Kowalewski-2017}. In our system, $\boldsymbol{\alpha}_n^{\text{\tiny CSA}}$ and $\boldsymbol{\alpha}_n^{\text{\tiny D}}$ should be noticeable, $\boldsymbol{\alpha}_n^{\text{\tiny SR}}$ negligible, and $\boldsymbol{\alpha}_n^{\text{\tiny E}}$ absent. The field $\boldsymbol{\alpha}_n^{\text{\tiny J}}$ is, normally, neglected in $\mathcal{H}_{\text{fluct}}$, because the time average of $\boldsymbol{\alpha}_n^{\text{\tiny J}}$ is included in time-averaged Hamiltonians, such as $\mathcal{H}_{\text{av}}$, while its fluctuating component is much smaller than $\boldsymbol{\alpha}_n^{\text{\tiny D}}$. We neglect $\boldsymbol{\alpha}_n^{\text{\tiny J}}$ until Section~\ref{additional}, where we argue that $\boldsymbol{\alpha}_n^{\text{\tiny J}}$ makes a measurable contribution to the relaxation rates not because of the fast fluctuations of the coupling constants $J_{nk}$, but rather because of the slow fluctuations of distant spins $\mathbf{S}_k$ themselves.

We adopt the assumption that the fluctuations of all local fields $\boldsymbol{\alpha}_n(t)$ are independent of the spin configuration of the central pair. Such an assumption is accurate, when the dynamics behind these fluctuations is determined by the energy/frequency scales that are much larger than those associated with the coupling between the central pair and the sources of the local fields. 
Such an assumption is reliable in the case of CSA fluctuations, spin-rotation coupling and, usually, electronic paramagnetic fluctuations. It should also work well for the dipolar and $J$-coupling fields due to distant nuclear spins, when each distant spin has other nearby nuclear spins that would relax it faster than the correlations with the central pair build up --- as is the case for our molecule.  (On the contrary, the above approximation would become problematic when the interaction of a distant spin with the central pair is the dominant relaxation mechanism for that spin.)

The equilibrium density matrix of the central spin pair in both the laboratory and the double-rotating frames is
\begin{equation}
\label{rho-eq}
    \rho_{eq} =  \frac{1}{4} \mathbb{I} + \varepsilon_1 S_{1z} + \varepsilon_2 S_{2z}  ,
\end{equation}
where $\mathbb{I}$ is a $4 \times 4$ identity operator,  $\varepsilon_1 = \hbar\Omega_1/(4 k_B T)  $,  \mbox{$\varepsilon_2 = \hbar\Omega_2/(4 k_B T) $}, with $k_B$ and $T$ being respectively the Boltzmann constant and the temperature. Eq.(\ref{rho-eq}) implies the high-temperature approximation.
Since the term $\mathbb{I}/4$ in the density matrix remains invariant under time evolution and does not contribute to the NMR signal, it is mostly omitted in all formulas for the density matrices  in the rest of this article. The general $4 \times 4$ density matrix of the two-spin system without $\mathbb{I}/4$ is to be denoted as $\rho$. 

We will perform conventional inversion recovery experiments initialized by $\pi$ radio-frequency (rf) pulses on one of the two spins, resulting in the initial density matrices \mbox{$\rho(0) = - \varepsilon_1 S_{1z} + \varepsilon_2 S_{2z} $} or $\rho(0) = \varepsilon_1 S_{1z} - \varepsilon_2 S_{2z} $, and also conduct less conventional measurements initialized by Bell PPSs, which we introduce in the next subsection.

\subsection{Pseudo-pure Bell states}
\label{Bell}

The spin pair is defined to be in a  pseudo-pure state (PPS) when its full density matrix has form \mbox{$\mathbb{I}/4 + \kappa |\Psi \rangle \langle\Psi | $}, where $|\Psi \rangle$ is a wave function of some pure quantum state and $\kappa$ is a coefficient. The equilibrium density matrix (\ref{rho-eq}) is not a PPS but rather a generic mixed state. As mentioned in the introductory section, we will use the term ``Bell pseudo-pure state'' (Bell PPSs), when $|\Psi \rangle$ is a maximally entangled wave function. For the very small values of $\kappa$ typical of NMR experiments,  the full density matrix $\mathbb{I}/4 + \kappa |\Psi \rangle \langle\Psi | $  is not entangled in the strict quantum-informational sense, even when $|\Psi \rangle \langle\Psi |$ is maximally entangled, because the former  can be represented as a sum of the density matrices of nonentangled pure states\cite{Braunstein-1999,Fine-2005}. Such a mathematical definition of nonentanglement of a mixed state has, however, rather limited utility in terms of measurable properties of condensed matter systems: A physical system does not on its own explore all possible unravelings of a mixed state in terms of entangled or nonentangled pure states, and, as a result, the abrupt appearance or disappearance of thereby defined mixed-state entanglement does not, normally, affect other physical correlations in the system\cite{Yu-2009}. On the other hand, the entanglement of a pure quantum state is a more robust physical variable, which often correlates with other observable physical properties due to the entangled character of the energy eigenstates and/or due to the symmetry of the coupling to the environment. An NMR experiment is only sensitive to the entangled pure part $|\Psi \rangle \langle\Psi | $ of a Bell PPS independent of the value of $\kappa$. In this work, the maximum entanglement of the pure part of Bell PPSs will signify  particularly symmetric initial quantum states, which will streamline the theoretical analysis and also initialize some of the relaxational eigenmodes of the system. 

As the quantum basis of the spin pair, we use one singlet and three triplet states (defined in the double rotating frame):  
\begin{eqnarray}
        |T_{1,z}\rangle &=& |\uparrow_z \uparrow_z \, \rangle, 
        \label{T1z} \\
        |T_{0,z}\rangle &=& \frac{1}{\sqrt{2}}(|\uparrow_z \downarrow_z \,\rangle  + |\downarrow_z \uparrow_z \,\rangle),
        \label{T0z} \\
         |S_{0}\rangle &=& \frac{1}{\sqrt{2}}(|\uparrow_z \downarrow_z \,\rangle  - |\downarrow_z \uparrow_z \,\rangle),
         \label{S0z}\\
        |T_{-1,z}\rangle &=& |\downarrow_z \downarrow_z \,\rangle. 
        \label{T-1z}
\end{eqnarray} 
where spin projections are quantized along the $z$-axis, $|T_{s,z}\rangle$ are the symmetric triplet states and $|S_{0}\rangle$ is the antisymmetric singlet state. In the following, we will often refer to the decomposition of the above Hilbert space into the so-called ``zero quantum" (ZQ) space with the basis $|T_{0,z}\rangle$ and 
$|S_{0}\rangle$, and the ``double quantum" (DQ) space with the basis $|T_{+1,z}\rangle $ and $|T_{-1,z}\rangle $. 

There exist four mutually orthogonal Bell states of the spin pair: two in the ZQ space, namely, $|S_{0}\rangle$ and $|T_{0,z}\rangle$ defined above, and also two more in the DQ space: 
\begin{equation} 
        \vert \psi_{+,z}\rangle = \frac{1}{\sqrt{2}} (\vert \uparrow_z \uparrow_z \, \rangle +\vert \downarrow_z \downarrow_z \, \rangle)
        \label{psi+z} 
\end{equation}
and
\begin{equation} 
        \vert \psi_{-,z}\rangle = \frac{1}{\sqrt{2}} (\vert \uparrow_z \uparrow_z \, \rangle - \vert \downarrow_z \downarrow_z \, \rangle). 
        \label{psi-z}
\end{equation} 
Density matrices of the four Bell states can be represented in the operator form as:
\begin{eqnarray}
                \vert S_0 \, \rangle \langle \, S_0 \vert &=&
          \frac{1}{4} \mathbb{I} - S_{1z}S_{2z} - (S_{1x}S_{2x} + S_{1y}S_{2y}), \, 
          \label{rhoBell-S0} \\
          \vert T_{0,z} \, \rangle \langle \, T_{0,z} \vert &=&
          \frac{1}{4} \mathbb{I} - S_{1z}S_{2z} + (S_{1x}S_{2x} + S_{1y}S_{2y}), \,
          \label{rhoBell-T0} \\ 
          \vert \psi_{+,z} \, \rangle \langle \, \psi_{+,z} \vert &=&
          \frac{1}{4} \mathbb{I} + S_{1z}S_{2z} + (S_{1x}S_{2x} - S_{1y}S_{2y}), 
          \label{rhoBell-psi+}\\
          \vert \psi_{-,z} \, \rangle \langle \, \psi_{-,z} \vert &=&
          \frac{1}{4} \mathbb{I} + S_{1z}S_{2z} - (S_{1x}S_{2x} - S_{1y}S_{2y}), \ \ \ \ \ \ \ \ 
          \label{rhoBell-psi-}
  \end{eqnarray}
The nontrivial terms of  the above density matrices are $\pm S_{1z}S_{2z}$ and $\pm (S_{1x}S_{2x} \pm S_{1y}S_{2y})$. When the above Bell states constitute a pure part of a PPS, they lead to measurable non-zero averages  $\langle S_{1z}S_{2z} \rangle$, $\langle S_{1x}S_{2x} \rangle$ and $\langle S_{1y}S_{2y} \rangle$. The relaxation behavior of $\langle S_{1z}S_{2z} \rangle (t)$ and $\langle S_{1x}S_{2x} \rangle(t)$ is to be described in the next subsection in a broader theoretical context.

\subsection{Redfield theory for the relaxation of two-spins-1/2 density matrix in the operator-expansion formalism}
\label{relaxation-theory}

\subsubsection{General aspects}
\label{G-aspects}

As mentioned in the introduction, the density matrix $\rho$ of  a two-spin-1/2 system is determined by 15 real numbers that can be chosen as 15 coefficients in the product operator formalism \cite{SORENSEN1984163}:
\begin{equation}
\rho = \sum_{l=1}^{15} v_l P_l,
\label{rho-expansion}
\end{equation}
where $P_l$ is one of 15 ``basis operators'': 
\begin{equation}
\begin{split}
\{S_{1x}, \, S_{1y}, \, S_{1z}, \, S_{2x}, \, S_{2y}, \, S_{2z}, \\
2S_{1x}S_{2x}, \, 2S_{1x}S_{2y}, \, 2S_{1x}S_{2z},  \\
2S_{1y}S_{2x}, \, 2S_{1y}S_{2y}, \, 2S_{1y}S_{2z}, \\ 
2S_{1z}S_{2x}, \, 2S_{1z}S_{2y}, \, 2S_{1z}S_{2z}\}
\end{split}
\label{operators}
\end{equation}
obeying the orthonormality condition $\Tr \{P_l P_m\} = \delta_{lm}$, and $v_l = \Tr \{P_l \rho\}$ is the corresponding expansion coefficient. The coefficients $v_l$  are thus simultaneously the average values $ \langle P_l \rangle_{\rho}$ of the respective basis operators. According to Redfield's theory\cite{Redfield}, the relaxation 
of the parameters $v_l$ obeys a set of  coupled first-order linear differential equations \cite{Slichter-96}
\begin{equation}
\frac{d v_m}{dt} = \sum_l \Gamma_{ml} (v_l - v_l^{\text{eq}}) ,
\label{v-relaxation}
\end{equation}
where $v_l^{\text{eq}}$ are the equilibrium values of $v_l$, and $\Gamma_{ml}$ are the relaxation rates that can be computed using the master equation \cite{Redfield,Slichter-96}: 
\begin{equation}
      \label{master}
      \frac{\partial \rho}{\partial t} = - \frac{1}{\hbar^2} \int\limits_0^t \overline{\bigl[[\rho, \mathcal{H}(t^\prime)], \mathcal{H}(t) \bigl] }\, dt^\prime .
  \end{equation}
(Note: the traditional notations of the Redfield's theory use two-index variables $\rho_{\alpha \beta}$ for the elements of the density matrix  instead of  $v_l$, and hence four-index rates $\Gamma_{\alpha \alpha' \beta \beta'}$ instead of $\Gamma_{ml}$.)

In order to compute the rates, let us transform the fluctuating Hamiltonians $\mathcal{H}^{d}$ and $\mathcal{H}^{\alpha}$ from their respective laboratory-frame representations (\ref{Hd}) and (\ref{Halpha})  to the double rotating reference frame.  For $\mathcal{H}^{d}$, we obtain  
\begin{equation}
    \label{dipolar}
    \mathcal{H}^{d}(t) = \hbar \left[ \mathcal{K}_0(t) + \mathcal{K}_1(t) + \mathcal{K}_2(t) \right], 
\end{equation} 
where
\begin{equation}
\label{A(t)}
 \mathcal{K}_0(t) \! = \! F_0(t) \! \left( \! S_{1z}S_{2z} \! - \! \frac{1}{4}\left[ e^{i(\Omega_1-\Omega_2)t}
 S_{1+}S_{2-} \! + \! \text{h.c.}\right]\right)   
\end{equation}
\begin{equation}
\label{B(t)}
\mathcal{K}_1(t) = F_1(t) (e^{i\Omega_1 t} S_{1+}S_{2z} + e^{i\Omega_2 t} S_{1z} S_{2+}) + \text{h.c.} 
\end{equation}
\vspace{-0.4cm}
\begin{equation}
\label{C(t)}
\mathcal{K}_2(t) = F_2(t) e^{i(\Omega_1 + \Omega_2)t}S_{1+}S_{2+} + \text{h.c.} 
\end{equation}
with $ S_{n\pm} \equiv S_{nx} \pm i S_{ny}$. The coefficients in Eqs.(\ref{A(t)}-\ref{C(t)}) are \cite{Solomon}:
\begin{equation}   
       F_0(t) = k (1 - 3\cos^2 \theta(t)),  
\label{F0}
\end{equation}
\begin{equation}  
       F_1(t) = - \frac{3}{2} k \sin \theta(t) \cos \theta(t) e^{i\varphi(t)}, 
\label{F1}
\end{equation}
\begin{equation}
       F_2(t) = -\frac{3}{4} k \sin^2\theta(t) e^{2i\varphi(t)}, 
\label{F2}
\end{equation}
where 
\begin{equation}
    k = \gamma_1 \gamma_2 \hbar /r^3,
    \label{k}
\end{equation} 
$\theta(t), \varphi(t)$ and $r$ are the two spherical angles and the length of the displacement vector $\mathbf{r}$ between the two nuclei. The transformation of $\mathcal{H}^{\alpha}$ into the double-rotating frame converts (\ref{Halpha}) into
\begin{equation}
\mathcal{H}^{\alpha}(t) = \hbar \sum\limits_{n = 1, 2} [\alpha_{n\perp}(t) e^{i\Omega_n t}S_{n+} + \text{h.c.}] + \alpha_{nz}(t) S_{nz}.
\label{CSA}
\end{equation}
where $\alpha_{n\perp} \equiv \frac{1}{2} (\alpha_{nx} - i \alpha_{ny})$.

We assume, until stated otherwise, that the primary source of time-dependent fluctuations in the system is the rapid random reorientations  of the molecule, which control the coefficients $F_0(t)$, $F_1(t)$, $F_2(t)$  and also $\boldsymbol{\alpha}_n^{\text{\tiny CSA}}(t)$ and $\boldsymbol{\alpha}_n^{\text{\tiny D}}(t)$. (Thereby, we neglect $\boldsymbol{\alpha}_n^{\text{\tiny J}}$, $\boldsymbol{\alpha}_n^{\text{\tiny SR}}$ and $\boldsymbol{\alpha}_n^{\text{\tiny E}}$). 
Under this assumption, the time correlators of all relevant quantities are proportional to the same correlation function $C(\tau)$ 
decaying on the timescale $\tau_c$ characterizing molecular reorientations and normalized such that $C(0)=1$, which means that:  
$\langle F_j(t) F_j^*(t + \tau) \rangle = \langle \vert F_j\vert^2\rangle C(\tau)$,  \mbox{$\langle \alpha_{nj}(t) \alpha_{nj}^*(t + \tau) \rangle = \langle \vert \alpha_{nj}\vert^2\rangle C(\tau)$}, and also $\langle \alpha_{nj}(t) F_l^*(t + \tau) \rangle = \langle \alpha_{nj} F_l^*\rangle C(\tau)$. (The absence of the time argument inside $\langle ... \rangle$ implies same-time average.)
The Fourier transform of $C(\tau)$ is  the spectral function
\begin{equation}
    \mathcal{J}(\omega) \equiv \int\limits_{- \infty}^{+ \infty} C(\tau) e^{- i \omega \tau}  d\tau.
    \label{Jomega}
\end{equation}
(For the common choice $C(\tau) =  e^{-|\tau|/\tau_c}$ \cite{BPP, Abragam-61}, $\mathcal{J}(\omega) = \frac{2 \tau_c}{1 + \omega^2 \tau_c^2}$.)  
We also define variable $\mathcal{J}_0 \equiv \mathcal{J}(0)$ to simplify the formulas in the limit $\Omega_1, \Omega_2 \ll 1/\tau_c$ representative of not~too~large molecules, such as the one we investigate.


Our experiments will directly monitor the time evolution (i.e. relaxation) of  four of the 15 expansion coefficients $\{v_l\}$, namely, those associated with operators $S_{1z}$, $S_{2z}$, $2 S_{1z}S_{2z}$ and $2 S_{1x}S_{2x}$. 
Since we are not pursuing the complete characterization of the density matrix, we now switch to an intuitive representation of the respective coefficients $v_l$ as $\langle S_{1z} \rangle$, $\langle S_{2z} \rangle$,  $\langle 2 S_{1z}S_{2z} \rangle$ and $\langle 2 S_{1x}S_{2x} \rangle$. 
The time dependencies $\langle S_{1z} \rangle(t)$, $\langle S_{2z} \rangle(t)$ and  $\langle 2 S_{1z}S_{2z} \rangle(t)$ will characterize the relaxation of the diagonal terms of $\rho$ in the $z$-basis - to be referred to as the ``diagonal relaxation", while $\langle 2 S_{1x}S_{2x} \rangle(t)$ will partially characterize the ``off-diagonal relaxation".
We will also replace the rate matrix $\Gamma_{ml}$ by individual variables for each rate affecting the measured quantities. 

\subsubsection{Diagonal relaxation}
\label{D-relaxation}

In a general setting where both ${\cal H}^{d}$ and ${\cal H}^{\alpha}$ are present, $\langle S_{1z} \rangle$,  $\langle S_{2z} \rangle$ and $\langle 2 S_{1z}S_{2z} \rangle$ do not relax independently --- rather they are linearly coupled according to the following system of equations~\cite{Goldman1984,KUMAR2000191,GHOSH2005125,Kowalewski-2017}: 
\begin{widetext}
 \begin{equation}
     \label{C17}
     \frac{d}{dt}\begin{pmatrix} \langle S_{1z} \rangle(t) \\ \langle S_{2z} \rangle(t) \\ \langle 2 S_{1z}S_{2z} \rangle(t) \end{pmatrix} = -
\begin{pmatrix} \mu_1 & \sigma_{12} & \delta_1 \\ \sigma_{12} & \mu_2 &\delta_2 \\ \delta_1 & \delta_2 & \mu_{12} \end{pmatrix}
\begin{pmatrix} \langle S_{1z} \rangle(t) - \langle S_{1z} \rangle_{\text{eq}} \\ \langle S_{2z} \rangle(t) - \langle S_{2z} \rangle_{\text{eq}} \\ \langle 2S_{1z}S_{2z} \rangle(t) \end{pmatrix},
  \end{equation}
\end{widetext}
where 
\begin{equation}
\langle S_{nz} \rangle_{\text{eq}} = \varepsilon_n
\label{Snz_eq}
\end{equation}
are the equilibrium spin polarisations determined by Eq.(\ref{rho-eq}), while the matrix consists of the following relaxation rates: $\mu_1$ and $\mu_2$ characterize the usual ``$T_1$-relaxation" of the respective nuclear spins, $\mu_{12}$ is the diagonal relaxation rate of the so-called ``two-spin ZZ-coherence", $\sigma_{12}$ is the cross-relaxation rate associated with NOE,  $\delta_1$ and $\delta_2$ are the cross-relaxation rates connecting single-spin and two-spin coherences. Here and below index $n$ takes values 1 or 2. 

The rates in Eq.(\ref{C17}) can be computed as follows (see Appendix~\ref{derivations}): 
\begin{equation}
\mu_n = \mu_n^d + \mu_n^{\alpha}, 
    \label{mu_n_tot}
\end{equation}
where
\begin{equation}
\mu_n^{d} =  \frac{\langle F_0 ^2\rangle}{16} \mathcal{J}(\Omega_1 - \Omega_2) + \frac{\langle\vert F_1 \vert^2\rangle}{2} \mathcal{J}(\Omega_n) + \langle\vert F_2 \vert^2\rangle \mathcal{J}(\Omega_1 + \Omega_2)
\label{mun_dd}
\end{equation}
and
\begin{equation}
\mu_n^{\alpha} =  2\langle\vert \alpha_{n\perp} \vert^2\rangle \mathcal{J}(\Omega_n);
\label{mun_alpha} 
\end{equation}
\begin{equation}
\sigma_{12} =  -\frac{\langle F_0 ^2\rangle}{16} \mathcal{J}(\Omega_1 - \Omega_2)  + \langle\vert F_2 \vert^2\rangle \mathcal{J}(\Omega_1 + \Omega_2);
\label{sigma12}
\end{equation}
\begin{equation}
\mu_{12} = \mu_{12}^d + \mu_{12}^{\alpha}, 
    \label{mu_12_tot}
\end{equation}
where  
\begin{equation}
\mu_{12}^{d} = \frac{1}{2} \langle\vert F_1 \vert^2\rangle \left[\mathcal{J}(\Omega_1) + \mathcal{J}(\Omega_2)\right],
\label{mu12}
\end{equation}
\begin{equation}
\mu_{12}^{\alpha} =  2 \langle\vert \alpha_{1\perp} \vert^2\rangle \mathcal{J}(\Omega_1) + 2 \langle\vert \alpha_{2\perp} \vert^2\rangle \mathcal{J}(\Omega_2);
\label{mu12_alpha}
\end{equation}
and also
\begin{equation}
\delta_{n} =  \langle F_1 \alpha_{n\perp}^* + F_1^*\alpha_{n\perp}\rangle \mathcal{J}(\Omega_n).
\label{delta_n}
\end{equation}

As one can see from Eq.(\ref{delta_n}), the cross-relaxation  rates $\delta_{n}$  originate from the so-called ``cross-correlations" between the Hamiltonians  ${\cal H}^{d}$ and ${\cal H}^{\alpha}$ representing different relaxation mechanisms\cite{Goldman1984,KUMAR2000191,GHOSH2005125,Kowalewski-2017}. Without these cross-correlations,  $\langle 2 S_{1z}S_{2z} \rangle$ would be completely decoupled from $\langle S_{1z} \rangle$ and  $\langle S_{2z} \rangle$, which would lead to the Solomon equations\cite{Solomon} for the latter two quantities.  Among the two kinds of local fields retained so far, namely, $\boldsymbol{\alpha}_n^{\text{\tiny CSA}}$ and $\boldsymbol{\alpha}_n^{\text{\tiny D}}$, only the former can make non-zero contribution to $\langle F_1 \alpha_{n\perp}^* + F_1^*\alpha_{n\perp}\rangle$, because both $\boldsymbol{\alpha}_n^{\text{\tiny CSA}}(t)$ and $F_1(t)$ are  determined by the second order angular harmonics and averaged over the same molecular rotations.  The contribution to that average  from $\boldsymbol{\alpha}_n^{\text{\tiny D}}$ is always zero, because, according to Eq.(\ref{alpha-D}), each term in $F_1 {\alpha_{n\perp}^{\text{\tiny D}^*}} + F_1^*\alpha_{n\perp}^{\text{\tiny D}}$ is linear in terms of projections of distant spins $\mathbf{S}_k$, which are not correlated with molecular reorientations, while $\langle \mathbf{S}_k \rangle = \mathbf{0}$ in the infinite temperature limit. 

Let us also remark here that, independent of a concrete mechanism, the cross-correlation average in Eq.(\ref{delta_n}) is the subject to the mathematical inequality \begin{equation}
    |\langle F_1 \alpha_{n\perp}^* + F_1^*\alpha_{n\perp}\rangle|  \ \leq \ 
    2 \sqrt{\langle |F_1|^2 \rangle \ \langle |\alpha_{n\perp}|^2    \rangle},  \label{F1_alpha-n_inequality}
\end{equation} 
which we will test experimentally.

We further note that, according to Eqs.(\ref{mu_n_tot}-\ref{mu12_alpha}), the combination $\mu_1  \,  +  \, \mu_2  \,  - \, \mu_{12}$ is controlled exclusively by the intra-pair magnetic dipole Hamiltonian ${\cal H}^{d}$. Since the same is also true for $\sigma_{12}$, one can use Eqs.(\ref{F0}-\ref{F2}) in the standard limit $\Omega_1, \Omega_2 \ll 1/\tau_c$ to obtain  the following fundamental parameter-free relation for spin pairs where the intra-pair coupling is dominated by IPMDI (see Appendix~\ref{derivations}): 
\begin{equation}
    \frac{\mu_1  \,  +  \, \mu_2  \,  - \, \mu_{12}}{\sigma_{12}} = 2.8 .
    \label{mu-sigma_ratio}
\end{equation}

Extracting the six rates entering the relaxation matrix in (\ref{C17}) from the overall fit to  relaxation measurements is rather impractical due to difficult-to-quantify uncertainties.  Therefore, instead of the overall fits, we will be extracting the rates $\mu_1$, $\mu_2$, $\sigma_{12}$,  $\delta_1$ and $\delta_2$ from the initial slopes of specially selected measurable relaxation functions, such that each of these slopes is determined by only one of the above rates.  Extracting the remaining rate $\mu_{12}$ will require the initial slopes of two more relaxation functions. Our overall scheme of measurements is summarized in Table~\ref{tbl:rates}. We now explain the content of this scheme in detail.

\setlength{\tabcolsep}{6pt} 
\renewcommand{\arraystretch}{2} 
\begin{table*}[t]
     \begin{center}
     \begin{tabular}{ | c | c | c | c | c | c | }   
     \hline
     Rate & \makecell{Initial \\ conditions} & \makecell{Monitored \\ variable} & \makecell{Experimental pulse and\\ measurement sequence} & \makecell{Figure \\ number} & \makecell{Measured  \\ value $\left[ \text{s}^{-1} \right]$} \\
     \hline
     \multicolumn{6}{|c|}{Inversion recovery and NOE} \\
     \hline
     $\mu_1$ & \multirow{3}{*}{\makecell{ $\ \ \ \ \ \ \ \ \ \langle S_{1z} \rangle(0) = - \varepsilon_1$ \vspace{4pt} \\ $\ \ \ \ \ \ \ \langle S_{2z} \rangle(0) =  \varepsilon_2$ \vspace{4pt} \\ $\langle 2 S_{1z}S_{2z} \rangle(0) = 0$ }} & $\langle S_{1z} \rangle(t)$ & $(\pi)_{1y}  \  \text{--} \ t \text{ --}   \  (\pi/2)_{1y} \ \text{-- S1} $  & \ref{fig:mu1-mu2-sig12}(a) & $0.50 \pm 0.01$  \\
     \cline{1-1} \cline{3-6}  
     $\delta_1$ & & $\langle 2 S_{1z}S_{2z} \rangle(t)$ & $(\pi)_{1y}  \  \text{--} \ t \text{ --}   \  (\pi/2)_{1y} \ \text{-- A1}$ & \ref{fig:delta}(a1) & $0.0159 \pm 0.0008$  \\
     \cline{1-1} \cline{3-6}
     $\sigma^{(1)}_{12}$ & & $\langle S_{2z} \rangle(t)$ & $(\pi)_{1y}  \  \text{--} \ t \text{ --}   \  (\pi/2)_{2y} \ \text{-- S2} $ & \ref{fig:mu1-mu2-sig12}(d) & $0.179 \pm 0.003$  \\
     \hline
     $\mu_2$ & \multirow{3}{*}{\makecell{$\ \ \ \ \ \ \langle S_{1z} \rangle(0) = \varepsilon_1$ \vspace{4pt} \\ $\ \ \ \ \ \ \ \ \ \langle S_{2z} \rangle(0) =  - \varepsilon_2$ \vspace{4pt} \\ $\langle 2 S_{1z}S_{2z} \rangle(0) = 0$ }} & $\langle S_{2z} \rangle(t)$ & $(\pi)_{2y}  \  \text{--} \ t \text{ --}   \  (\pi/2)_{2y} \ \text{-- S2} $ & \ref{fig:mu1-mu2-sig12}(b) & $0.41 \pm 0.02$  \\
     \cline{1-1} \cline{3-6}  
     $\delta_2$ & & $\langle 2 S_{1z}S_{2z} \rangle(t)$ & $(\pi)_{2y}  \  \text{--} \ t \text{ --}   \  (\pi/2)_{2y} \ \text{-- A2} $ & \ref{fig:delta}(b1) & $-0.026 \pm 0.004$  \\
     \cline{1-1} \cline{3-6}
     $\sigma^{(2)}_{12}$ & & $\langle S_{1z} \rangle(t)$ & $(\pi)_{2y}  \  \text{--} \ t \text{ --}   \  (\pi/2)_{1y} \ \text{-- S1} $ & \ref{fig:mu1-mu2-sig12}(c) & $0.199 \pm 0.007 $  \\
     \hline 
     \makecell{ \vspace{3pt} $\displaystyle \sigma_{12} = \frac{\sigma^{(1)}_{12}+\sigma^{(2)}_{12}}{2}$} & \multicolumn{4}{|c|}{} & $0.19 \pm 0.02$ \\
     \hline
     \multicolumn{6}{|c|}{Relaxation starting from Bell pseudo-pure states } \\
     \hline
      & \multirow{4}{*}{\makecell{$ \langle S_{1z} \rangle(0) = 0$ \vspace{4pt}\\ $\langle S_{2z} \rangle(0) =  0$ \vspace{4pt}\\ $\langle 2 S_{1z}S_{2z} \rangle(0) = - \frac{\varepsilon_1 + \varepsilon_2}{4}$ \vspace{4pt}\\ $\langle 2 S_{1x}S_{2x} \rangle(0) = \pm \frac{\varepsilon_1 + \varepsilon_2}{4}$ }} & & & & \\ 
     $\mu_{\text{\tiny ZQ}}$ &  & $\langle 2 S_{1z}S_{2z} \rangle(t)$ &  $ | S_0 \rangle$ or $| T_0 \rangle  \  \text{--} \ t \text{ --}   \  (\pi/2)_{1y} \ \text{-- A1} $ & \ref{fig:muZQ-muDQ}(a) & $0.37 \pm 0.01$   \\
     \cline{1-1} \cline{3-6}
     & & & \multirow{2}{*}{  \makecell{ \\ CPMG \ \ \  \\ $ | S_0 \rangle$ or $| T_0 \rangle  \  \text{--} \ t \  \text{--}   \  (\pi/2)_{2y} \ \text{-- A1} $  }}  & & \\
     $\lambda_{\text{\tiny ZQ}}$ & & $\langle 2 S_{1x}S_{2x} \rangle(t)$ &   & \ref{fig:lambdaZQ-lambdaDQ}(b) & $0.326 \pm 0.002$  \\
     \hline
     & \multirow{4}{*}{\makecell{$ \langle S_{1z} \rangle(0) = 0$ \vspace{4pt}\\ $ \langle S_{2z} \rangle(0) =  0$ \vspace{4pt}\\ $\langle 2 S_{1z}S_{2z} \rangle(0) = \frac{\varepsilon_1 + \varepsilon_2}{4}$ \vspace{4pt}\\ $\langle 2 S_{1x}S_{2x} \rangle(0) = \pm \frac{\varepsilon_1 + \varepsilon_2}{4}$ }} & & & & \\ 
     $\mu_{\text{\tiny DQ}}$ &  & $\langle 2 S_{1z}S_{2z} \rangle(t)$ & $ | \psi_+ \rangle$ or $| \psi_- \rangle  \  \text{--} \ t \text{ --}   \  (\pi/2)_{1y} \ \text{-- A1} $ & \ref{fig:muZQ-muDQ}(a) & $0.30 \pm 0.04$  \\
     \cline{1-1} \cline{3-6}
     & & & \multirow{2}{*}{  \makecell{ \\ CPMG \ \   \\ $ | \psi_+ \rangle$ or $| \psi_- \rangle  \  \text{--} \ t \  \text{--}   \  (\pi/2)_{2y} \ \text{-- A1} $  }} & & \\
     $\lambda_{\text{\tiny DQ}}$ & & $\langle 2 S_{1x}S_{2x} \rangle(t)$ & & \ref{fig:lambdaZQ-lambdaDQ}(b) & $0.568 \pm 0.008$  \\
     \hline
     \makecell{ \vspace{3pt} $\displaystyle \mu_{12} = \frac{\mu_{\text{\tiny ZQ}}+\mu_{\text{\tiny DQ}}}{2}$} & \multicolumn{4}{|c|}{} & $0.34 \pm 0.02$  \\
     \hline 
     \end{tabular}
     
     \vspace{5pt}
     \begin{tabular}{ | c | c | c || c | c | c | }   
     \hline 
     \multicolumn{6}{|c|}{\makecell{\vspace{3pt} Tests of microscopic theory}} \\
     \hline
     Dimensionless ratio   & Theory: Eq.(\ref{mu-sigma_ratio})  & Experiment & Dimensionless ratio   & Theory: Eq.(\ref{rate-difference})   & Experiment \\
     \hline
      $\ \ \ \displaystyle    \frac{\mu_1  \,  +  \, \mu_2  \,  - \, \mu_{12}}{\sigma_{12}} $ & 2.8  & $3.0 \pm 0.4$ & \makecell{\vspace{3pt} $ \displaystyle \frac{(\mu_{\text{\tiny ZQ}} - \mu_{\text{\tiny DQ}})(\varepsilon_1 + \varepsilon_2)}{\delta_1\, \varepsilon_1  \,  +  \, \delta_2 \,  \varepsilon_2}$ }   & 8 & $9 \pm 6$   \\
     \hline 
      \end{tabular}
      \caption{Relaxation rates for the density matrix of two spins 1/2, $\mathbf{S}_1$ and $\mathbf{S}_2$, representing, respectively, $^1$H and $^{13}$C nuclei,  together with two parameter-free tests of microscopic theory involving some of the measured rates. Rate notations, initial conditions, monitored variables, pulse sequences and the tests are explained in Section~\ref{relaxation-theory} (except for the values of $\langle 2 S_{1z}S_{2z} \rangle(0)$ and $\langle 2 S_{1x}S_{2x} \rangle(0)$ for the Bell PPSs, which are explained in Section~\ref{preparation}). Figure numbers  refer to the experimental plots, from which the measured rate values were extracted.   }
      \label{tbl:rates}
      \end{center}
      \end{table*}

A relaxation function is determined by the preparation of the initial state and by the variable monitored in the course of the relaxation. The first three columns of Table~\ref{tbl:rates} list the rates of interest, the relevant initial conditions and the monitored variable. The fourth column then gives the experimental pulse sequence delivering the desired relaxation function. Labels S1, S2, A1, or A2 at the end of each pulse sequence indicate the index of the measured spin (1 or 2) and whether we extract the symmetric  or antisymmetric (S or A) spectral component of that spin. The symmetric and the antisymmetric components of the NMR free induction decay (FID) spectrum for the $n$th spin ($\mathbf{S}_1$ or $\mathbf{S}_2$) are to be obtained, respectively, as $I_{n+} + I_{n-}$ and $I_{n+} - I_{n-}$, where $I_{n+}$ and $I_{n-}$ are the intensities of two spectral peaks  originating from the $J$-coupling between the $n$th spin and the other spin of the central pair;  $I_{n+}$ corresponds to the $z$-projection of the other spin equal to  1/2, and $I_{n-}$ to $-1/2$.  

The averages $\langle S_{nz} \rangle(t)$ are to be monitored as the symmetric spectral components $I_{n+}(t) + I_{n-}(t)$ of the FIDs induced by the $(\pi/2)_{ny}$-pulses at time $t$. The average $\langle 2 S_{1z}S_{2z} \rangle(t)$ is  proportional to the antisymmetric component of either of the above FID spectra. Indeed, let us consider the spectrum of the first spin, so that the peak intensities corresponding to   $S_{2z}= 1/2$ and $S_{2z}= -1/2$ are $I_{1+}$ and $I_{1-} $ respectively. While $I_{1+} + I_{1-} \propto \langle S_{1z}  \rangle $, one can find  the individual peak intensities $I_{1+}$ and $I_{1-}$   by multiplying $S_{1z}$ by the operators $(\frac{1}{2}\mathbb{I} \pm S_{2z})$, which make quantum projections to  states with a definite value of $S_{2z}$. This gives  $I_{1\pm} \propto \langle S_{1z}  (\frac{1}{2}\mathbb{I} \pm S_{2z}) \rangle $, which implies that $I_{1+} - I_{1-} \propto\langle 2 S_{1z}S_{2z} \rangle$. (We note that $\langle 2 S_{1z}S_{2z} \rangle$ can also be monitored using the so-called ``double-quantum filter"\cite{Jaccard-1987,Kowalewski-2017}.)  


The measurements of $\mu_1$, $\mu_2$, $\sigma_{12}$,  $\delta_1$ and $\delta_2$ are to be based on the conventional inversion recovery experiments initialized by a $\pi$-pulse on one of the two spins:

(i) Rates $\mu_n$, will be obtained as $\mu_n = - \frac{\langle \Dot{S}_{nz} \rangle(0)}{2 \langle S_{nz} \rangle(0) }$ after the initial pulse $(\pi)_{ny}$,  with $\langle S_{nz} \rangle(t)$ measured as the symmetric spectral component of the FID induced by the $(\pi/2)_{ny}$-pulse at time $t$. [Here and below, dot above the variable indicates the time derivative.]

(ii) Rate $\sigma_{12}$ characterizes NOE and, as such, can be obtained in two ways --- by either applying the \mbox{$(\pi)_{1y}$-pulse} on the first spin and then monitoring $\langle S_{2z} \rangle(t)$, thereby obtaining \mbox{$\sigma_{12}^{(1)} = \frac{\langle \Dot{S}_{2z} \rangle(0)}{\langle S_{2z} \rangle_{\text{eq}} } $},  or, {\it vice versa}, by applying a \mbox{$(\pi)_{2y}$-pulse} on the second spin and then monitoring $\langle S_{1z} \rangle(t)$, which gives $\sigma_{12}^{(2)} = \frac{\langle \Dot{S}_{1z} \rangle(0)}{\langle S_{1z} \rangle_{\text{eq}} } $ . The values of both $\sigma_{12}^{(1)}$ and $\sigma_{12}^{(2)}$ are given in Table~\ref{tbl:rates}; the difference between them characterizes the experimental error, while their average is listed as $\sigma_{12}$.


(iii) Rates $\delta_n$ will be obtained as $\delta_n = \frac{\Dot{\langle 2 S_{1z}S_{2z} \rangle}(0)}{2 \langle S_{nz} \rangle(0) }$, with $\langle 2 S_{1z}S_{2z} \rangle(t)$ monitored using the antisymmetric spectral components in the same experiments that give, respectively, $\mu_1$ and $\mu_2$.  

In order to determine the rate $\mu_{12}$, we will be measuring the initial slope of $\langle 2 S_{1z}S_{2z} \rangle(t)$ for the relaxation following the preparation of the Bell PPSs. In that case, $\langle S_{1z} \rangle (0) = \langle S_{2z} \rangle(0)=0$, while $\langle 2 S_{1z}S_{2z} \rangle(0) \neq 0$, which, according  to (\ref{C17}), gives
\begin{equation}
      \label{T0}
          \frac{\Dot{\langle 2 S_{1z}S_{2z} \rangle}(0)}{ \langle  2 S_{1z}S_{2z} \rangle(0) }  = \mu_{12} 
          - \delta_1 \frac{\langle S_{1z} \rangle_{\text{eq}}}{\langle 2 S_{1z}S_{2z} \rangle(0)} 
          - \delta_2 \frac{\langle S_{2z} \rangle_{\text{eq}}}{\langle 2 S_{1z}S_{2z} \rangle(0)}
          .
  \end{equation}
When $\delta_1 \neq 0$ and/or $\delta_2 \neq 0$, Eq.(\ref{T0}) implies that the initial relaxation rate of $\langle 2 S_{1z}S_{2z} \rangle$ depends on the sign of $\langle 2 S_{1z}S_{2z} \rangle(0)$. We note here that, according to \mbox{Eqs.(\ref{rhoBell-S0}-\ref{rhoBell-psi-}),} $\langle 2 S_{1z}S_{2z} \rangle$ is equal to $-1/2$ for the pure ZQ Bell states and $+1/2$ for the pure DQ states. For the pseudo-pure Bell states to be prepared in our experiments (see Section~\ref{Pulse_sequences} below) \mbox{$\langle 2 S_{1z}S_{2z} \rangle = \pm \frac{\varepsilon_1 + \varepsilon_2}{4}$} --- positive for the DQ and negative for the ZQ states. Using Eqs.(\ref{T0}) and (\ref{Snz_eq}), we thus express the measurable initial relaxation rates appearing in Table~\ref{tbl:rates} as:
\begin{equation}
\mu_{\text{\tiny ZQ}} = \mu_{12} + 4 \  \frac{
\delta_1\, \varepsilon_1  \,  +  \, 
           \delta_2 \,  \varepsilon_2
           }{\varepsilon_1 + \varepsilon_2}
    \label{muZQ}
\end{equation}
and 
\begin{equation}
\mu_{\text{\tiny DQ}} = \mu_{12} - 4 \  \frac{
\delta_1\, \varepsilon_1  \,  +  \, 
           \delta_2 \,  \varepsilon_2
           }{\varepsilon_1 + \varepsilon_2}
    \label{muDQ}
\end{equation}
for the ZQ and DQ states respectively, which then allows us to extract the value of $\mu_{12}$ as
\begin{equation}
\mu_{12} = \frac{\mu_{\text{\tiny ZQ}}+\mu_{\text{\tiny DQ}}}{2} .
    \label{mu12-ZQ-DQ}
\end{equation}
Once all the rates entering Eq.(\ref{C17}) are measured, we will use the difference 
\begin{equation}
\mu_{\text{\tiny ZQ}} - \mu_{\text{\tiny DQ}} \ = \  8 \  \frac{
\delta_1\, \varepsilon_1  \,  +  \, 
           \delta_2 \,  \varepsilon_2
           }{\varepsilon_1 + \varepsilon_2} 
           \label{rate-difference}
\end{equation}
to perform a parameter-free test of the theory --- see Table~\ref{tbl:rates}. 


\subsubsection{Off-diagonal relaxation}
\label{OD-relaxation}

We now focus on  the off-diagonal relaxation associated with $\langle 2 S_{1x}S_{2x} \rangle$, which has been considered in the literature in the context of differential multiple quantum relaxation\cite{Konrat-1993,Tessari-2000,Kowalewski-2017}.  

In principle, many relaxation eigenmodes of the system of equations~(\ref{v-relaxation}) could contribute to $\langle 2 S_{1x}S_{2x} \rangle (t)$. However, given the master equation (\ref{master}) and the concrete form of the Hamiltonians (\ref{Hd}) and (\ref{Halpha}), only two eigenmodes become involved, namely, $S_{1x}S_{2x} \pm S_{1y}S_{2y}$ (see Appendix~\ref{derivations}). The relevant sector of the rate equations (\ref{v-relaxation}) then reads
\begin{widetext}
 \begin{equation}
     \label{T2}
     \frac{d}{dt}\begin{pmatrix} \langle 2 S_{1x}S_{2x}  + 2 S_{1y}S_{2y}\rangle(t) \\ \langle 2 S_{1x}S_{2x}  - 2 S_{1y}S_{2y} \rangle(t)  \end{pmatrix} = -
\begin{pmatrix} \lambda_{\text{\tiny ZQ}} & 0  \\ 0 & \lambda_{\text{\tiny DQ}}   \end{pmatrix}
\begin{pmatrix} \langle 2 S_{1x}S_{2x}  + 2 S_{1y}S_{2y} \rangle(t) \\ \langle 2 S_{1x}S_{2x}  - 2 S_{1y}S_{2y} \rangle(t)  \end{pmatrix},
  \end{equation}
\end{widetext}
where $\lambda_{\text{\tiny ZQ}}$ and $\lambda_{\text{\tiny DQ}}$ are relaxation rates that can be expressed, according to Appendix~\ref{derivations}, as:
\begin{equation}
\lambda_{\text{\tiny ZQ/DQ}} = \lambda_{\text{\tiny ZQ/DQ}}^d + \lambda_{\text{\tiny ZQ/DQ}}^{\alpha},
    \label{lambda_pm_tot}
\end{equation}
with
\begin{equation}    \label{lambda_plus_dd}
    \lambda_{\text{\tiny ZQ}}^{d} =  \frac{\langle F_0^2\rangle}{16} \mathcal{J}(\Omega_1 - \Omega_2) + \frac{\langle \vert F_1 \vert ^2 \rangle}{4} \left[ \mathcal{J}(\Omega_1) + \mathcal{J}(\Omega_2)\right],
\end{equation}
\begin{equation}
  \label{lambda_minus_dd}
    \lambda_{\text{\tiny DQ}}^{d} =  \frac{\langle \vert F_1 \vert ^2 \rangle}{4} \left[ \mathcal{J}(\Omega_1) + \mathcal{J}(\Omega_2)\right] +  \langle \vert F_2 \vert ^2 \rangle \mathcal{J}(\Omega_1 + \Omega_2),
\end{equation}
\begin{equation}
    \label{lambda_plus_alpha}
    \lambda_{\text{\tiny ZQ}}^{\alpha} \! = \! \frac{\langle (\alpha_{1z} - \alpha_{2z})^2 \rangle}{2} \mathcal{J}(0) +  \langle \vert \alpha_{1\perp} \vert^2 \rangle \mathcal{J}(\Omega_1) + \langle \vert \alpha_{2\perp}\vert^2 \rangle \mathcal{J}(\Omega_2),
\end{equation}
and
\begin{equation}
    \label{lambda_minus_alpha}
    \lambda_{\text{\tiny DQ}}^{\alpha}  \! = \!  \frac{\langle (\alpha_{1z} + \alpha_{2z})^2 \rangle}{2} \mathcal{J}(0) +  \langle \vert\alpha_{1\perp}\vert^2 \rangle \mathcal{J}(\Omega_1) + \langle \vert\alpha_{2\perp}\vert^2 \rangle \mathcal{J}(\Omega_2).
\end{equation}
We note that, in the limit $\Omega_1, \Omega_2 \ll 1/\tau_c$, Eqs.(\ref{lambda_plus_dd},\ref{lambda_minus_dd}), together with (\ref{F0}-\ref{F2}),  lead to
\begin{equation}
    \frac{\lambda_{\text{\tiny DQ}}^{d}}{\lambda_{\text{\tiny ZQ}}^{d}} \  = \ \frac{9}{4}
    \label{lambda_d_ratio}
\end{equation}
(see Appendix~\ref{derivations}).

While both eigenmodes in Eq.(\ref{T2}) can, in general, contribute to $\langle S_{1x}S_{2x} \rangle(t)$, the initial conditions associated with the Bell PPSs are exceptional in this regard,
because $S_{1x}S_{2x} + S_{1y}S_{2y}$ exclusively contributes to the ZQ density matrices $\vert S_0 \, \rangle \langle \, S_0 \vert$ and $\vert T_{0,z} \, \rangle \langle \, T_{0,z} \vert$ in Eqs.(\ref{rhoBell-S0},\ref{rhoBell-T0}), while $S_{1x}S_{2x} - S_{1y}S_{2y}$ exclusively contributes to their DQ counterparts $\vert \psi_{+,z} \, \rangle \langle \, \psi_{+,z} \vert$ and $\vert \psi_{-,z} \, \rangle \langle \, \psi_{-,z} \vert$ given by Eqs.(\ref{rhoBell-psi+},\ref{rhoBell-psi-}). Therefore, the relaxation of $\langle S_{1x}S_{2x} \rangle $ initiated by Bell PPSs  should exhibit two different \mbox{{\it monoexponential}} decays --- one with the rate  $\lambda_{\text{\tiny ZQ}}$ for the ZQ states  and the other with the rate $\lambda_{\text{\tiny DQ}}$ for the DQ states. For this reason, we will extract the rates $\lambda_{\text{\tiny ZQ}}$ and $\lambda_{\text{\tiny DQ}}$ not from the initial fits, as done for the diagonal relaxation, but rather from the overall single-exponential fits to the experimental plots of $\langle S_{1x}S_{2x} \rangle(t)$.

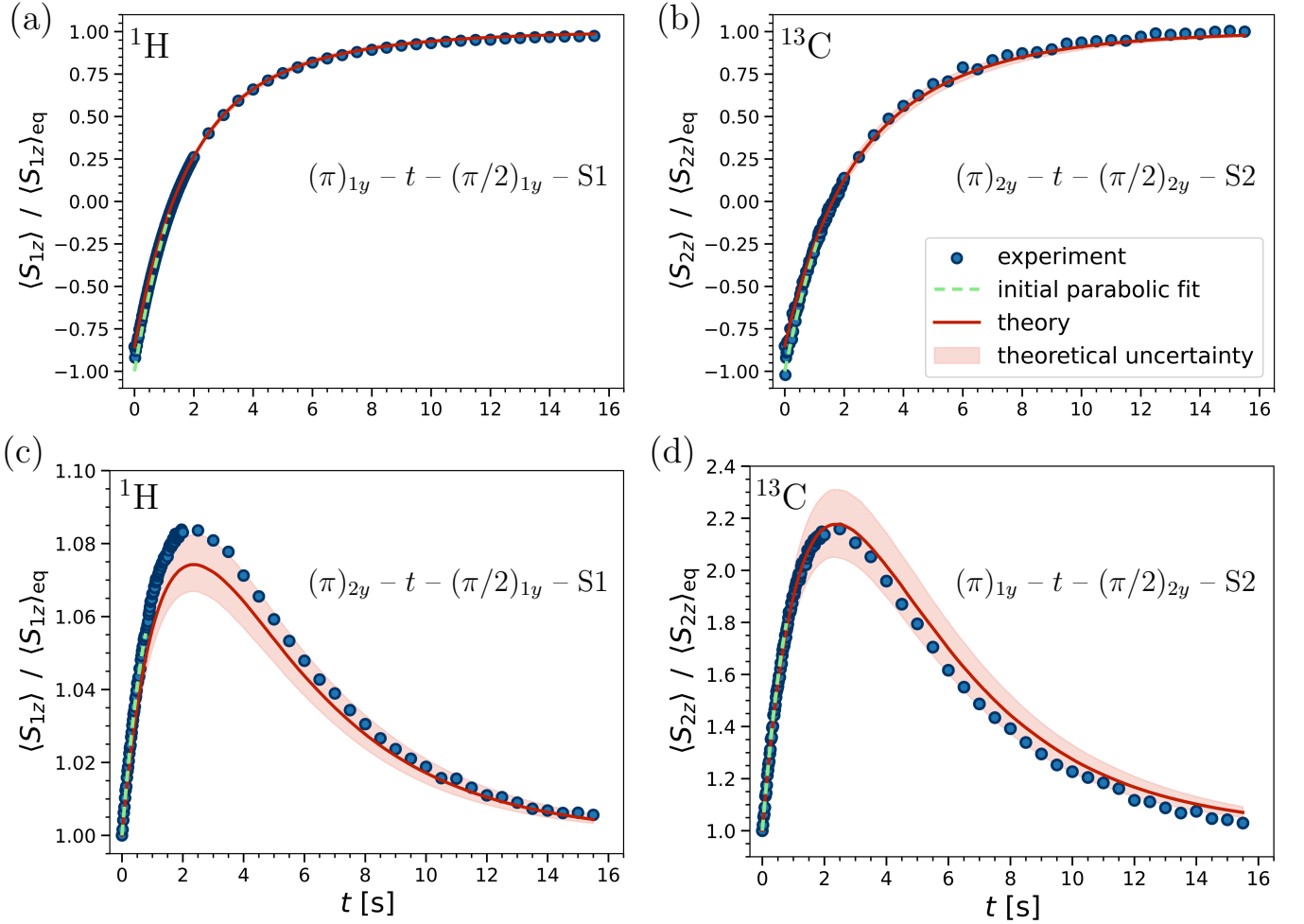
\begin{figure*}[t]
      \setlength{\unitlength}{0.1cm}
      \begin{picture}(200,140)
       \put(0,0){\includegraphics[width=1.0\textwidth]{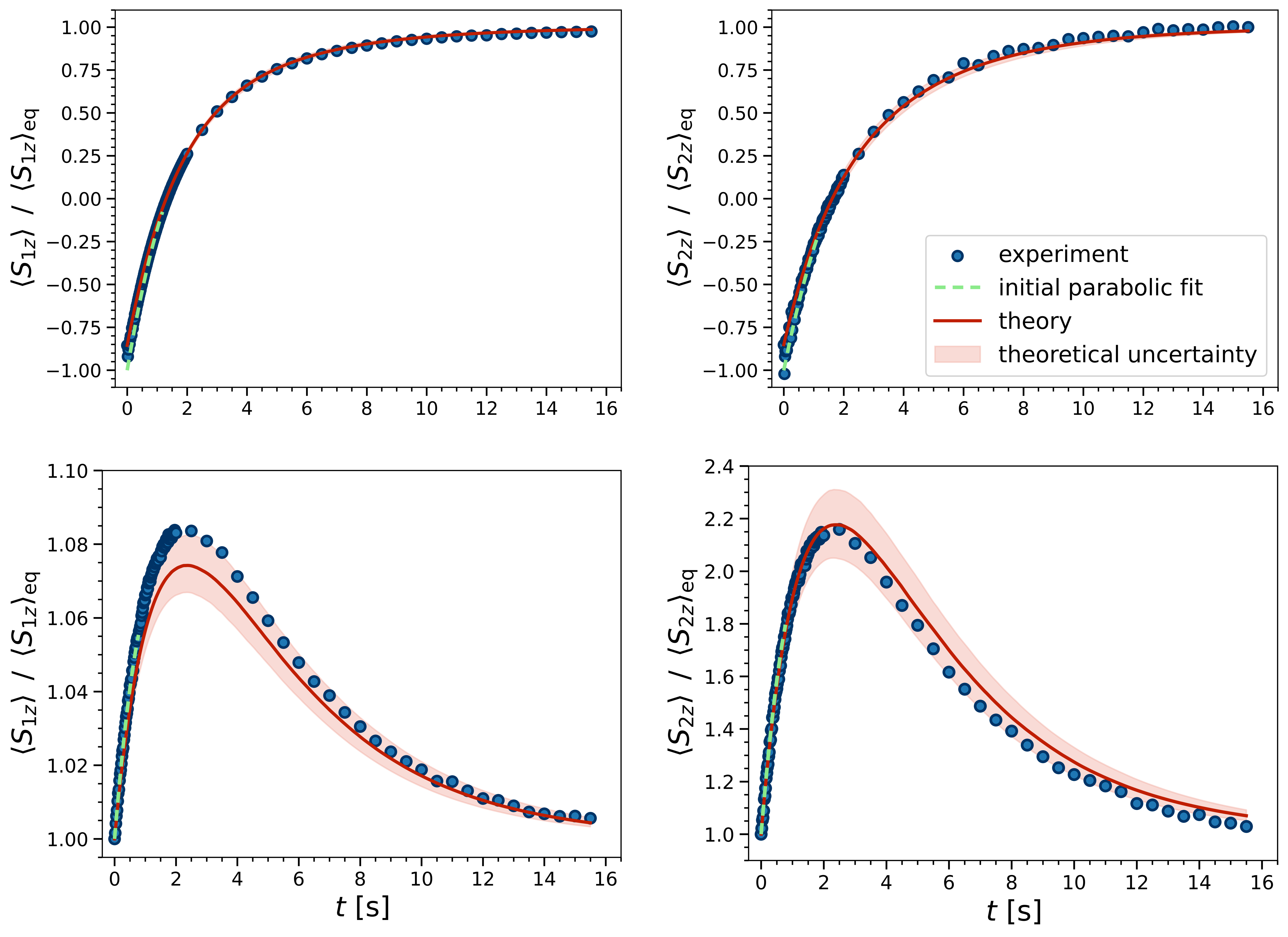}}
       \put(0,127){{\Large(a)}}
       \put(91,127){{\Large(b)}}
       \put(-1,66){{\Large(c)}}
       \put(90,66){{\Large(d)}}

       \put(17,123){{\Large $^1$H}}
       \put(108.5,123){{\Large $^{13}$C}}
       \put(15.5,59.5){{\Large $^1$H}}
       \put(105,59.5){{\Large $^{13}$C}}

       \put(42,105){{\large $(\pi)_{1y}  \  \text{--} \ t \text{ --}   \  (\pi/2)_{1y} \ \text{-- S1} $}}
       \put(133,105){{\large $(\pi)_{2y}  \  \text{--} \ t \text{ --}   \  (\pi/2)_{2y} \ \text{-- S2}  $}}
       \put(42,48){{\large $(\pi)_{2y}  \  \text{--} \ t \text{ --}   \  (\pi/2)_{1y} \ \text{-- S1}  $ }}
       \put(133,48){{\large $(\pi)_{1y}  \  \text{--} \ t \text{ --}   \  (\pi/2)_{2y} \ \text{-- S2}  $}}
      \end{picture}
    \caption{Plots of longitudinal relaxation (a,b) and NOE (c,d) obtained in the course of inversion recovery experiments for  the nuclei indicated in the upper left corners of each panel. Dots are the experimentally measured symmetric spectral components  of the FIDs of $^1$H and $^{13}$C following the rf-pulse sequence given in each plot (see Section~\ref{D-relaxation}).   Green dashed lines are the initial parabolic fits to the data used to extract the following initial rates listed in Table~\ref{tbl:rates}: (a) $\mu_1$, (b) $\mu_2$, (c) $\sigma^{(2)}_{12}$, and (d) $\sigma^{(1)}_{12}$. These and other initial rates from Table~\ref{tbl:rates} are then used to compute are the theoretical relaxation curves (solid red lines) based on the solutions of the system of equations (\ref{C17}) with the initial conditions also indicated in Table~\ref{tbl:rates}. The pink area around each curve represents the theoretical uncertainty of the solutions (the standard deviation) obtained by propagating the uncertainties of the initial rates.     
    }
    \label{fig:mu1-mu2-sig12}
\end{figure*}

The experimental pulse sequences for observing $\langle 2 S_{1x}S_{2x} \rangle(t)$ are nearly the same as those of $\langle 2 S_{1z}S_{2z} \rangle(t)$. The main difference between the two is in the choice of the measurement channel after the final $(\pi/2)_{y}$-pulse. Specifically, we measure $\langle 2 S_{1x}S_{2x} \rangle(t)$ by applying the final  $(\pi/2)_{2y}$-pulse on the second spin, which converts $\langle 2 S_{1x}S_{2x} \rangle$ into $-\langle 2 S_{1x}S_{2z} \rangle$, which is then measurable as the antisymmetric spectral component of the first spin. Another difference is that, unlike $\langle 2 S_{1z}S_{2z} \rangle(t)$, the relaxation of $\langle 2 S_{1x}S_{2x} \rangle(t)$ is affected by the inhomogeneity of the static magnetic field inside the sample and by the $J$-coupling of the $^1$H--$^{13}$C spin pair to the nearby $^{15}$N spin. In order to compensate both of these extrinsic effects, our pulse sequences will include Carr-Purcell-Meiboom-Gill (CPMG) echo trains\cite{Carr-Purcell-1954,Meiboom-Gill-1958}.

\section{Experiments}
\label{Experiments}

\subsection{Relaxation of non-Bell states}
\label{non-Bell}

\begin{figure*}[t]
    \setlength{\unitlength}{0.1cm}
      \begin{picture}(200,155)
       \put(0,0){\includegraphics[width=0.95\textwidth]{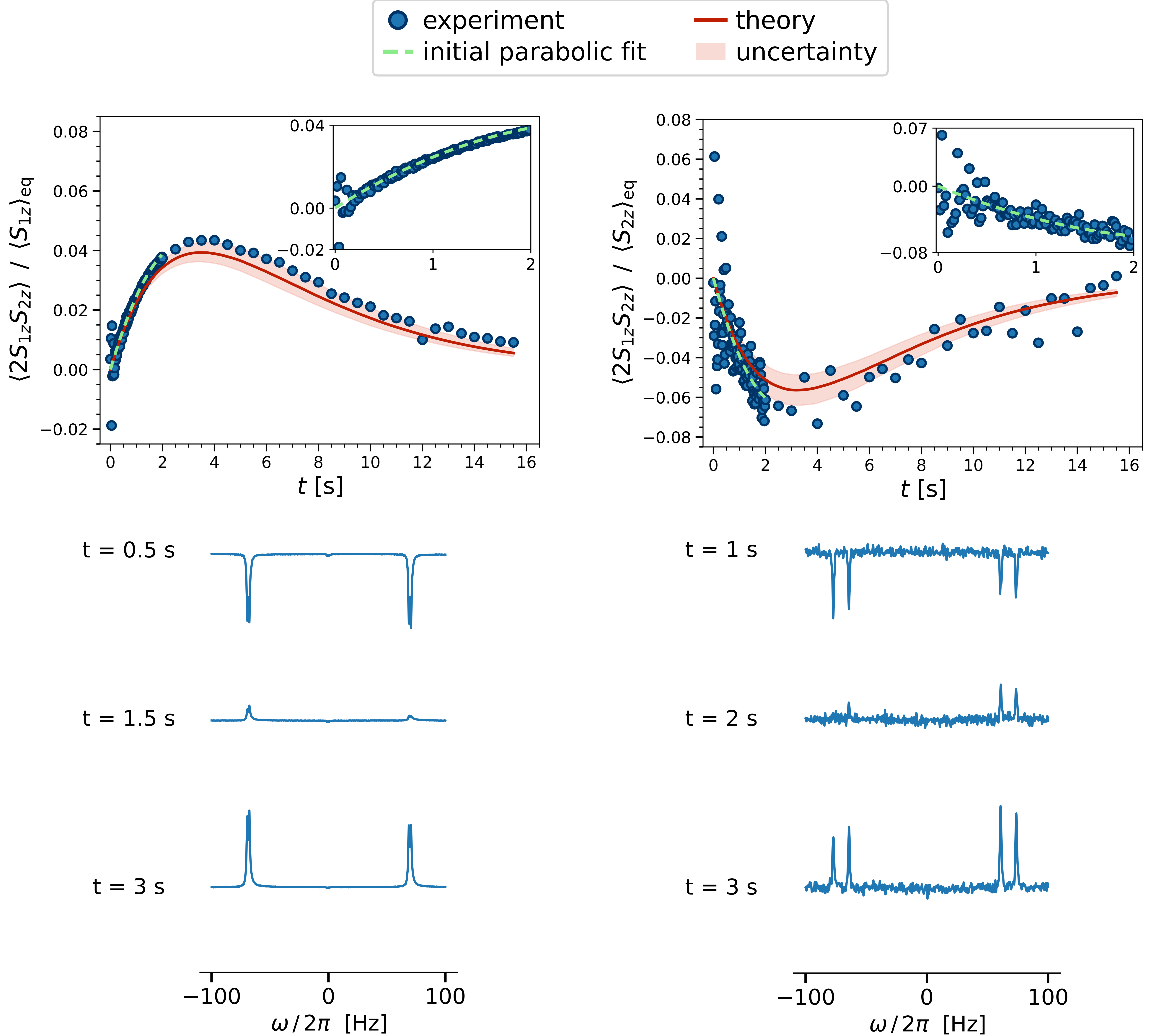}}

    \put(-2,135){{\Large(a1)}}
       \put(86,135){{\Large(b1)}}
       \put(-2,76){{\Large(a2)}}
       \put(86,76){{\Large(b2)}}

       \put(17,130){{\Large $^1$H}}
       \put(108.5,130){{\Large $^{13}$C}}

       \put(35,89.5){{\large $(\pi)_{1y}  \  \text{--} \ t \text{ --}   \  (\pi/2)_{1y} \ \text{-- A1} $}}
       \put(125,89){{\large $(\pi)_{2y}  \  \text{--} \ t \text{ --}   \  (\pi/2)_{2y} \ \text{-- A2}  $}}
       
      \end{picture}
    \caption{(a1, b1) Plots of correlators $\langle 2 S_{1z}S_{2z} \rangle(t)$ obtained in the course of inversion recovery experiments  from the antisymmetric spectral components of the FID spectra of $^1$H (a1) and $^{13}$C (b1) following the pulse sequences given in each panel and explained in Section~\ref{D-relaxation}. [The symmetric components of the same FID spectra are plotted in Figs.~\ref{fig:mu1-mu2-sig12}(a,b).] Dots are the experimental data.  Green dashed lines are the initial parabolic fits used to extract the following initial rates listed in Table~\ref{tbl:rates}: (a1) $\delta_1$, and (b1) $\delta_2$. These and other initial rates from Table~\ref{tbl:rates} are then used to compute the theoretical relaxation curves (solid red lines) based on the solutions of the system of equations (\ref{C17}) with the initial conditions indicated in Table~\ref{tbl:rates}. The pink area around each curve represents the theoretical uncertainty of the solutions (the standard deviation) obtained by propagating the uncertainties of the initial rates. (a2, b2) Examples of experimental asymmetric FID spectra, whose antisymmetric components  are plotted in (a1) and (b1), respectively. The opposite asymmetry of the spectra in (a2) and (b2) implies the opposite signs of $\delta_1$ and $\delta_2$.
    }
    \label{fig:delta}
\end{figure*}

In this subsection, we describe the basic inversion recovery NMR experiments on the selected $^1$H and $^{13}$C nuclear spins --- done as a part of the overall measurement scheme presented in Table~\ref{tbl:rates}.

Figure~\ref{fig:mu1-mu2-sig12} shows rather conventionally looking  results for $\langle S_{1z}\rangle(t)$ and $\langle S_{2z}\rangle(t)$ characterizing the longitudinal relaxation and the NOE of both $^1$H or $^{13}$C, while Fig.~\ref{fig:delta} shows less commonly measured $\langle 2 S_{1z}S_{2z} \rangle (t)$ extracted from the antisymmetric components of the $J$-coupling-split peaks. 

These experiments followed the pulse sequences indicated in each plot (same as in Table~\ref{tbl:rates}). Hard pulses were used, which, in the case of $^1$H, implies that all $^1$H spins belonging to our molecule were acted upon as well. The response of the $^1$H spin belonging to the central pair was distinguished from the rest by its chemical shift placing the relevant part of the FID spectrum  within the frequency window indicated by the gray background in the inset of Fig.~\ref{fig1}(b). (We did not use selective pulses on $^1$H of the central pair, because they were causing larger uncertainties in the measurements of the initial relaxation rates.)  The FID spectra of $^{13}$C were dominated by the $^{13}$C of the central pair due to the site-selective isotope enrichment. 

In each of the spectra, we digitized pairs of peaks separated by the intra-pair $J$-coupling between $^1$H and $^{13}$C (see Figs.~\ref{fig1}(b) and (c)) and then obtained the intensities of these peaks, $I_{n+}$ and $I_{n-}$, by integrating the spectra within the window of 20~Hz around the centers of, respectively,  the left and the right peaks. As illustrated in Fig.~\ref{fig1}(c), the spectra of $^{13}$C have an additional visible doublet structure due the $J$-coupling to a nearby $^{15}$N --- each such a doublet was treated as a single peak for the purpose of obtaining $I_{n+}$ and $I_{n-}$. As explained in Section~\ref{D-relaxation}, the plotted values of $\langle S_{1z}\rangle$ and $\langle S_{2z}\rangle$ were obtained as sums $I_{n+} + I_{n-}$ characterizing the symmetric components of the respective spectra, while $\langle 2 S_{1z}S_{2z} \rangle $ was obtained as the difference $I_{n+} - I_{n-}$ representing the antisymmetric spectral component. 

Each plot in Fig.~\ref{fig:mu1-mu2-sig12} indicates the initial time range from which the rates listed in Table~\ref{tbl:rates} were extracted with the help of the parabolic fits of the form $a + bt + ct^2$, were $a$, $b$ and $c$ were the fitting parameters.  The rates themselves were then obtained from the normalized initial slopes $|b/a|$. The experimental uncertainties listed in Table~\ref{tbl:rates} include both the statistical errors of the fitting parameters and the systematic errors associated with the deviation between the best parabolic fit in a given finite interval and the actual function being fitted. The systematic errors were {\it estimated} on the basis of the difference between the initial slopes of parabolic fits and cubic fits. 

We note that the two NOE rates $\sigma_{12}^{(1)}$ and $\sigma_{12}^{(2)}$ extracted from  Figs.~\ref{fig:mu1-mu2-sig12} (d) and (c) and listed in Table~\ref{tbl:rates} deviate from each other by about 10 percent. Theoretically, they are supposed to be the same. The 10-percent difference, while being relatively small, is still much larger than the experimental errors with which $\sigma_{12}^{(1)}$ and $\sigma_{12}^{(2)}$ were determined. This difference can, presumably, be attributed to the fact that the experiments used hard pulses, and, as a result, distant $^1$H spins could affect the NOE response of the central pair. The experimental value of the intrinsic rate $\sigma_{12}$ listed in Table~\ref{tbl:rates} is the average of $\sigma_{12}^{(1)}$ and $\sigma_{12}^{(2)}$, with the error being determined mostly by the difference between the two. The rate $\sigma_{12}$ is positive as expected for for IPMDI, because, in this case, the double-flips of the two spins dominate over flip-flops\cite{Fine-1997}.

The experimental plots in Figs.~\ref{fig:delta}(a1) and (b1), from which the cross-relaxation rates $\delta_1$ and $\delta_2$ were  extracted, appear to exhibit more statistical noise overall and, especially, around $t=0$ in comparison with the plots in Fig.~\ref{fig:mu1-mu2-sig12}. This should not be surprising, given that the antisymmetric spectral components ploted in Figs.~\ref{fig:delta}(a1, b1)   were significantly smaller than the symmetric ones.  Yet, as the examples of the spectra in Figs.\ref{fig:delta}(a2) and (b2) indicate, the spectral asymmetry and hence the presence of the antisymmetric components was clearly identifiable and quantifiable: the insets in Figs.\ref{fig:delta}(a1) and (b1) show that the initial fits used to extract $\delta_1$ and $\delta_2$ were quite reasonable. We note here the opposite asymmetry of the spectra for $^1$H and $^{13}$C in Figs.~\ref{fig:delta}(a2) and (b2), which implies the opposite signs of $\delta_1$ and $\delta_2$.

Finally, figures~\ref{fig:mu1-mu2-sig12} and \ref{fig:delta} also include theoretical plots obtained from the solution of the system of equations (\ref{C17}) with rates determined experimentally from the initial fits and with the initial conditions listed in Table~\ref{tbl:rates}. The pink background in the figures indicates the theoretical uncertainty of the above solutions due to the uncertain knowledge of the initial rates. 


 \subsection{Preparation and characterization of Bell pseudo-pure states}
\label{preparation}

\subsubsection{Preparation of Bell PPSs using detuned Hartmann-Hahn double-resonance condition}
 \label{Pulse_sequences}
  
\begin{figure*} 
    \centering
    \begin{tikzpicture}
    \node[inner sep=0pt] (duck) at (0,0)
    {\includegraphics[width=1.8\columnwidth]{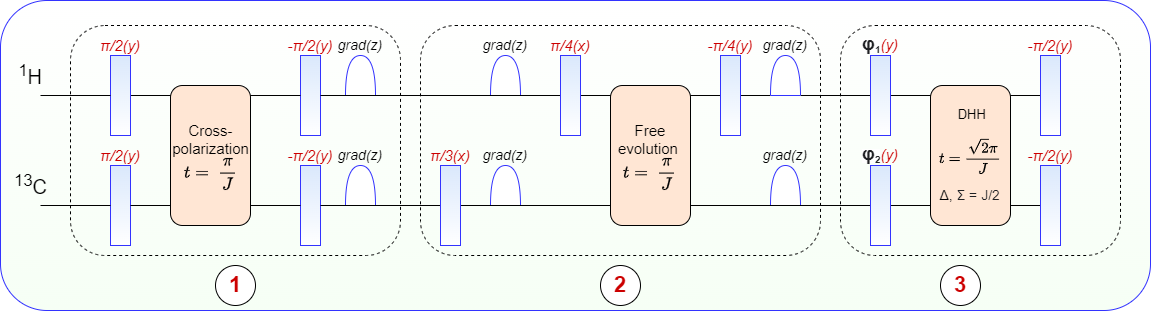}}; 
    \end{tikzpicture}

    \caption{
\label{fig2}
Experimental pulse sequence described in the Section~\ref{Pulse_sequences} for creating Bell PPSs: steps 1, 2, and 3 are indicated by the respective circled numbers. Step 3 implements the DHH condition appropriate for the target Bell PPS with either $\Delta = J/2$ or $\Sigma = J/2$ and with pulse parameters $\varphi_1$ and  $\varphi_2$ given in Table~\ref{table1}. 
}
\end{figure*}

\begin{table}[h!] 
\begin{tabular}{ c|c|c|c} 
\hline
Pulse parameters & States after  & DHH & Resulting\\
$\varphi_1$ ,  $\varphi_2$ & $\varphi_1, \varphi_2$  pulses & param. & Bell state\\
\hline
$ (-\pi/2)_{1y}$, $ (-\pi/2)_{2y}$  & $\vert \downarrow_x \downarrow_x \, \rangle $  & $\Sigma = J/2$ & $\vert \psi_{-,x} \,\rangle $\\   
$ ( -\pi/2)_{1y}$, $ ( \pi/2)_{2y}$  & $\vert \downarrow_x \uparrow_x \, \rangle $  & $\Delta = J/2$  & $\vert S_0 \,\rangle  $\\ 
$ ( \pi/2)_{1y}$, $ (-\pi/2)_{2y}$  & $\vert \uparrow_x \downarrow_x \, \rangle $  & $\Delta = J/2$  & $\vert T_{0,x} \,\rangle $\\  
$ ( \pi/2)_{1y}$, $ ( \pi/2)_{2y}$  & $\vert \uparrow_x \uparrow_x \, \rangle $  & $\Sigma = J/2$  & $\vert \psi_{+,x} \,\rangle  $\\  
\hline 
\end{tabular}
\caption{ 
Implementation of Step 3 in Fig.~\ref{fig2} for different resulting Bell states. The third column refers to the parameters of the detuned Hartmann-Hahn (DHH) pulses.
} 
\label{table1}
\end{table}

Our experimental procedure for preparing Bell PPSs requires the application of two resonant rf-fields at frequencies $\Omega_1 $ and $\Omega_2$ with amplitudes $h_1$ and $h_2$ respectively.  
In the presence of these rf fields, the Hamiltonian of our spin pair in the double-rotating reference frame becomes 
\begin{equation}
    \label{3}
    \mathcal{H}  = -\hbar\omega_1  S_{1x} - \hbar\omega_2  S_{2x} + JS_{1z}S_{2z},
\end{equation}
where $\omega_1  = \gamma_1 h_1 $ and $\omega_2  = \gamma_2 h_2$. 
The procedure starts with the equilibrium density matrix (\ref{rho-eq}) and then proceeds as follows 
 \begin{equation} \begin{split}
    \varepsilon_1 S_{1z} + \varepsilon_2 S_{2z}  \  \xrightarrow {\quad \text{Step 1} \quad}   \ \frac{\varepsilon_1 + \varepsilon_2}{2} (S_{1z} + S_{2z})
\label{17}
    \\ 
    \xrightarrow {\quad \text{Step 2} \quad} 
    \frac{\varepsilon_1 + \varepsilon_2}{2}
    |\uparrow_z \uparrow_z \, \rangle \langle \, \uparrow_z \uparrow_z  \! | \xrightarrow {\quad \text{Step 3} \quad} \mbox{\  Bell PPS},
    \end{split} 
\end{equation}
 where the formulas before and between the arrows represent the density matrices. The corresponding pulse sequence is summarized in Fig.~\ref{fig2}.

The goal of step 1  is to equalize the spin polarizations of $^1$H and $^{13}$C thereby preparing the system for step 2. Here, in order to minimize the loss of the signal, we do not use the simplest routine, namely, the partial rotation of $^1$H spin followed by a gradient pulse on $^1 $H, but rather we first apply $(\pi/2)_{1y}$,$(\pi/2)_{2y}$ pulses followed by the resonant Hartmann-Hahn pulse with $\omega_1 = \omega_2 = \frac{\sqrt{15}}{4} J$ during time $\pi/J$ to partially switch the polarizations of $^1$H and $^{13}$C, then apply $(-\pi/2)_{1y}$,$(-\pi/2)_{2y}$, and, at last, implement the gradient pulse removing the transverse polarization. 

Step 2 employs the sequence proposed in Ref.~\cite{CORY199882}
that starts with an equally polarized mixed state and then generates a non-entangled PPS proportional to ${|\uparrow_z \uparrow_z \, \rangle \langle \, \uparrow_z \uparrow_z  \! | }$.

In step 3, we convert the above non-entangled PPS into entangled Bell PPSs using detuned Hartmann-Hahn (DHH) double resonance condition, which we introduce below. 

The Bell PPSs are  to be initially obtained in the $x$-quantization basis, where they have pure parts  proportional to $\vert S_0 \, \rangle$,  $\vert T_{0,x} \, \rangle$, $\vert \psi_{-,x} \, \rangle$, and  $\vert \psi_{+,x} \, \rangle$, which are  defined similarly to Eqs.~(\ref{T0z},\ref{S0z},\ref{psi+z},\ref{psi-z}). The Hamiltonian~(\ref{3}) does not mix ZQ and DQ spaces in the $x$-basis. 
In that basis, the dynamics in the ZQ space is controlled by parameter $\Delta \equiv \omega_1 - \omega_2$, while the dynamics in the DQ space is controlled by $\Sigma \equiv \omega_1 + \omega_2$. In order to obtain $|S_{0}\rangle$ and $|T_{0,x}\rangle$, one needs the Hamiltonian~(\ref{3}) with $\Delta  = J/2$ to act on, respectively,
$\vert \downarrow_x \uparrow_x \, \rangle $ and 
$\vert \uparrow_x \downarrow_x \, \rangle $ during time $t = \pi\sqrt{2}/J $. The value of $\Sigma$ does not matter here.
On the other hand, the states $\vert \psi_{+,x}\rangle$ and  $\vert \psi_{-,x}\rangle$ can be obtained once the Hamiltonian~(\ref{3}) with ${\Sigma  = J/2}$ acts, respectively, on $\vert \uparrow_x \uparrow_x \, \rangle $ and
$\vert \downarrow_x \downarrow_x \, \rangle $ 
during time $t = \pi\sqrt{2}/J$. The value of $\Delta$ is arbitrary in this case.

To implement such a procedure, we rotate the state $|\uparrow_z \uparrow_z \, \rangle  $ obtained at the end of step 2 into one of the initial non-entangled states listed in Table~\ref{table1}  and then apply the DHH pulses defined above to generate the desired Bell states.  Finally, we apply two
$ ( -\pi/2)_{1y}$, $ ( -\pi/2)_{2y}$ pulses rotating the states   $\vert T_{0,x} \, \rangle$, $\vert \psi_{-,x} \, \rangle$, and  $\vert \psi_{+,x} \, \rangle$ into, respectively,
$\vert T_{0,z} \, \rangle$, $\vert \psi_{-,z} \, \rangle$, and  $\vert \psi_{+,z} \, \rangle$. (The  state $\vert S_0 \, \rangle$ was not supposed to be affected by these pulses.) The initial conditions  for the relaxation after the above pulse sequences  are: \mbox{$\langle S_{1z} \rangle(0)=0$,} $\langle S_{2z} \rangle(0)=0$, $\langle 2 S_{1z}S_{2z} \rangle(0) = \pm \frac{\varepsilon_1 + \varepsilon_2}{4}$ and $\langle 2 S_{1x}S_{2x} \rangle(0) = \pm \frac{\varepsilon_1 + \varepsilon_2}{4}$. The signs of $\langle 2 S_{1z}S_{2z} \rangle$ and $\langle 2 S_{1x}S_{2x} \rangle$  for the individual Bell PPSs can be found in Table~\ref{tbl:rates}.

Let us remark here that the DHH procedure does not require $\omega_1, \omega_2 \gg J$  \cite{Pelupessy-2000}. On the contrary, the condition $\Sigma  = J/2$ in the DQ space implies that $\omega_1, \omega_2 \sim J$. The condition $\Delta  = J/2$ can be satisfied with either $\omega_1, \omega_2 \sim J$ or $\omega_1, \omega_2 \gg J$. We also note that somewhat similar approach to the preparation of the singlet state is used by the ``spin-lock induced crossing'' (SLIC) technique\cite{DeVience-2013}, which is devised for pairs of spins with the same gyromagnetic ratios but very small differences of Larmor frequencies.


\subsubsection{Characterisation of Bell PPSs}
\label{tomography}

\begin{figure}[t!]
    \centering
    \begin{tikzpicture}  
    \node[inner sep=0pt] (duck) at (0,0)
    {\includegraphics[width=0.88\columnwidth]{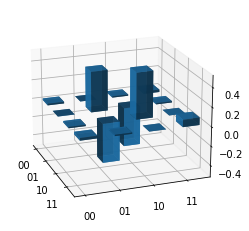}};
    \node[align=center,fill=white] at (-3.8, 1.8) {\textbf{(a)}};
    \end{tikzpicture}

    \begin{tikzpicture}
    \label{fig4b}
    \node[inner sep=0pt] (duck) at (0,0)
    {\includegraphics[width=0.88\columnwidth]{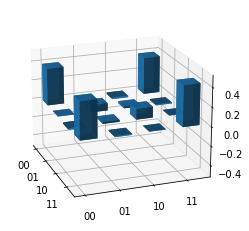}};
    \node[align=center,fill=white] at (-3.8, 1.8) {\textbf{(b)}};
    \end{tikzpicture}

    \caption{
\label{fig4.1}
  Real parts of the normalized density matrix of Bell PPSs obtained by quantum tomography for the central spin pair $^1$H~--$^{\ 13}$C: (a) PPS containing $\vert S_0  \, \rangle $, (b) PPS containing $\vert \psi_{+,z}  \, \rangle$. Labels $0$ and $1$ denote, respectively, the states $|\uparrow_z\rangle$ and $|\downarrow_z\rangle$.
}
\end{figure}

To verify experimentally that the prepared states are indeed the entangled Bell PPSs, we performed on them the quantum-state tomography according to the methodology of Ref.~\cite{PhysRevA.69.052302}. The experimentally measured tomograms of the real parts for the density matrices $\vert S_0  \, \rangle \langle  S_0 |$ and  $\vert \psi_{+,z} \, \rangle \langle \psi_{+,z} | $ are shown in Fig.~\ref{fig4.1}. They are in satisfactory agreement with the theoretical expectations, in particular,  as far as the entanglement-related off-diagonal elements are concerned.
The deviations of the measured density matrices seen in Fig.~\ref{fig4.1} from the idealized theoretical expressions indicate the overall error of roughly 10 percent. There are two main reasons for it: (i) the spatial inhomogeneity of the applied rf field, and (2) instrumental limitations constraining the control of the rf field amplitude to a discrete set of values.



\begin{figure}
    \centering
    \begin{tikzpicture}  
    \node[inner sep=0pt] (duck) at (0,0)
    {\includegraphics[width=0.85\columnwidth]{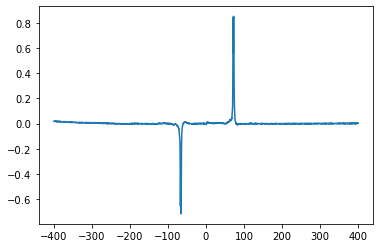}};
    \node[align=center,fill=white] at (-3.8, 2.1) {\textbf{(a)}};
    \node[align=center,fill=white] at (-3.9, 0.3) {{\rotatebox {90}{\text{$g(\omega)$ [a.u.]} }}};
    \end{tikzpicture}
\text{\qquad \quad $\omega/2\pi$ [Hz] }

    \begin{tikzpicture}
    \label{3b}
    \node[inner sep=0pt] (duck) at (0,0)
    {\includegraphics[width=0.85\columnwidth]{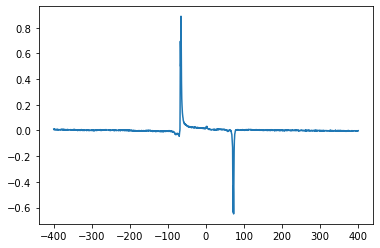}};
    \node[align=center,fill=white] at (-3.8, 2.1) {\textbf{(b)}};
    \node[align=center,fill=white] at (-3.9, 0.3) {{\rotatebox {90}{\text{$g(\omega)$ [a.u.]} }}};
    \end{tikzpicture}
\text{\qquad \quad $\omega/2\pi$ [Hz] }
    \caption{
\label{fig3}
Spectra $g(\omega)$ of the measured $^1$H as functions of the frequency offset $\omega$ for the Bell PPSs containing:  (a) $\vert S_0 \, \rangle$ and (b)  $\vert \psi_{+,z} \, \rangle$.
}
\end{figure}
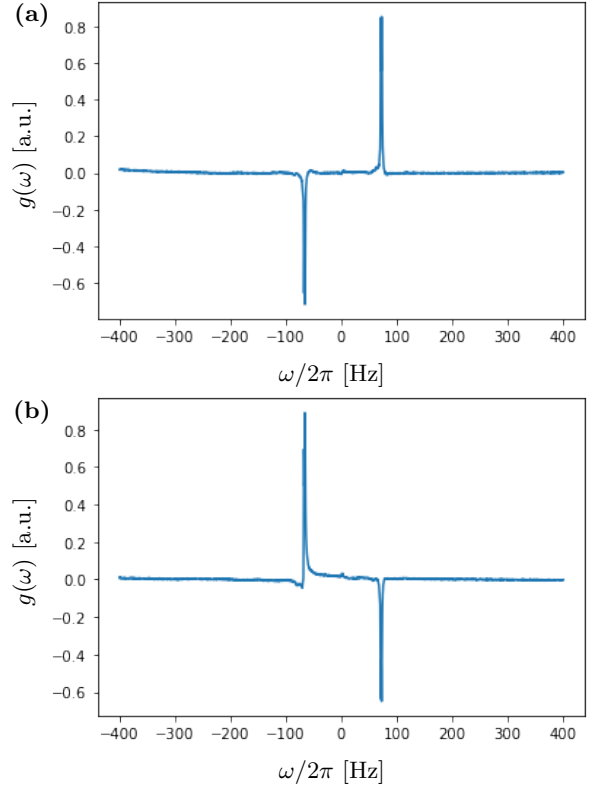


\subsection{Relaxation of Bell pseudo-pure states}
\label{relaxation-Bell}

\begin{figure}[t]
    \setlength{\unitlength}{0.1cm}
      \begin{picture}(100,142)
       \put(0,0){\includegraphics[width=1.0\columnwidth]{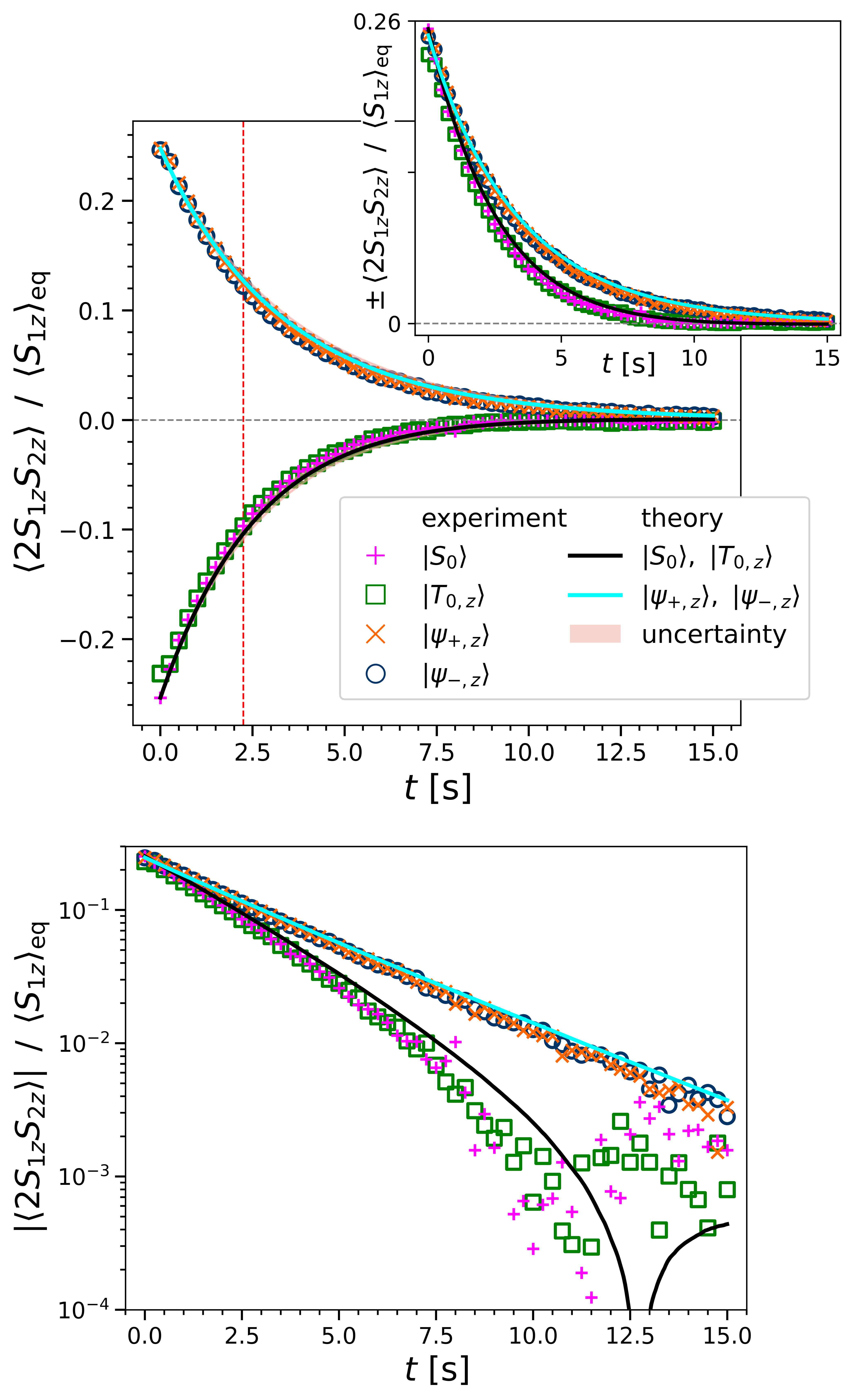}}
       
      \put(0,128){{\Large(a)}}
       \put(0,56){{\Large(b)}}

       \put(27,123){{\Large $^1$H}}
       
       \put(40,93.5){\colorbox{white}{$|\text{Bell PPS} \rangle \  \text{--} \ t \text{ --}   \  (\pi/2)_{1y} \ \text{-- A1} $}}

       \put(39,129){\colorbox{white}{\tiny\ }}
       \put(33.5,129){\colorbox{white}{\tiny\ }}

       \put(74.5,106.5){\colorbox{white}{\ }}
       \put(74.5,104.5){\colorbox{white}{\ }}
       \put(74.5,102.5){\colorbox{white}{\ }}
       \put(74.5,100.5){\colorbox{white}{\ }}
       \put(74.5,98.5){\colorbox{white}{\ }}
       \put(74.5,96.5){\colorbox{white}{\ }}
       
      \end{picture}
    \caption{(a) Relaxation of $\langle  S_{1z}S_{2z} \rangle$ starting from four Bell PPSs  and measured as the antisymmetric component of $^1$H FID spectrum following the rf-pulse sequence given in the plot. Vertical dashed red line indicates the end of the time interval for the parabolic fits used to extract the initial rates $\mu_{\text{\tiny ZQ}}$ and $\mu_{\text{\tiny DQ}}$ listed in Table~\ref{tbl:rates}. These and other initial rates from Table~\ref{tbl:rates} are then used to compute the theoretical relaxation curves (black and cyan solid lines) based on the solutions of the system of equations (\ref{C17}) with the initial conditions indicated in Table~\ref{tbl:rates}.  Inset: the data points from the main plot with the sign of $\langle S_{1z}S_{2z} \rangle$  for the ZQ states inverted to expose the difference between $\mu_{\text{\tiny ZQ}}$ and $\mu_{\text{\tiny DQ}}$. (b) Semilogarithmic plots of the data from (a). 
    }
    \label{fig:muZQ-muDQ}
\end{figure}
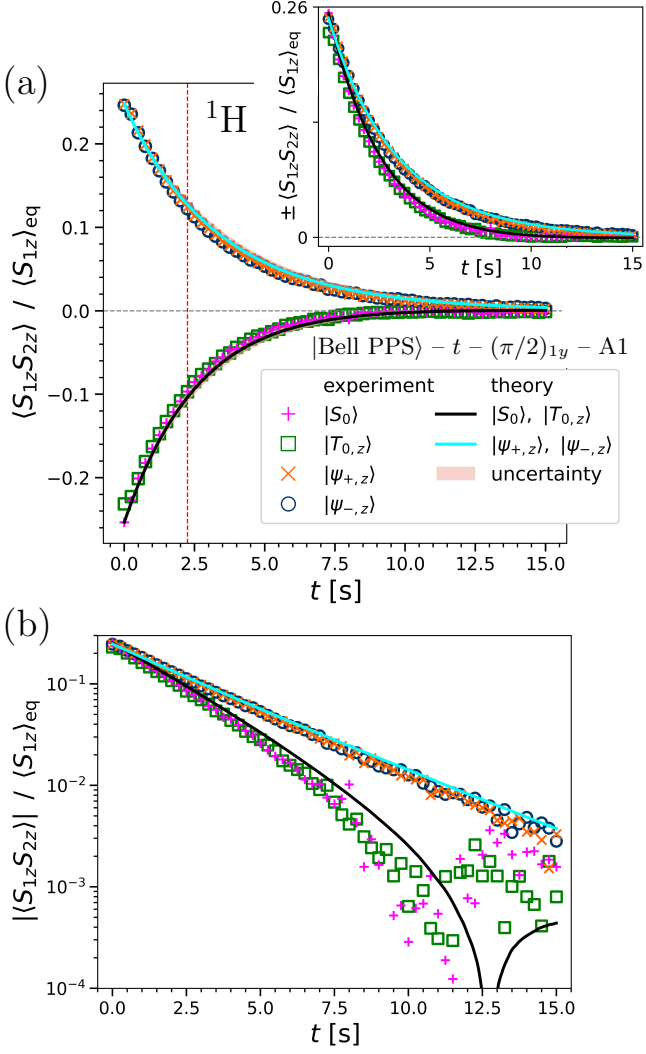

\begin{figure}[t]
    \setlength{\unitlength}{0.1cm}
      \begin{picture}(100,142)
       \put(0,0){\includegraphics[width=1.0\columnwidth]{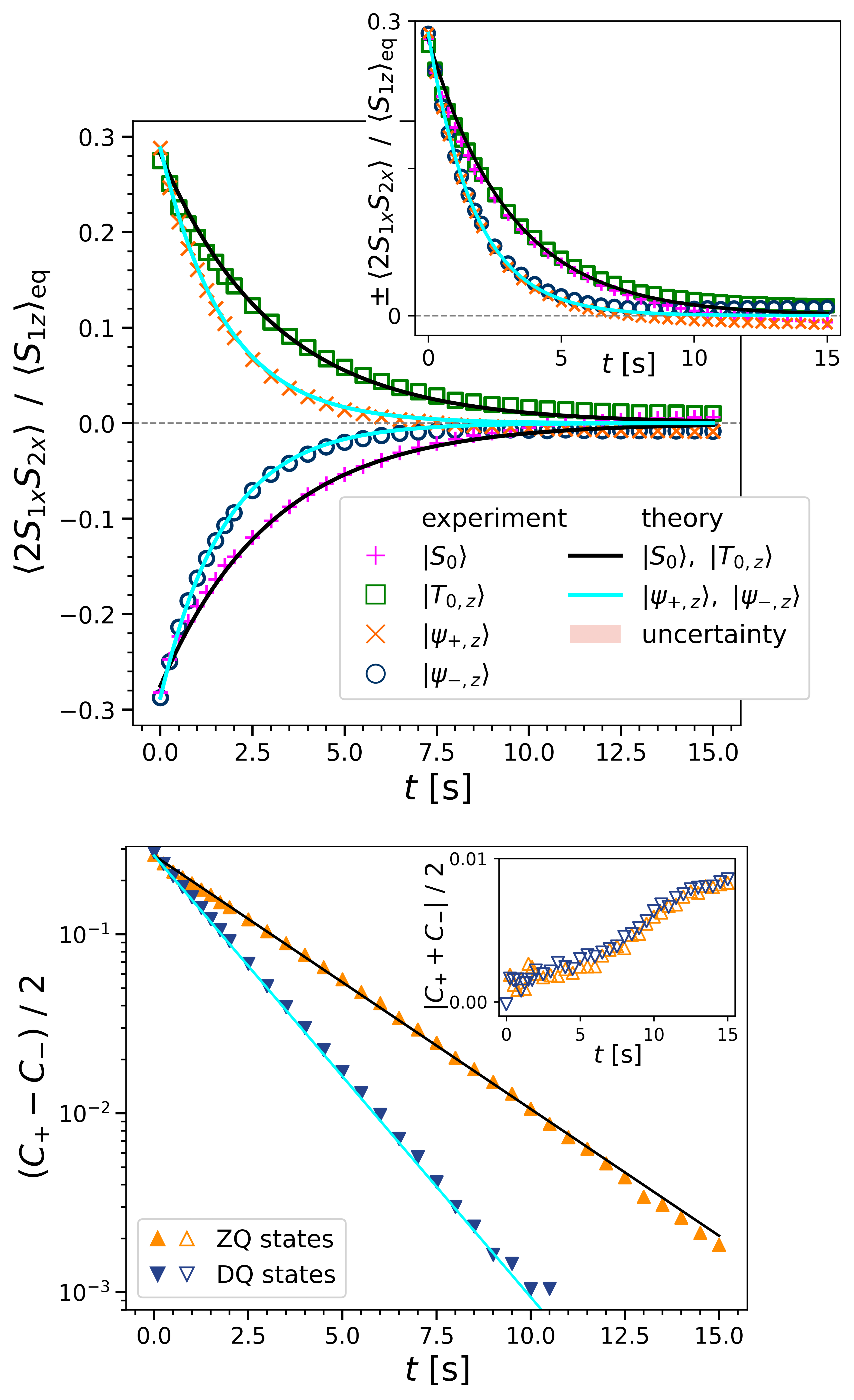}}

       \put(0,128){{\Large(a)}}
       \put(0,56){{\Large(b)}}

       \put(23,123){{\Large $^1$H}}

       \put(74.5,94.5){\colorbox{white}{\ }}
       \put(74.5,92.5){\colorbox{white}{\ }}
       \put(42.5,92.3){$|\text{Bell PPS} \rangle \  \text{--} \ t \text{ --}   \  (\pi/2)_{2y} \ \text{-- A1} $}
       \put(58,95){\tiny CPMG}

       \put(39.8,129){\colorbox{white}{\tiny }}
       \put(34,129){\colorbox{white}{\tiny\ }}

       \put(74.5,106.5){\colorbox{white}{\ }}
       \put(74.5,104.5){\colorbox{white}{\ }}
       \put(74.5,102.5){\colorbox{white}{\ }}
       \put(74.5,100.5){\colorbox{white}{\ }}
       \put(74.5,98.5){\colorbox{white}{\ }}
       \put(74.5,96.5){\colorbox{white}{\ }}
       
      \end{picture}
    \caption{(a) Relaxation of $\langle S_{1x}S_{2x} \rangle$ starting from four Bell PPSs  and measured as the antisymmetric component of $^1$H FID spectrum following the rf-pulse sequence given in the plot. 
    Inset in (a): the data points from the main plot with the sign of $\langle S_{1x}S_{2x} \rangle$ inverted for  $|S_0\rangle$ and $|\psi_-\rangle$ (b) Semilogarithmic plots of half-differences of  the data for the two ZQ states and and for the two DQ states, with $C_+$ being the data set for either $|T_0\rangle$ or $|\psi_+\rangle$  and $C_-$ for either $|S_0\rangle$ or $|\psi_-\rangle$. Black and cyan solid lines in both (a) and (b) are the theoretical relaxation curves for the ZQ and DQ states obtained as the best single-exponential fits to the semilog data points. Rates $\lambda_{\text{\tiny ZQ}}$ and $\lambda_{\text{\tiny DQ}}$ given in Table~\ref{tbl:rates} are the parameters of these fits.  Inset in (b) shows the same small drift  of  \mbox{$|C_+ + C_-|/2$} away from zero for both ZQ and DQ states.
    }
    \label{fig:lambdaZQ-lambdaDQ}
\end{figure}

As explained in Section~\ref{relaxation-theory}, we use each of the four Bell PPSs to monitor separately $\langle 2 S_{1z}S_{2z} \rangle(t)$ and $\langle 2 S_{1x}S_{2x} \rangle(t)$. Once Bell PPSs are prepared, the spectra of both $^1$H  and $^{13}$C  after a $(\pi/2)_y$ pulse on either of them are supposed to be completely antisymmetric. Let us consider as an example, the spectra of $^1$H after it is subjected to $(\pi/2)_{1y}$ pulse: the antisymmetric component of that spectrum would be proportional to $\langle 2 S_{1z}S_{2z} \rangle(0)$.  In such a case, Bell PPSs containing ZQ states   $\vert S_0 \, \rangle $ or $\vert T_{0,z} \, \rangle $ exhibit, respectively, negative left peak (lower frequency) and positive right peak (higher frequency), while Bell PPSs containing DQ states $\vert \psi_{+,z} \, \rangle$ or $\vert \psi_{-,z} \, \rangle$ show the opposite asymmetry --- see the two examples shown in Fig.~\ref{fig3}. When, however, the $^1$H spectra are measured after the $(\pi/2)_{2y}$ pulse on $^{13}$C, their antisymmetric components give $\langle 2 S_{1x}S_{2x} \rangle(0)$.  In this case, PPSs containing $\vert T_{0,z} \, \rangle $ and $\vert \psi_{-,z} \, \rangle $ result in the same spectral asymmetries as those they had after the $(\pi/2)_{1y}$ pulse, while the  spectral asymmetries for $\vert S_0 \, \rangle $ and $\vert \psi_{+,z} \, \rangle $ reverse their signs.

Figure~\ref{fig:muZQ-muDQ} presents the results for $\langle 2 S_{1z}S_{2z} \rangle(t)$ obtained after the pulse sequence given in that figure. As explained in Section~\ref{D-relaxation}, the initial conditions and hence the subsequent relaxation for the two ZQ PPSs $\vert S_0 \, \rangle $ or $\vert T_{0,z} \, \rangle $ are expected theoretically to coincide. The same is true for the two DQ PPSs $\vert \psi_{+,z} \, \rangle$ or $\vert \psi_{-,z} \, \rangle$. These expectations are fully consistent with the experimental plots in Fig.~\ref{fig:muZQ-muDQ}. 

On the other hand, according to Eqs.(\ref{muZQ}) and (\ref{muDQ}), the initial relaxation rates for the ZQ Bell PPSs are supposed to be different from those for the DQ Bell PPSs when $\delta_1 \neq 0$ and/or $\delta_2 \neq 0$, which is, indeed, the case for our system --- see Fig.~\ref{fig:delta} and Table~\ref{tbl:rates}. This difference is clearly observable in the inset of Fig.~\ref{fig:muZQ-muDQ}(a) and in Fig.~\ref{fig:muZQ-muDQ}(b). Two different initial rates $\mu_{\text{\tiny ZQ}}$ and $\mu_{\text{\tiny DQ}}$ extracted from these plots are listed in Table~\ref{tbl:rates}. 


The experimental data in the linear plots of Fig.~\ref{fig:muZQ-muDQ}(a) are compared with the plotted theoretical solutions of Eqs.(\ref{C17}) that use the measured initial rates from Table~\ref{tbl:rates}.  The plotted range of the theoretical uncertainty in this figure is almost hidden behind the plotted experimental data points. According to Eqs.(\ref{C17}), the relaxation of $\langle 2 S_{1z}S_{2z} \rangle(t)$ is not supposed,  to be monoexponential, which is especially clear for the ZQ states in Fig.~\ref{fig:muZQ-muDQ}(b). That figure also exhibits deviation  between the long-time tails of the theoretical and the experimental plots for the ZQ states but only at the level of about 1 percent of the initial value. Even then, the theory correctly predicts that the plot changes sign and also predicts the zero-crossing point with accuracy of about 10 percent.

The experimental results for the off-diagonal relaxation associated with $\langle 2 S_{1x}S_{2x} \rangle(t)$ for the four Bell PPSs are presented in Fig.~\ref{fig:lambdaZQ-lambdaDQ}. These results are obtained using  the pulse sequence given in the figure, which includes the CPMG sequence $(\pi)_{1y}, (\pi)_{2y} \ \text{---} \ \tau \ \text{---} \ (-\pi)_{1y}, (-\pi)_{2y} \ \text{---} \ \tau \ \text{---} ...$ with \mbox{$\tau = 1$~ms}. The CPMG sequence was used to suppress the dephasing effects due to the inhomogeneity of the static field and due to the $J$-coupling of the central spin pair to the nearby $^{15}$N spin.

According to Eq.(\ref{T2}) and the related discussion, the two ZQ states are expected to exhibit the same monoexponential decay characterized by the rate $\lambda_{\text{\tiny ZQ}}$. However, as apparent from Eqs.(\ref{rhoBell-S0}) and  (\ref{rhoBell-T0}), the initial values $\langle 2 S_{1x}S_{2x} \rangle(0)$ should be positive for the Bell PPS containing $\vert T_{0,z} \, \rangle $ and negative for the one containing $\vert S_0 \, \rangle $. Likewise, the two DQ Bell PPSs containing $\vert \psi_{+,z} \, \rangle$ or $\vert \psi_{-,z} \, \rangle$ should exhibit monoexponential decay characterized by the rate $\lambda_{\text{\tiny DQ}}$, with $\langle 2 S_{1x}S_{2x} \rangle(0)$ being positive for $\vert \psi_{+,z} \, \rangle$ and negative for $\vert \psi_{-,z} \, \rangle$.

All experimental plots of $\langle 2 S_{1x}S_{2x} \rangle(t)$ in Fig.~\ref{fig:lambdaZQ-lambdaDQ}(a) are generally consistent with monoexponential decays used for the theoretical fitting functions. However, the long-time tails of these plots exhibit a slight positive drift. This drift is the same for all four Bell PPSs, which suggests that a non-ideal implementation of the CPMG sequence is the likely culprit. It is possible to refine the plots from the above drift by exploiting the fact that the two ZQ PPSs  are supposed to exhibit the same exponential decays, but one with the positive and the other with the negative values of $\langle 2 S_{1x}S_{2x} \rangle(t)$. The subtraction of the two measured curves from each other cancels their overall positive drift, while preserving the exponentially decaying parts. At the same time, the averaging  of the two curves cancels the monoexponential decays, while revealing the drift. The same can also be done for the two DQ states. The results of the subtractions for the ZQ and DQ states are shown in the semilog plot of Fig.~\ref{fig:lambdaZQ-lambdaDQ}(b), revealing two clearly monoexponential decays, from which the values of $\lambda_{\text{\tiny ZQ}}$ and $\lambda_{\text{\tiny DQ}}$ listed in Table~\ref{tbl:rates} were extracted. The inset in Fig.~\ref{fig:lambdaZQ-lambdaDQ}(b) shows the averages of $\langle 2 S_{1x}S_{2x} \rangle(t)$ for for two ZQ and for two DQ states, thereby revealing the same drift at the level of one percent.

Finally, we have also performed a control experiment measuring $\langle 2 S_{1z}S_{2z} \rangle(t)$ in the presence of the CPMG sequence. The results of this experiment and the underlying theoretical description are presented in Appendix~\ref{SzSz_CPMG}.

\section{Analysis of experimental results }
\label{discussion}

\subsection{Extracting microscopic characteristics from the measured rates}

\begin{table}[t]
    \begin{tabular}{ | c | c |  }   
     \hline 
     Microscopic characteristic   & Value [s$^{-1}$] \\
     \hline
     $ k^2 \mathcal{J}_0 $ & $0.76 \pm 0.08$ \\
     $ \langle\vert \alpha_{1\perp} \vert^2\rangle \ \mathcal{J}_0 $ & $0.06 \pm 0.02$ \\
     $ \langle\vert \alpha_{2\perp} \vert^2\rangle \ \mathcal{J}_0 $ & $0.02 \pm 0.02$ \\
     $ \left[  \langle \alpha_{1z}^2 \rangle + \langle \alpha_{2z}^2 \rangle \right] \mathcal{J}_0 $ & $0.25 \pm 0.04 $  \\
     $ \langle \alpha_{1z} \alpha_{2z}  \rangle \  \mathcal{J}_0 $  & $0.026 \pm 0.011$ \\
     $\langle F_1 \alpha_{1\perp}^* + F_1^*\alpha_{1\perp}\rangle \ \mathcal{J}_0$ & $0.0159 \pm 0.0008$ \\
     $\langle F_1 \alpha_{2\perp}^* + F_1^*\alpha_{2\perp}\rangle \ \mathcal{J}_0$ & $- 0.026 \pm 0.004$  \\
     \hline
      \end{tabular}
    \caption{Microscopic characteristics extracted from the rate measurements with the help of Eqs.(\ref{k_sigma12}, \ref{alpha_perp_cor}, \ref{alpha_z_quadrat_cor}, \ref{alpha_z1_z2_cor}) and from Eq.(\ref{delta_n}).}
    \label{tbl:microscopic}
\end{table}

Let us start by noting that the NOE rate $\sigma_{12}$ is supposed to be determined exclusively by the intra-pair $^1$H~--$^{\ 13}$C spin-spin coupling. We assume that (i) the fluctuating part of the intra-pair Hamiltonian is completely dominated by IPMDI given by Eq.(\ref{Hd}), and that (ii) the limit $\Omega_1, \Omega_2 \ll 1/\tau_c$ is applicable. As shown in Appendix~\ref{derivations}, this implies the relation
\begin{equation}
    k^2 \mathcal{J}_0 = 4 \sigma_{12},
    \label{k_sigma12}
\end{equation}
which then allows us to express the IPMDI contributions to all other rates, as being proportional to $\sigma_{12}$, namely: $\mu_n^d = 2 \sigma_{12}$, $\mu_{12}^d = \frac{6}{5} \sigma_{12}$, $\lambda_{\text{\tiny ZQ}}^d = \frac{4}{5} \sigma_{12}$, and $\lambda_{\text{\tiny DQ}}^d = \frac{9}{5} \sigma_{12}$.

Using the above relations together with  Eq.(\ref{mu_12_tot}) and (\ref{mun_alpha})), we first obtain
\begin{equation}
 \langle\vert \alpha_{n\perp} \vert^2\rangle \ \mathcal{J}_0 \  = \  \frac{1}{2} \ \mu_n \ - \  \sigma_{12},
   \label{alpha_perp_cor} 
\end{equation}
and then, with the help of Eqs.(\ref{lambda_pm_tot}, \ref{lambda_plus_alpha}, \ref{lambda_minus_alpha}) arrive at
\begin{equation}
\! \! \! \! \langle (\alpha_{1z} + \alpha_{2z})^2 \rangle \mathcal{J}_0 = 2 \lambda_{\text{\tiny DQ}} - \frac{18}{5} \sigma_{12} - 2 \left[ \langle \vert \alpha_{1\perp} \vert^2 \rangle + \langle \vert \alpha_{2\perp}\vert^2 \rangle \right] \mathcal{J}_0   
    \label{alpha_z_plus_cor}
\end{equation}
\begin{equation}
\! \! \! \!  \langle (\alpha_{1z} - \alpha_{2z})^2 \rangle \mathcal{J}_0 = 2 \lambda_{\text{\tiny ZQ}} - \frac{8}{5} \sigma_{12} - 2 \left[ \langle \vert \alpha_{1\perp} \vert^2 \rangle + \langle \vert \alpha_{2\perp}\vert^2 \rangle \right] \mathcal{J}_0  . 
    \label{alpha_z_minus_cor}
\end{equation}
By adding and subtracting Eqs.(\ref{alpha_z_plus_cor}) and (\ref{alpha_z_minus_cor}), and also using Eq.(\ref{alpha_perp_cor}), we, finally, obtain
\begin{equation}
 \left[  \langle \alpha_{1z}^2 \rangle + \langle \alpha_{2z}^2 \rangle \right] \mathcal{J}_0 
= 
\lambda_{\text{\tiny DQ}} + \lambda_{\text{\tiny ZQ}} - \mu_1 - \mu_2 + \frac{7}{5} \sigma_{12} 
    \label{alpha_z_quadrat_cor}
\end{equation}
and
\begin{equation}
   \langle \alpha_{1z} \alpha_{2z}  \rangle \  \mathcal{J}_0 
= 
\frac{\lambda_{\text{\tiny DQ}} - \lambda_{\text{\tiny ZQ}} - \sigma_{12}}{2} .
    \label{alpha_z1_z2_cor}
\end{equation}

The values of the microscopic characteristics in the left-hand-sides of Eqs.(\ref{k_sigma12}, \ref{alpha_perp_cor}, \ref{alpha_z_quadrat_cor}, \ref{alpha_z1_z2_cor}) obtained from the experimentally measured rates are listed in Table~\ref{tbl:microscopic}, which, in addition, includes $\langle F_1 \alpha_{n\perp}^* + F_1^*\alpha_{n\perp}\rangle \mathcal{J}_0$ obtained from Eq.(\ref{delta_n}) using the experimental values of $\delta_1$ and $\delta_2$.

We, finally, use Eq.(\ref{k}) with the estimate $r \approx 1.1$~\AA \ for  the bond length of the $^1$H~--$^{\ 13}$C pair to obtain $k \approx 1.4 \cdot 10^5 \ \text{s}^{-1}$, which, in turn, given the value for $k^2 \mathcal{J}_0$ from Table~\ref{tbl:microscopic}, yields \mbox{$\mathcal{J}_0 \approx 3.9 \cdot 10^{-11} \ \text{s}$}. These numbers for $k$ and $\mathcal{J}_0$ are entirely consistent with the assumptions made throughout the article.

\subsection{Quantitative tests of theoretical predictions}
\label{tests}

Here we summarize the parameter-free experimental tests of the density-matrix relaxation theory that was presented Section~\ref{relaxation-theory}:

(i) {\it Overall shape of the relaxation functions:}

While the six rates $\mu_1$, $\mu_2$, $\mu_{12}$, $\sigma_{12}$, $\delta_1$ and $\delta_2$ of the diagonal relaxation were extracted from the initial behavior  of the relaxation functions in Figs.~\ref{fig:mu1-mu2-sig12}, \ref{fig:delta} and \ref{fig:muZQ-muDQ}, the comparison of the overall experimental plots in those figures with the solutions of Eqs.(\ref{C17}) amount to a set of parameter-free tests of the theory. All these tests exhibited a very good agreement between the experiments and the theory.

For the off-diagonal relaxation, the theory predicted mono-exponential decays starting from Bell PPSs, which was confirmed in Fig.~\ref{fig:lambdaZQ-lambdaDQ}(b). 

Furthermore, the theory predicted that the two ZQ Bell PPSs exhibit the same relaxation, which is different from the one exhibited by the two DQ Bell PPSs. This prediction was made for both the diagonal and the off-diagonal relaxation and confirmed, respectively, in Figs.~\ref{fig:muZQ-muDQ} and \ref{fig:lambdaZQ-lambdaDQ}. 

(ii) {\it Test of the rate ratio (\ref{mu-sigma_ratio}) determined by the IPMDI contributions to the relaxation rates:}

This test is presented in Table~\ref{tbl:rates}: it exhibits a very good agreement between theory and experiment. (The broader significance of the ratio (\ref{mu-sigma_ratio}) is to be further discussed in Section~\ref{sec:ratio}.)

(iii) {\it Test of the rate relation (\ref{rate-difference}) associated with the cross-correlation effects:}

The test of the relation (\ref{rate-difference}) is presented in Table~\ref{tbl:rates} in terms of the dimensionless ratio $\displaystyle \frac{(\mu_{\text{\tiny ZQ}} - \mu_{\text{\tiny DQ}})(\varepsilon_1 + \varepsilon_2)}{\delta_1\, \varepsilon_1  \,  +  \, \delta_2 \,  \varepsilon_2}$, which is supposed to be equal to 8 theoretically. The experimental value for this ratio, $9 \pm 6$,  is consistent with the theory, but the conclusiveness of the test is undermined by the large experimental error. The latter is due to the subtraction of two close-valued quantities $\mu_{\text{\tiny ZQ}}$ and $\mu_{\text{\tiny DQ}}$ in the numerator, and, in addition, due to  the opposite signs of $\delta_1$ and $\delta_2$ in the denominator, which also imply a subtraction. It should be mentioned, however, that the experimental ratio would be off by a factor of two, if the opposite signs of $\delta_1$ and $\delta_2$ (apparent in Fig.~\ref{fig:delta}) were overlooked.

(iv) {\it Inversion recovery of $^1$H in a sample with natural abundance of carbon isotopes.}

\begin{figure}[!t]
    \centering
    \begin{tikzpicture}
    \node[inner sep=0pt] (duck) at (0,0)    {\includegraphics[width=0.98\columnwidth]{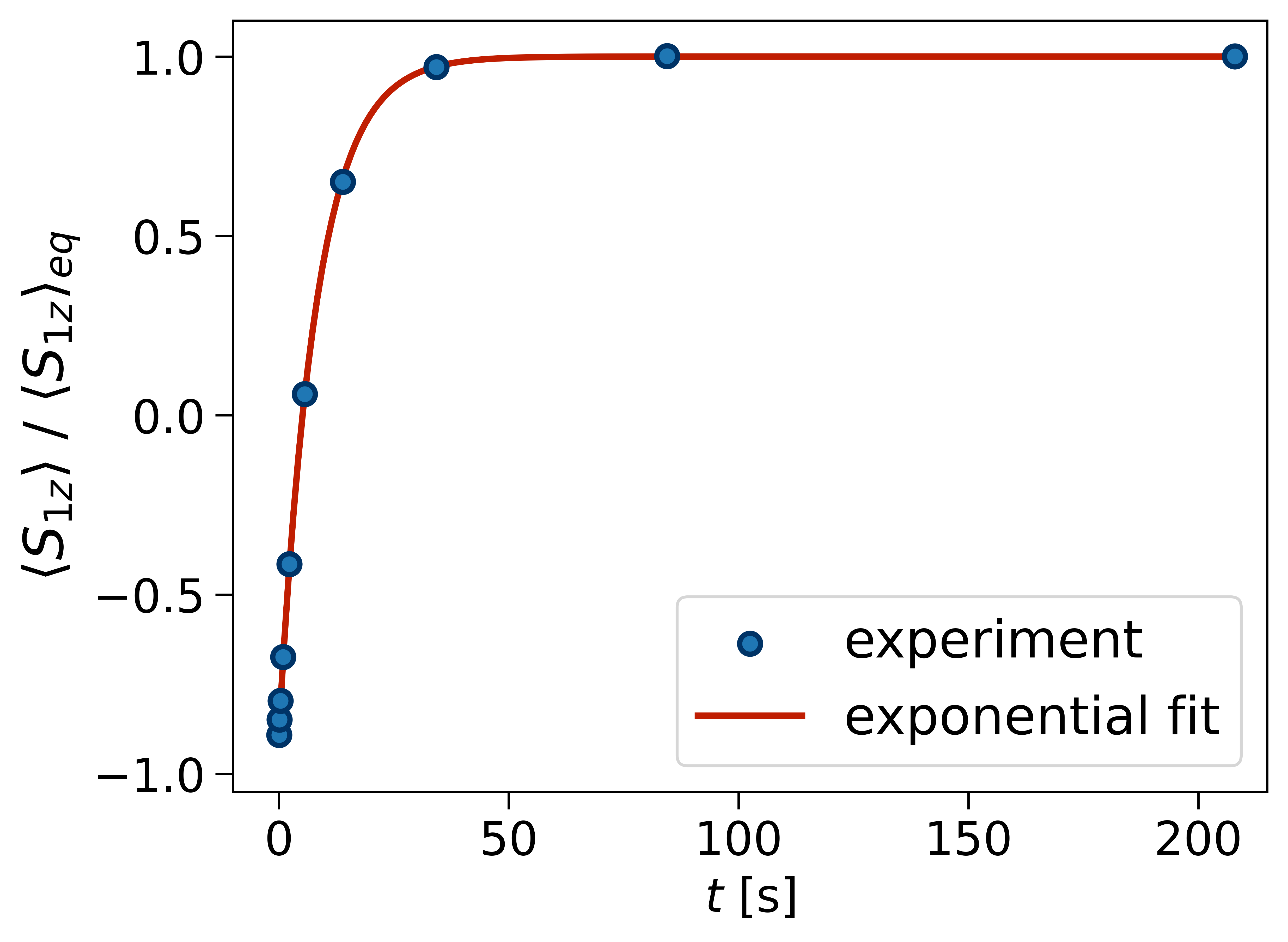}};
     
    \end{tikzpicture}
  
    \caption{
\label{fig:12C}
Control experiment on a sample with the natural abundance of carbon isotopes: inversion recovery of $^{1}H$ spin polarization $\langle S_{1z} \rangle$ within the selected C--H pair  after a \mbox{$\pi$-pulse} on  $^{1}H$. Blue dots are the experimental points. Red line is a fit of the form $\langle S_{1z} \rangle (t) = 1 - A \cdot   \exp{(- \mu_{\text{\tiny H}} t)}$, where $\mu_{\text{\tiny H}} = 0.122 \pm 0.001 \ \text{s}^{-1} $ and $A = 1.887 \pm 0.005$.
}
\end{figure}

This test is based on the control experiment presented in Fig.\ref{fig:12C}, where the inversion recovery ($T_1$ relaxation) of the $^1$H spin belonging to the selected C--H pair was measured in a sample with the natural abundance of carbon isotopes (98.9 percent of non-magnetic $^{12}$C). 
In this case the intra-pair magnetic dipolar mechanism is predominantly absent due to the absence of the magnetic $^{13}$C isotope. According to Eqs.(\ref{mu_n_tot}) and (\ref{mun_alpha}), the relaxation of $^1$H  is then supposed to be characterized by the rate $\mu_1 \approx 2 \langle\vert \alpha_{n\perp} \vert^2\rangle \mathcal{J}_0 = 0.12 \pm 0.04 \ \text{s}^{-1}$, where the value of $\langle\vert \alpha_{n\perp} \vert^2\rangle \mathcal{J}_0$ is substituted from Table~\ref{tbl:microscopic}. This prediction is in an excellent agreement with the rate $\mu_{\text{\tiny H}} = 0.122 \pm 0.001 \ \text{s}^{-1} $ extracted from the experimental plot in Fig.~\ref{fig:12C}.

(v) {\it Test of inequality (\ref{F1_alpha-n_inequality}) for the cross-correlation averages}

The numbers from Table~\ref{tbl:microscopic} together with Eq.(\ref{F1av}) can be used to test inequality (\ref{F1_alpha-n_inequality}). Those numbers, indeed, result in valid numerical inequalities, giving  for  $^1$H and $^{13}$C (with factor $\mathcal{J}_0$ omitted) \mbox{$ 0.0159 \leq 0.33 $} and \mbox{$ 0.026 \leq 0.19 $}, respectively. 

Finally, using the values from Table~\ref{tbl:rates}, we obtain  the experimentally determined ratio $\frac{\lambda_{\text{\tiny DQ}}}{\lambda_{\text{\tiny ZQ}}}  = 1.74 \pm 0.03$, which is noticeably smaller than the value $2.25$ appearing in Eq.(\ref{lambda_d_ratio}) for the relaxation caused exclusively by IPMDI. This is not a parameter-free test of the theory but rather an indicator of significant contributions of other relaxation mechanisms to $\lambda_{\text{\tiny ZQ}}$ and/or $\lambda_{\text{\tiny DQ}}$.

\subsection{Contributions of different relaxation mechanisms to the experimental results}
\label{contributions}

Here we analyze the microscopic parameters listed in Table~\ref{tbl:microscopic} in terms of assessing the contributions of various relaxation mechanisms.

Let us start by noting that the relatively large value of $k^2 \mathcal{J}_0$ implies the dominance of the IPMDI relaxation mechanism, as expected. Below we use the results in Table~\ref{tbl:microscopic} to separate various contributions to the local fields $\boldsymbol{\alpha}_n$. 

At least two contributions to $\boldsymbol{\alpha}_n$ are definitely expected in our case, namely, $\boldsymbol{\alpha}_n^{\text{\tiny CSA}}$ and $\boldsymbol{\alpha}_n^{\text{\tiny D}}$. We discriminate them using the fact, explained after Eq.~(\ref{delta_n}), that  $\boldsymbol{\alpha}_n^{\text{\tiny D}}$ cannot contribute to the cross-correlation averages $\langle F_1 \alpha_{n\perp}^* + F_1^*\alpha_{n\perp}\rangle \mathcal{J}_0$ determining rates $\delta_n$, while $\boldsymbol{\alpha}_n^{\text{\tiny CSA}}$ can. 

It is, generally, expected that the chemical shifts and hence the CSA fluctuations are larger for $^{13}$C than for $^1$H. On the other hand, the local fields $\boldsymbol{\alpha}_n^{\text{\tiny CSA}}$  are defined in units of frequency, implicitly including the gyromagnetic ratios of 
$^{13}$C and $^1$H, which boosts the CSA contribution for the latter by a factor $\gamma_1/\gamma_2 \approx 4$.
The two trends should roughly compensate each other, and, indeed, we observe in Table~\ref{tbl:microscopic} that $|\langle F_1 \alpha_{2\perp}^* + F_1^*\alpha_{2\perp}\rangle \mathcal{J}_0|$ is only by factor 1.6 larger than $|\langle F_1 \alpha_{1\perp}^* + F_1^*\alpha_{1\perp}\rangle \mathcal{J}_0|$.

Given the preceding considerations, it is natural to expect that the fluctuations of $\alpha_{1\perp}^{\text{\tiny CSA}}$ are comparable or smaller than those of $\alpha_{2\perp}^{\text{\tiny CSA}}$. At the same time, according to Table~\ref{tbl:microscopic}, the value of $\langle\vert \alpha_{1\perp} \vert^2\rangle$ is 3 times larger than $\langle\vert \alpha_{2\perp} \vert^2\rangle$, which suggests that $\langle\vert \alpha_{1\perp} \vert^2\rangle$ is dominated by $\langle\vert \alpha_{1\perp}^{\text{\tiny D}} \vert^2\rangle$.  Furthermore, the fact that  $\langle\vert \alpha_{1\perp}^{\text{\tiny D}} \vert^2\rangle$ cannot exceed the measured value of $\langle\vert \alpha_{1\perp} \vert^2\rangle$  
implies that $\frac{\gamma_2^2}{\gamma_1^2}\langle\vert \alpha_{1\perp} \vert^2\rangle \mathcal{J}_0 \approx 0.004$ is the upper limit for $\langle\vert \alpha_{2\perp}^{\text{\tiny D}} \vert^2\rangle$. This limit is by factor of 5 smaller than the measured value of $\langle\vert \alpha_{2\perp} \vert^2\rangle$. Thereby, we conjecture that $\langle\vert \alpha_{2\perp} \vert^2\rangle$ is dominated by $\langle\vert{\alpha_{2\perp}^{\text{\tiny CSA}}}|^2\rangle$. 

The preceding analysis leaves us, however, with the following puzzle: How to explain the relatively large experimental value of $ \left[  \langle \alpha_{1z}^2 \rangle + \langle \alpha_{2z}^2 \rangle \right] \mathcal{J}_0$ listed in Table~\ref{tbl:microscopic}.   If $\boldsymbol{\alpha}_1$ is dominated by $\boldsymbol{\alpha}_1^{\text{\tiny D}}$ representing the magnetic dipolar fields of distant $^1$H spins, then the isotropic average of these fields should imply $\langle \alpha_{1x}^2 \rangle = \langle \alpha_{1y}^2 \rangle = \langle \alpha_{1z}^2 \rangle $ and, hence, $\langle \alpha_{1z}^2 \rangle = 2 \langle |\alpha_{1\perp}|^2    \rangle$ [we recall that  $\alpha_{1\perp} \equiv \frac{1}{2} (\alpha_{1x} - i \alpha_{1y})$]. At the same time, if $\boldsymbol{\alpha}_2$ is dominated by the CSA fluctuations, then, we can take for an estimate the case, where the bond direction of the $^1$H---$^{13}$C pair coincides with one of the principal axes of the $^{13}$C CSA tensor, which would imply\cite{Goldman1984} that $\langle \alpha_{2z}^2 \rangle = \frac{8}{3} \langle |\alpha_{2\perp}|^2    \rangle$. Substituting the values for 
$\langle |\alpha_{1\perp}|^2 \rangle$ and $\langle |\alpha_{2\perp}|^2 \rangle$ from Table~\ref{tbl:microscopic} into the above relations, we then obtain $ \left[  \langle \alpha_{1z}^2 \rangle + \langle \alpha_{2z}^2 \rangle \right] \mathcal{J}_0 = 0.17 \pm 0.07$, which is noticeably smaller than the experimental value $0.25 \pm 0.04$ listed in Table~\ref{tbl:microscopic}.

With the above discrepancy in mind, we have looked for an additional mechanism that would preferentially create fluctuating local fields $\boldsymbol{\alpha}_1$ and $\boldsymbol{\alpha}_2$ oriented along the direction of the static magnetic field determining the $z$-axis. 

Spin rotation relaxation has been neglected so far, because we assumed that the molecule was too large for that mechanism to significantly contribute. It can also be discarded in the present context, because it would be supposed to produce an isotropic distribution of fields $\boldsymbol{\alpha}_1$ and $\boldsymbol{\alpha}_2$ without any noticeable preference for the direction of the static magnetic field. 

In the next subsection, we propose a relaxation mechanism that would create the fluctuating local fields exclusively along the $z$-axis.

\subsection{Additional relaxation mechanism involving very weak $J$-coupling to distant $^1$H spins}
\label{additional}

The general idea of this mechanism is that it originates from very weak $J$-couplings of the $^1$H---$^{13}$C spin pair to  distant ${^1}H$ spins on the same molecule. Since $J$-couplings originate from the isotropic averaging of the transferred hyperfine couplings, the local fields caused by the $J$-couplings to distant spins do not fluctuate as a result of fast molecular reorientations.  Rather they fluctuate on much slower timescale associated with the $T_1$-relaxation of the distant spins. As a result, despite having very small amplitude, these fields can make noticeable contribution to the relaxation of the central pair because of their very long correlation time.  

To be specific, the molecule in Fig.~\ref{fig1} contains four $^1$H spins separated from the $^{13}$C by four molecular bonds and six more $^1$H spins separated by 5 bonds. Those distant spins are supposed to be coupled to $^{13}$C by very small $J$-couplings, while their $J$-couplings to the intra-pair $^1$H should be even smaller due to the additional C--H bond to bridge.  

The Hamiltonian of the $J$-coupling  between the central \mbox{$^1$H\,--\,$^{13}$C} spin pair and the distant $^1$H spins has form:
\begin{equation}
\mathcal{H}^{\text{\tiny J}} = \hbar \left(\alpha_{1z}^{\text{\tiny J}} S_{1z} + \alpha_{2z}^{\text{\tiny J}} S_{2z} \right),
\label{Hamiltonian-J}
\end{equation}
where
\begin{equation}
    \alpha_{nz}^{\text{\tiny J}} \ = \sum_{k}^{\text{distant spins}} J_{nk} S_{kz},
    \label{alpha-J_z}
\end{equation}
are the $z$-projections of the local fields $\boldsymbol{\alpha}_n^{\text{\tiny J}}$ defined by Eq.(\ref{alpha-J}). These projections, in turn, fluctuate because of the random flipping of $S_{kz}$ on the time scale $T_{1,\text{dist}}$ of the longitudinal relaxation of distant spins.  Since the fields $\boldsymbol{\alpha}_n^{\text{\tiny J}}$ are always directed along the $z$ spin axis, they do not affect the diagonal relaxation rates $\mu_n$, $\mu_{12}$, $\sigma_{12}$ and $\delta_n$. However, they do contribute to the off-diagonal rates $\lambda_{\text{\tiny ZQ/DQ}}$. 
[We note in this regard that the hard $\pi$-pulses used for the CPMG sequence during our measurements of the off-diagonal relaxation flip not only both spins of the central \mbox{$^1$H\,--\,$^{13}$C} pair but also the distant $^1$H spins. Such a CPMG sequence is supposed to leave the Hamiltonian (\ref{Hamiltonian-J}) unchanged.]

The applicability of the Redfield's theory based on Eqs.(\ref{v-relaxation},\ref{master}) to the present relaxation mechanism requires that $|\alpha_{nz}^{\text{\tiny J}}| \ll 1/T_{1,\text{dist}}$, which may, at first sight, appears problematic, given that $T_{1,\text{dist}}$ is rather long. However,  this mechanism is, by construction, about very small $J$-couplings, which can easily be much smaller than $1/T_{1,\text{dist}}$. Even if not, for a local field fluctuating along the $z$-axis only, one can compute the relaxation function using the stochastic model of Anderson and Weiss\cite{Anderson-Weiss-1953}, which would reproduce the predictions of the Redfield's theory in the limit $|\alpha_{nz}^{\text{\tiny J}}| \ll 1/T_{1,\text{dist}}$ and would also be applicable beyond that limit.  The Anderson-Weiss relaxation function has the form
\begin{equation}
    G(t) = \exp \left[\,- \int_0^t (t-t^{\prime}) \Tilde{C}(t^{\prime}) d t^{\prime}\right],
    \label{AW-relaxation}
\end{equation}
where $\Tilde{C}(t) = \langle h(t) h(0) \rangle$ is the correlation function of the fluctuating field. In the present context, this field is either \mbox{$h = \alpha_{1z}^{\text{\tiny J}} - \alpha_{2z}^{\text{\tiny J}}$} in the ZQ space, or $h = \alpha_{1z}^{\text{\tiny J}} + \alpha_{2z}^{\text{\tiny J}}$ in the DQ space. 

The expressions for the off-diagonal relaxation in the ZQ and DQ spaces due to all other mechanisms should now be multiplied by function $G(t)$. This function has exponential tail characterized by the rate \mbox{$\Tilde{\lambda} = \int_0^{\infty}  \Tilde{C}(t^{\prime}) d t^{\prime} = \frac{1}{2}\langle h^2 \rangle \Tilde{\mathcal{J}}_0 $}, where $\Tilde{\mathcal{J}}_0$ is the zero-frequency value of the spectral function for $\Tilde{C}(t)/\Tilde{C}(0)$. If $\Tilde{C}(t)$ decays exponentially with time constant $T_{1,\text{dist}}$, then $\Tilde{\mathcal{J}}_0 = 2 T_{1,\text{dist}}$. In a departure from the Redfield's theory, the onset of the exponential decay for $G(t)$ is delayed by the time $\sim T_{1,\text{dist}}$. We note in this regard that the time range of exponential fits in Fig.~\ref{fig:lambdaZQ-lambdaDQ}(b) is significantly longer than the expected value of $ T_{1,\text{dist}}$ of the order of a few seconds. 

Once the above considerations are taken into account, the expression (\ref{lambda_pm_tot}) for the off-diagonal relaxation rates should be modified to become  
\begin{equation}
\lambda_{\text{\tiny ZQ/DQ}} = \lambda_{\text{\tiny ZQ/DQ}}^d + \lambda_{\text{\tiny ZQ/DQ}}^{\alpha} + \Tilde{\lambda}_{\text{\tiny ZQ/DQ}}^{\text{\tiny J}},
    \label{lambda_pm_tot_tilda}
\end{equation}
where $\lambda_{\text{\tiny ZQ/DQ}}^d$ and $\lambda_{\text{\tiny ZQ/DQ}}^{\alpha}$ have been defined by Eqs. (\ref{lambda_plus_dd}-\ref{lambda_minus_alpha}), and the newly added third term is either
\begin{equation}
    \label{lambda_plus_alpha_tilde}
    \Tilde{\lambda}_{\text{\tiny ZQ}}^{\text{\tiny J}} =  \frac{\langle (\alpha_{1z}^{\text{\tiny J}} - \alpha_{2z}^{\text{\tiny J}})^2 \rangle}{2} \ \Tilde{\mathcal{J}}_0 ,
\end{equation}
or
\begin{equation}
    \label{lambda_minus_alpha_tilde}
    \Tilde{\lambda}_{\text{\tiny DQ}}^{\text{\tiny J}} =  \frac{ \langle (\alpha_{1z}^{\text{\tiny J}} + \alpha_{2z}^{\text{\tiny J}})^2 \rangle }{2} \  \Tilde{\mathcal{J}}_0 .
\end{equation}

Correspondingly, one should revisit the fourth and the fifth entries of Table~\ref{tbl:microscopic}, which were based on Eqs.(\ref{alpha_z_quadrat_cor}) and (\ref{alpha_z1_z2_cor}). The detailed form of the fourth entry now becomes
\begin{equation}
\begin{split}
    \left[  \left\langle {\alpha_{1z}^{\text{\tiny CSA}}}^2 \right\rangle + \left\langle {\alpha_{2z}^{\text{\tiny CSA}}}^2 \right\rangle 
    +
    \left\langle {\alpha_{1z}^{\text{\tiny D}}}^2 \right\rangle + \left\langle {\alpha_{2z}^{\text{\tiny D}}}^2 \right\rangle
    \right] \mathcal{J}_0
    \\
    +
    \left[  \left\langle {\alpha_{1z}^{\text{\tiny J}}}^2 \right\rangle + \left\langle {\alpha_{2z}^{\text{\tiny J}}}^2 \right\rangle \right] \Tilde{\mathcal{J}}_0 = 0.25 \pm 0.04 \  \text{s}^{-1}.
\end{split}
    \label{alpha_z_quadrat_new}
\end{equation}
The analysis in the preceding subsection gave the value 0.17~s$^{-1}$ for the contribution of the terms proportional to $\mathcal{J}_0$.
We, therefore, assume that the terms proportional to $\Tilde{\mathcal{J}}_0$ are responsible for the rest, i.e. $\left[  \left\langle {\alpha_{1z}^{\text{\tiny J}}}^2 \right\rangle + \left\langle {\alpha_{2z}^{\text{\tiny J}}}^2 \right\rangle \right] \Tilde{\mathcal{J}}_0 \approx 0.08 \ \text{s}^{-1}$.

The detailed form of the modified fifth entry of Table~\ref{tbl:microscopic} is
\begin{equation}
\begin{split}
    \left[ 
    \langle \alpha_{1z}^{\text{\tiny CSA}} \alpha_{2z}^{\text{\tiny CSA}}  \rangle
    +
    \langle \alpha_{1z}^{\text{\tiny D}} \alpha_{2z}^{\text{\tiny D}}  \rangle
    \right] \mathcal{J}_0
    +
    \langle \alpha_{1z}^{\text{\tiny J}} \alpha_{2z}^{\text{\tiny J}}  \rangle
    \Tilde{\mathcal{J}}_0 
    \\
    = 0.026 \pm 0.011 \  \text{s}^{-1}.
\end{split}
    \label{alpha_z1_z2_new}
\end{equation}
The experimental value in the rhs of Eq.(\ref{alpha_z1_z2_new}) is ten times smaller than that in Eq.(\ref{alpha_z_quadrat_new}). This is consistent with our expectation that $ \alpha_{2z}^{\text{\tiny D}}$ is significantly smaller than $\alpha_{1z}^{\text{\tiny D}}$, while also $\alpha_{1z}^{\text{\tiny J}}  $ is significantly smaller than $ \alpha_{2z}^{\text{\tiny J}}$. As a result, the larger value within each pair would dominate in Eq.(\ref{alpha_z_quadrat_new}), while the respective cross-correlation term in Eq.(\ref{alpha_z1_z2_new}) would necessarily be reduced due to the smaller of the two values. In principle, all three correlators appearing in Eq.(\ref{alpha_z1_z2_new}) can make comparable contributions to their sum. They cannot be further discriminated on the basis of our measurements.

We now make a concrete estimate, which neglects $\alpha_{1z}^{\text{\tiny J}}  $, while assuming that $\alpha_{2z}^{\text{\tiny J}}  $ originates from $N=4$ distant $^1$H spins coupled to the $^{13}$C spin of the central pair by the $J$-coupling $J_{\text{dist}}/(2 \pi \hbar) = 0.06$~Hz and fluctuating with the characteristic time \mbox{$T_{1,\text{dist}} = 3$~s} (which we also measured). This implies  \mbox{$ \left\langle {\alpha_{2z}^{\text{\tiny J}}}^2 \right\rangle  \Tilde{\mathcal{J}}_0  = N \left( \frac{1}{2} J_{\text{dist}}/ \hbar\right)^2 (2T_{1,\text{dist}})  \approx 0.08 \ \text{s}^{-1}$}, which would account for the contribution of the $J$-coupling fields to Eq.(\ref{alpha_z_quadrat_new}).

\section{Dimensionless ratio $\displaystyle \frac{\mu_1  \,  +  \, \mu_2  \,  - \, \mu_{12}}{\sigma_{12}} $ }
\label{sec:ratio}

\begin{table*}[t]
     \begin{center}
     \begin{tabular}{ | c | r || c | c | c | c || c | }   
     \hline      
     Ref.  & System \& parameters  & $\mu_1$ [s$^{-1}$] & $\mu_2$ [s$^{-1}$]  & $\mu_{12}$ [s$^{-1}$]   & $\sigma_{12}$ [s$^{-1}$] & \makecell{ \vspace{3pt } $\displaystyle    \frac{\mu_1  \,  +  \, \mu_2  \,  - \, \mu_{12}}{\sigma_{12}} $  }    \\
     \hline
     \cite{Maeler-1992}  &  \multirow{1}{*}{\makecell{ $^{13}$C\,--$^1$H$^{^{\textcolor{white}{A}}}$ \ \ \ \ \ \ \ \ \ \ \ \ \ \ \  \\ in methyl formate: \ \ }}                     & & & & &   \\
                         & $B = 4.7 \ \text{T}$, $\tau_m = 15 \ \text{s}$  & $0.0275 \pm 0.0006$ & $0.0446 \pm 0.0008$ & $0.0415 \pm 0.0004$ & $0.0107 \pm 0.0004$  & $2.86 \pm 0.15$   \\
                         & $\tau_m = 30 \ \text{s}$                        & $ 0.0270 \pm 0.0005 $ & $ 0.0438 \pm 0.0008 $ & $ 0.0411 \pm 0.0003 $ & $ 0.0107 \pm 0.0003 $ & $ 2.78 \pm 0.12 $ \\
                         & $B = 9.4 \ \text{T}$, $\tau_m = 15 \ \text{s}$  & $ 0.0310 \pm 0.0011 $ & $ 0.0439 \pm 0.0013 $ & $ 0.0448 \pm 0.0006 $ & $0.0098 \pm 0.0006 $ & $3.1 \pm 0.3$ \\ 
                         & $\tau_m = 30 \ \text{s}$                        & $ 0.0313 \pm 0.0009 $ & $ 0.0446 \pm 0.0013 $ & $ 0.0447 \pm 0.0006 $ & $ 0.0101 \pm 0.0006 $ & $ 3.2 \pm 0.3 $ \\ 
                         \cline{2-7}
                         &  \multirow{1}{*}{\makecell{ $^{13}$C\,--$^1$H$^{^{\textcolor{white}{A}}}$ \ \ \ \ \ \ \ \ \ \ \ \ \ \ \  \\ in chloroform:  \ \ \ \ \ \ \ \ \   }}               & & & & &   \\
                         & $B = 4.7 \ \text{T}$,  $\tau_m = 1 \ \text{s}$ \ \    & $ 0.463 \pm 0.005 $ & $ 0.411 \pm 0.005 $ & $ 0.288 \pm 0.002 $ & $ 0.192 \pm 0.003 $ & $ 3.05 \pm 0.06 $   \\
                         & $\tau_m = 3 \ \text{s}$ \ \                         & $ 0.455 \pm 0.004 $ & $ 0.411 \pm 0.004 $ & $ 0.292 \pm 0.001 $ & $ 0.199 \pm 0.003 $ & $ 2.88 \pm 0.05 $ \\
                         & $B = 9.4 \ \text{T}$, $\tau_m = 1 \ \text{s}$ \ \    & $ 0.469 \pm 0.016 $ & $ 0.443 \pm 0.016 $ & $ 0.322 \pm 0.007 $ & $ 0.207 \pm 0.008 $ & $2.85 \pm 0.16 $ \\ 
                         & $\tau_m = 3 \ \text{s}$ \ \                         & $ 0.477 \pm 0.016 $ & $ 0.437 \pm 0.014 $ & $ 0.316 \pm 0.004 $ & $ 0.212 \pm 0.010 $ & $ 2.82 \pm 0.17 $ \\ 
     \hline
     \cite{Maeler-1995}  & \multirow{1}{*}{\makecell{ $^1$H$_A$\,--$^1$H$_X^{^{\textcolor{white}{A}}}$ in \ \ \ \ \ \ \ \ \ \ \ \  \\ $cis$-chloroacrylic acid:    }}                       & & & & &   \\
                         & without Ni$^{2+}$                               & $ 0.109 \pm 0.007 $ & $ 0.116 \pm 0.008 $ & $ 0.091 \pm 0.003 $ & $ 0.046 \pm 0.004 $ & $ 2.9 \pm 0.3 $   \\
                         & with Ni$^{2+}$                                  & $ 0.442 \pm 0.013 $ & $ 0.286 \pm 0.009 $ & $ 0.469 \pm 0.006 $ & $ 0.051 \pm 0.006 $ & $ 5.1 \pm 0.7 $ \\
     \hline
     \cite{Steiner-2015} & \multirow{1}{*}{\makecell{ $^{13}$C\,--$^1$H$^{^{\textcolor{white}{A}}}$ \ \ \ \ \ \ \ \ \ \ \ \ \ \ \  \\ in chloroform:  \ \ \ \ \ \ \ \ \   }}             &  &  &  &  &    \\
                         & free in solution                                & $ 0.09 $ & $ 0.11 $ & $ 0.050 \pm 0.004 $ & $ 0.045^{*} $ & $ 3.33 \pm 0.09 $   \\
                         & in cryptophane-D                                & $ 3.6 $ & $ 2.8 $ & $ 4.0 \pm 1.0 $ & $ 1.2 \pm 0.1 $ & $ 2.0 \pm 0.8 $ \\
     \hline 
      \end{tabular}
      \caption{Literature-based tests of the theoretical prediction (\ref{mu-sigma_ratio}) that $ \frac{\mu_1  \,  +  \, \mu_2  \,  - \, \mu_{12}}{\sigma_{12}} = 2.8$ in the limit $\Omega_1, \Omega_2 \ll 1/\tau_c$.   The tests use the experimental results of Refs.~\cite{Maeler-1992,Maeler-1995,Steiner-2015}. Nuclear spin pairs involved in the tests are given in the second column, together with the the relevant experimental details (composition, static magnetic field $B$ and mixing time $\tau_m$) allowing one to trace hereby listed numbers back to the original references. (The rate marked by $^*$ was not measured but rather deduced based on the assumption that $\sigma_{12} = \mu_1/2$.)}
      \label{tbl:ratios}
      \end{center}
      \end{table*}

In this section, we discuss the  applicability of the rate relation (\ref{mu-sigma_ratio}) beyond our concrete system. This relation predicted the value 2.8 for the dimensionless ratio $\frac{\mu_1  \,  +  \, \mu_2  \,  - \, \mu_{12}}{\sigma_{12}} $, which was consistent with our experiments (see Table~\ref{tbl:rates}). According to Section~\ref{D-relaxation}, the rate contributions from all mechanisms described in terms of fluctuating local fields $\boldsymbol{\alpha}_n$ cancel in the numerator of  the above ratio in the limit $ \Omega_1, \Omega_2 \ll 1/\tau_c $, while the denominator is independent of $\boldsymbol{\alpha}_n$. We further note that the rate contributions from fields $\boldsymbol{\alpha}_n$ cancel also when those fields have other correlation times, longer or shorter than $\tau_c$.  The ratio $\frac{\mu_1  \,  +  \, \mu_2  \,  - \, \mu_{12}}{\sigma_{12}} $ should thus have the universal value 2.8 for a broad class chemical-exchange-stable nuclear spin pairs experiencing  relaxation due to a combination of (i) the IPMDI  mechanism in the limit $ \Omega_1, \Omega_2 \ll 1/\tau_c $, and (ii) all other mechanisms describable in terms of local-fields uncorrelated with the spin state of the pair.


We, therefore, searched the literature for the examples of  nuclear spin pairs, for which $\mu_1$, $\mu_2$, $\mu_{12}$ and $\sigma_{12}$ were simultaneously measured. Such measurements were, indeed, reported by M{\"a}ler, Kowalewski and coauthors in Refs.\cite{Maeler-1992,Maeler-1995,Steiner-2015}. The experimental rates taken from Refs.\cite{Maeler-1992,Maeler-1995,Steiner-2015} together with the tests of the rate relation (\ref{mu-sigma_ratio}) are presented in Table~\ref{tbl:ratios}.

As one can see in Table~\ref{tbl:ratios}, the measurements of Ref.~\cite{Maeler-1992} involving $^{13}$C\,--$^1$H pairs in methyl formate and in chloroform agree very well with the rate ratio 2.8. 

The same table also presents the results from Ref.~\cite{Maeler-1995} for a pair of non-equivalent proton spins, $^1\text{H}_A$ and $^1\text{H}_X$ in $cis$-chloracrylic acid in a solution with and without Ni$^{2+}$ paramagnetic centers. The rates measured without   Ni$^{2+}$ are consistent with the ratio 2.8. In the presence of Ni$^{2+}$, there is a factor-of-two discrepancy between 2.8 and the observed ratio. The likely reason for this is that the condition $ \Omega_1, \Omega_2 \ll 1/\tau_c $ is violated once the ligand containing spin pair is attached to a paramagnetic center. In addition, the attachment-detachment process can modulate the intra-pair $J$-coupling on a relatively slow time scale, which, in turn, may be competing with \mbox{IPMDI}. 

Table~\ref{tbl:ratios} also contains two sets of rates from Ref.~\cite{Steiner-2015} that describe a solution, where chloroform fluctuates between a free state and the state encaged in cryptophane-D.  In the former state, the experiment-based ratio is  reasonably close to 2.8, yet the discrepancy is larger than the experimental error. Here, however, $\sigma_{12}$ was not measured directly but rather assumed to be equal to $\mu_1/2$. For the chloroform encaged in cryptophane-D, the experimental ratio is consistent with 2.8 but with rather large experimental uncertainty. On the theoretical ground, a large discrepancy here would not be surprising, because the condition $ \Omega_1, \Omega_2 \ll 1/\tau_c $ may be violated due to a relatively long reorientation time $\tau_c$.

To summarize, in all cases presented in Table~\ref{tbl:rates} where the condition $ \Omega_1, \Omega_2 \ll 1/\tau_c $ is supposed to be fulfilled, the experimental values of the ratio (\ref{mu-sigma_ratio}) are, indeed, well consistent with 2.8. 

Let us remark that the fact that the intra-pair relaxation mechanism methyl formate and in chloroform is dominated by the magnetic dipolar  interaction in the limit $ \Omega_1, \Omega_2 \ll 1/\tau_c $ was apparent to the authors of Ref.~\cite{Maeler-1992}. They, however, chose to test its validity not through the ratio (\ref{mu-sigma_ratio}) but rather using the combinations 
$\mu_1 + \mu_2 - \mu_{12} \pm \sigma_{12}$ to isolate and compare rate contribution proportional to $\mathcal{J}(\Omega_1 \pm  \Omega_2)$ thereby testing the expectation that $\mathcal{J}(\Omega_1 +  \Omega_2) \approx \mathcal{J}(\Omega_1 -  \Omega_2)$. The advantage of the test based on the ratio (\ref{mu-sigma_ratio}) is that one can use this ratio also to compare different systems. Furthermore, in suitable cases, ratio (\ref{mu-sigma_ratio}) can help to determine one of the four participating rates from measuring only three of them: e.g., $\mu_{12}$ from the measurements of $\mu_1$, $\mu_2$ and $\sigma_{12}$.




\section{Summary and conclusions} 
\label{conclusions}

In this work, we performed experimental and theoretical multi-rate characterization of the relaxation of two interacting nuclear spins 1/2 in a liquid. Some of our experiments involved Bell PPSs, which were prepared by the method exploiting detuned Hartmann-Hahn resonant condition.  Eight different relaxation rates were measured and decomposed in terms of contributions from various microscopic mechanisms. The outcomes of this analysis include, in particular: (i) the connection between eigenmodes of the off-diagonal relaxation and the initial conditions based on the maximally entangled Bell PPSs; (ii) identification of the contribution to the off-diagonal relaxation associated with very weak $J$-couplings between the selected spin pair and other distant spins.  A number of parameter-free tests of the theory resulted in either good or excellent agreement between the theory and the experiment. Particularly notable was the parameter-free test based on Eq.(\ref{mu-sigma_ratio}) that depended only on the contributions of the intra-pair magnetic dipolar interaction to the measured rates. Eq.(\ref{mu-sigma_ratio}) was also found to be in a very good agreement with other published experimental results.

Overall the present work demonstrates the utility and effectiveness of multirate relaxation measurements involving Bell PPSs. Such measurements can be used as a tool for discriminating the relaxation contributions from various mechanisms in liquid-state NMR, and, more generally, as additional relaxation markers in the studies of complex molecules. 


\begin{acknowledgements}  
The authors acknowledge helpful communication from J. Matysik. This work was performed using the equipment of the MIPT Shared Facilities Center.
\end{acknowledgements}
 
\appendix

\section{Effects of $^{15}$N spin on the $^1$H~--$^{\ 13}$C pair }
\label{N15}

\begin{figure}[!t]
    \centering
    \begin{tikzpicture}
    \node[inner sep=0pt] (duck) at (0,0)    {\includegraphics[width=0.98\columnwidth]{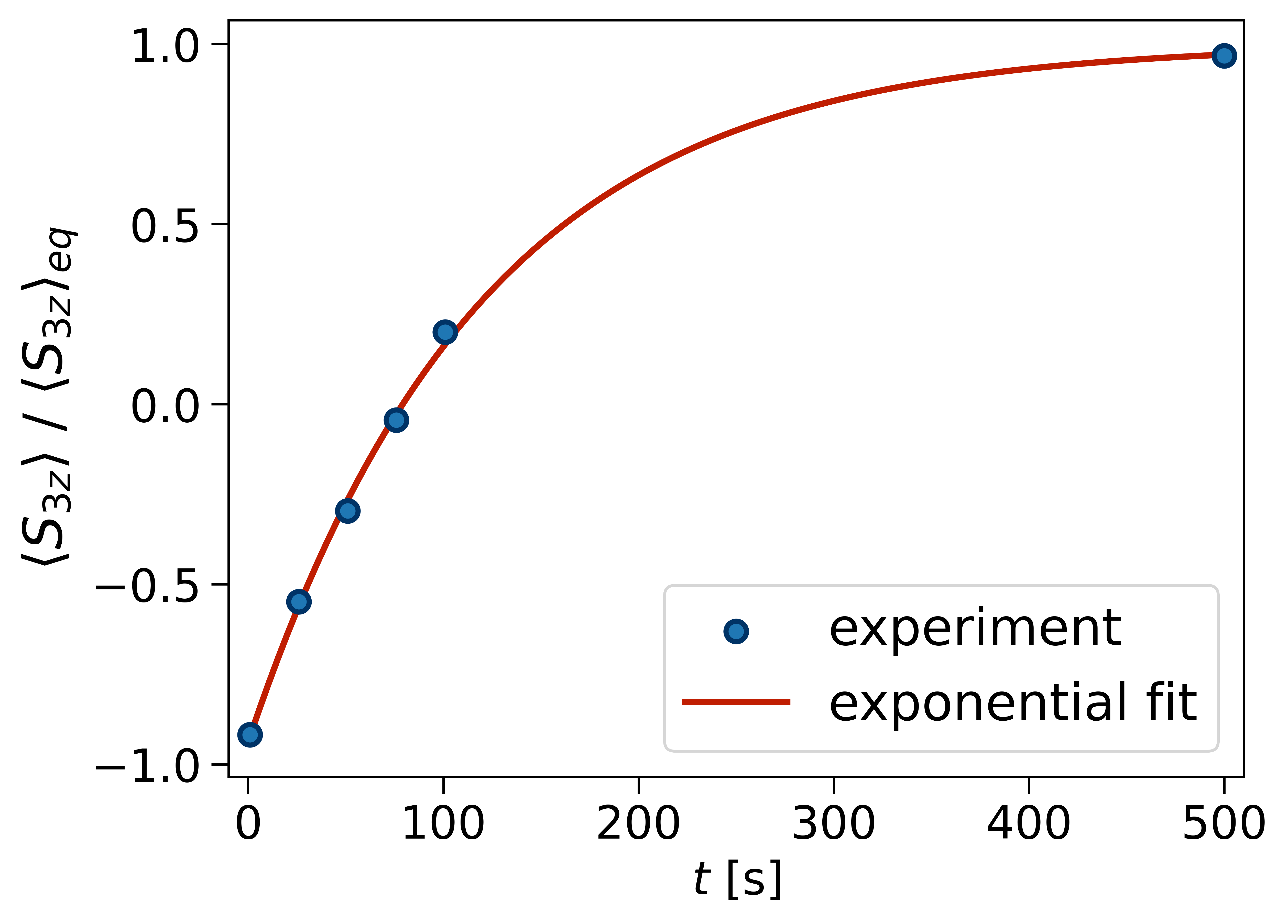}};
     
    \end{tikzpicture}
  
    \caption{
\label{fig_N}
Inversion recovery of $^{15}N$ spin polarization $\langle S_{3z} \rangle$ after a $\pi$ pulse on  $^{15}N$. Blue dots are the experimental points. Red line is a fit of the form $\langle S_{3z} \rangle (t) = 1 - A \cdot   \exp{(-t/T_1^N)}$, where $T_1^N = (1.20 \pm 0.04) \cdot 10^2$~s and $A = 1.94 \pm 0.02$.
}
\end{figure} 

The effects of $^{15}N$ spin on the relaxation of the central $^1$H~--$^{\ 13}$C pair  can be both incoherent and coherent. The incoherent effects in the present liquid-state setting are dominated by the magnetic dipolar coupling between $^{15}$N and the $^1$H~--$^{\ 13}$C pair, while the coherent effects come from the much weaker $J$-coupling. Here, we explain why these effects are neglected in the theoretical description of our experiments.


The incoherent relaxation rates for the chosen pair $^1$H~--$^{\ 13}$C are primarily determined by the square of their dipolar coupling to each other as opposed to the dipolar coupling to $^{15}$N:  The intra-pair rate is proportional to  $\gamma_1^2 \gamma_2^2$, while, for comparison, the incoherent rates due to the couplings of $^{15}$N to $^1$H and $^{13}$C  are proportional, respectively to  $\gamma_1^2 \gamma_{\text{N}}^2$ and $\gamma_2^2 \gamma_{\text{N}}^2$, where $\gamma_{\text{N}}$ is the gyromagnetic ratio of $^{15}$N. Given that  $\gamma_{\text{N}}^2 / \gamma_1^2 \approx 10^{-2}$ and $\gamma_2^2 / \gamma_{\text{N}}^2 \approx 1/6$, this aspect alone significantly weakens the incoherent rates due to the $^{15}$N~--$\ ^1$H and $^{15}$N~--$^{ \ 13}$C magnetic dipolar couplings.
In addition, the incoherent rate due to $^{15}$N~--$\ ^1$H coupling  is further weakened by the inverse sixth power of the distance between the two. 

Since the incoherent coupling rates for any pair of nuclei have the reciprocal character, the overall incoherent effect of $^{15}$N on the $^1$H~--$^{\ 13}$C pair can be quantified from the reverse action of the pair onto $^{15}$N, which, in turn, can be constrained from above by the experimentally measured overall longitudinal relaxation rate $ 1/T_1^N$ of the $^{15}N$ spin. This measurement is presented in Fig.~\ref{fig_N}, where, the exponential fit to the data gives $T_1^N = (1.20 \pm 0.04) \cdot 10^2$~s. The total rate $ 1/T_1^N$ can only be larger than the sum of the positive rate contributions from $^1$H and $^{ 13}$C. This implies that the times required for the $^1$H~--$^{\ 13}$C pair to feel  the incoherent action of $^{15}$N are of the order of or greater than $T_1^N$, which is, in turn, much longer than the measurement times during our experiments.  

Regarding the possible coherent influence of  $^{15}N$ on the $^1$H~--$^{\ 13}$C pair, let us recall that the truncated Hamiltonian of the $J$-coupling between $^{15}$N and the \mbox{$^1$H~--$^{\ 13}$C} pair has form  
\begin{equation}
{\cal H}' = J_{\text{HN}} S_{1z} S_{3z} + J_{\text{CN}} S_{2z} S_{3z},
\label{Hnch}
\end{equation}
where $S_{3z}$ is the $z$-projection of the $^{15}$N spin operator ${\mathbf S}_3$.  One may be concerned here that the coupling constants  $J_{\text{CN}}$ and $J_{\text{HN}}$ are larger than the measured relaxation rates (see Section~\ref{exp}). However, such an interaction can, in principle, cause coherent transitions  between two $z$-axis ZQ states $\vert S_0 \, \rangle$ and  $\vert T_{0,z} \, \rangle$ or between two DQ states $\vert \psi_{-,z} \, \rangle$ and  $\vert \psi_{+,z} \, \rangle$ but not between a ZQ state and a DQ state.  Within the ZQ pair or within the DQ pair, the interaction (\ref{Hnch}) has no effect on the diagonal relaxation -- monitored as $\langle S_{1z} \rangle (t)$, $\langle  S_{2z} \rangle (t)$, and $\langle S_{1z} S_{2z} \rangle (t)$.
The coherent effect of the  interaction (\ref{Hnch})   on the off-diagonal relaxation monitored in this work through $\langle S_{1x} S_{2x} \rangle (t)$ can, in principle, be noticeable, but it is suppressed in our off-diagonal relaxation experiments by the CPMG pulse sequence.

Let us, finally, note that, since $J_{\text{HC}} \gg J_{\text{CN}}, J_{\text{HN}}$, it is justifiable to treat the $^1$H~--$^{\ 13}$C pair as isolated on the time scale of the order of $1/J_{\text{HC}}$ required to prepare the Bell PPSs.

\

\section{Relaxation rates for microscopic Hamiltonians $\mathcal{H}^{d}$ and $\mathcal{H}^{\alpha}$}
\label{derivations}

In this Appendix, we compute the relaxation rates appearing in subsections~\ref{D-relaxation} and ~\ref{OD-relaxation}, and simultaneously substantiate  the statements made in those subsections regarding the cross-correlation effects and the relaxation eigenmodes.

The application of the master equation (\ref{master}) requires one to compute the double-commutators with the Hamiltonian. Below are the explicit expressions for the relevant double-commutators:
\begin{widetext}
\vspace{0cm}
\begin{equation}
 \begin{split}
     \label{S1z}
     \frac{1}{\hbar^2} \biggl[ \biggl[ S_{1z}, \mathcal{H}^{d}(t^\prime) +  \mathcal{H}^{\alpha}(t^\prime) \biggl], \mathcal{H}^{d}(t) +  \mathcal{H}^{\alpha}(t) \biggl] = 
     \ \ \ \ \ \ \ \ \ \ \ \ \ \ \ \ \ \ \ \ \ \ \ \ \ \ \ \ \ \ \ \ \ \ \ \ \ \ \ \ \ \ \ \ \ \ \ \ \ \ \ \ \ \ \ \ \ \ \ \ \ \ \ \ \ \ \ \ \ \ \ \ \ \ \ \ \ \ \ \ \ \ \ \ \ \ \ 
     \\
     =  (S_{1z} - S_{2z}) \ \biggl( \frac{1}{16}F_0(t^\prime)F_0(t) e^{i(\Omega_1 - \Omega_2) (t^\prime - t)} +  \frac{1}{16}F_0(t^\prime)F_0(t) e^{-i(\Omega_1 - \Omega_2) (t^\prime - t)} \biggl) +  
     \\ 
     + \ S_{1z} \biggl(\frac{1}{2}F_1(t^\prime)F_1^*(t) e^{i\Omega_1 (t^\prime - t)} +  \frac{1}{2}F_1^*(t^\prime)F_1(t) e^{-i\Omega_1 (t^\prime - t)} \biggl) +
     \\
     + \  (S_{1z} + S_{2z}) \ \biggl( F_2(t^\prime)F_2^*(t) e^{i(\Omega_1 + \Omega_2) (t^\prime - t)} +  F_2(t^\prime)F_2(t) e^{-i(\Omega_1 + \Omega_2) (t^\prime - t)} \biggl)  +  
     \\
     + \ S_{1z} \biggl(2\alpha_{1\perp}(t^\prime)\alpha_{1\perp}^*(t) e^{i\Omega_1 (t^\prime - t)} + 2\alpha_{1\perp}^*(t^\prime)\alpha_{1\perp}(t) e^{-i\Omega_1 (t^\prime - t)} \biggl) + 
     \\ 
     + \ 2S_{1z}S_{2z} \biggl( \alpha_{1\perp} (t^\prime) F_1^*(t) e^{i\Omega_1 (t^\prime - t)} + \alpha_{1\perp}^* (t^\prime) F_1(t) e^{-i\Omega_1 (t^\prime - t)} + \alpha_{1\perp}^* (t) F_1(t^\prime) e^{i\Omega_1 (t^\prime - t)} + \alpha_{1\perp} (t) F_1^*(t^\prime) e^{-i\Omega_1 (t^\prime - t)} \biggl) + 
     \\
     + \  \text{QOT};
     \end{split}
  \end{equation}
  \vspace{-1cm}
\begin{equation}
 \begin{split}
     \label{S1zS2z}
     \frac{1}{\hbar^2} \biggl[ \biggl[ 2 S_{1z}S_{2z}, \mathcal{H}^{d}(t^\prime) +  \mathcal{H}^{\alpha}(t^\prime) \biggl], \mathcal{H}^{d}(t) +  \mathcal{H}^{\alpha}(t) \biggl] = 
     \ \ \ \ \ \ \ \ \ \ \ \ \ \ \ \ \ \ \ \ \ \ \ \ \ \ \ \ \ \ \ \ \ \ \ \ \ \ \ \ \ \ \ \ \ \ \ \ \ \ \ \ \ \  \ \ \ \ \ \ \ \ \ \ \ \ \ \ \ \ \ \ \ \ \ \ \ \ \ \ \ \ \ \ \ \ \ \ \  \ \ \ \ 
     \\
     =  2 S_{1z}S_{2z} \ \biggl( \frac{1}{2}F_1(t^\prime)F_1^*(t) e^{i\Omega_1 (t^\prime - t)} +  \frac{1}{2}F_1^*(t^\prime)F_1(t) e^{-i\Omega_1 (t^\prime - t)} + + \frac{1}{2}F_1(t^\prime)F_1^*(t) e^{i\Omega_2 (t^\prime - t)} +  \frac{1}{2}F_1^*(t^\prime)F_1(t) e^{-i\Omega_2 (t^\prime - t)} + \ \ \ 
     \\ 
      + 2\alpha_{1\perp}(t^\prime)\alpha_{1\perp}^*(t) e^{i\Omega_1 (t^\prime - t)} 
     + 2\alpha_{1\perp}^*(t^\prime)\alpha_{1\perp}(t) e^{-i\Omega_1 (t^\prime - t)} + 2\alpha_{2\perp}(t^\prime)\alpha_{2\perp}(t) e^{i\Omega_2 (t^\prime - t)} + 2\alpha_{2\perp}^*(t^\prime)\alpha_{2\perp}(t) e^{-i\Omega_2 (t^\prime - t)}\biggl) + 
     \\ 
     + \ S_{1z} \ \bigl( 2 F_1(t^\prime)\alpha_{1\perp}^*(t) e^{i\Omega_1(t^\prime - t)} + 2 F_1^*(t^\prime)\alpha_{1\perp}(t) e^{-i\Omega_1(t^\prime - t)} \bigl) + 
     \\ 
     + \  S_{2z} \ \bigl(2 F_1(t^\prime)\alpha_{2\perp}^*(t) e^{i\Omega_2(t^\prime - t)} +  2 F_1^*(t^\prime)\alpha_{2\perp}(t) e^{-i\Omega_2(t^\prime - t)} \bigl) + 
     \\
     + \  \text{QOT};
     \end{split}
  \end{equation}
 \vspace{-1cm} 
 \begin{equation}
 \begin{split}
     \label{singlet/triplet}
     \frac{1}{\hbar^2} \biggl[ \biggl[ S_{1+}S_{2-} + S_{1-}S_{2+}, \mathcal{H}^{d}(t^\prime) +  \mathcal{H}^{\alpha}(t^\prime) \biggl], \mathcal{H}^{d}(t) +  \mathcal{H}^{\alpha}(t) \biggl] = \qquad \qquad \qquad \qquad \qquad \qquad \qquad \qquad \qquad \qquad \qquad 
     \\ 
     = \ S_{1+}S_{2-} \biggl(\frac{1}{2}F_1(t^\prime)F_1^*(t) e^{i\Omega_2(t^\prime - t)} +  \frac{1}{2}F_1^*(t^\prime)F_1(t) e^{-i\Omega_1(t^\prime - t)}  +  \frac{1}{8} F_0(t^\prime) F_0(t) e^{-i(\Omega_1 - \Omega_2)(t^\prime - t)} + 
     \ \ \ \ \ \ \ \ \ \ \ \ \ \ 
     \\ 
     + (\alpha_{1z} - \alpha_{2z})(t^\prime)(\alpha_{1z} -\alpha_{2z}) (t) + 2 \alpha_{1\perp}^*(t^\prime)\alpha_{1\perp}(t) e^{-i \Omega_1 (t^\prime - t)} + 2 \alpha_{2\perp}(t^\prime)\alpha_{2\perp}^*(t) e^{i \Omega_2 (t^\prime - t)}\biggl)  + 
     \\ 
     + \ S_{1-}S_{2+}\biggl(\frac{1}{2}F_1(t^\prime)F_1^*(t) e^{i\Omega_1(t^\prime - t)}  +  \frac{1}{2}F_1^*(t^\prime)F_1(t) e^{-i\Omega_2(t^\prime - t)}  + \frac{1}{8} F_0(t^\prime) F_0(t) e^{i(\Omega_1 - \Omega_2)(t^\prime - t)}  + 
     \ \ \ \ \ \ \ \ \ \ \ \ \ \ \ \ 
     \\   
     + (\alpha_{1z} - \alpha_{2z})(t^\prime)(\alpha_{1z} -\alpha_{2z}) (t)  +   + 2 \alpha_{1\perp}(t^\prime)\alpha_{1\perp}^*(t) e^{i \Omega_1 (t^\prime - t)} + 2 \alpha_{2\perp}^*(t^\prime)\alpha_{2\perp}(t) e^{-i \Omega_2 (t^\prime - t)}\biggl)  +  
     \\ 
     + \ \text{QOT};
     \end{split}
  \end{equation}

 \begin{equation}
 \begin{split}
     \label{plus/minus}
     \frac{1}{\hbar^2} \biggl[ \biggl[ S_{1+}S_{2+} + S_{1-}S_{2-}, \mathcal{H}^{d}(t^\prime) +  \mathcal{H}^{\alpha}(t^\prime) \biggl], \mathcal{H}^{d}(t) +  \mathcal{H}^{\alpha}(t) \biggl] = \qquad \qquad \qquad \qquad \qquad \qquad \qquad \qquad \qquad \qquad \qquad 
     \\ 
     = \ S_{1+}S_{2+}\biggl(\frac{1}{2}F_1^*(t^\prime)F_1(t) e^{-i\Omega_2(t^\prime - t)} + \frac{1}{2}F_1^*(t^\prime)F_1(t) e^{-i\Omega_1(t^\prime - t)} + 2F_2^*(t^\prime)F_2(t)e^{-i(\Omega_1 + \Omega_2)(t^\prime-t)} + 
     \ \ \ \ \ \ \ \ \ \ 
     \\  
     + (\alpha_{1z} + \alpha_{2z})(t^\prime)(\alpha_{1z} +\alpha_{2z}) (t) +  2 \alpha_{1\perp}^*(t^\prime)\alpha_{1\perp}(t) e^{-i \Omega_1 (t^\prime - t)}  + 2 \alpha_{2\perp}(t^\prime)\alpha_{2\perp}^*(t) e^{i \Omega_2 (t^\prime - t)}\biggl) + 
     \\ 
     + \ S_{1-}S_{2-} \biggl(\frac{1}{2}F_1(t^\prime)F_1^*(t) e^{i\Omega_1(t^\prime - t)}  +   \frac{1}{2}F_1(t^\prime)F_1^*(t) e^{i\Omega_2(t^\prime - t)} +  2F_2(t^\prime)F_2^*(t)e^{i(\Omega_1 + \Omega_2)(t^\prime-t)} + 
     \ \ \ \ \ \ \ \ \ \ \ \ \ \ \ \ 
     \\   
     + (\alpha_{1z} + \alpha_{2z})(t^\prime)(\alpha_{1z} + \alpha_{2z}) (t)  + 2 \alpha_{1\perp}(t^\prime)\alpha_{1\perp}^*(t) e^{i \Omega_1 (t^\prime - t)} + 2 \alpha_{2\perp}^*(t^\prime)\alpha_{2\perp}(t) e^{-i \Omega_2 (t^\prime - t)}\biggl) + 
     \\ 
     + \  \text{QOT}.
     \end{split}
  \end{equation}
\end{widetext}
where $S_{n\pm} = S_{nx} \pm i S_{ny}$. The abbreviation QOT  stands for ``quickly oscillating terms", implying the terms that contain factors such as, e.g.,  $e^{i\Omega_n(t^\prime + t)}$ or $e^{i(\Omega_1 t^\prime -  \Omega_2 t)}$. After integration in Eq.(\ref{master}), these factors cannot be fully absorbed into the spectral function $\mathcal{J}(\omega)$ associated with the Fourier transform with respect to $t^\prime - t$ \ --- \ rather they lead to quickly oscillating time-dependent prefactors in front of very small relaxation rates, thereby precluding those rates from having any effect on the solutions of Eq.(\ref{master}). (The factor $e^{i(\Omega_1 t^\prime -  \Omega_2 t)}$ is quickly oscillating only when $|\Omega_2 - \Omega_1|$ is much larger than the rate eventually multiplied by that factor, which is the basic assumption of the present article.)

To connect the above double-commutators to the relaxation rates in Eqs.(\ref{C17}) and (\ref{T2}), one considers the components of the density matrix $\rho$ associated with  the  basis operators $S_{1z}$, $S_{2z}$, $2 S_{1z}, S_{2z}$, $2 S_{1x}, S_{2x}$ and $2 S_{1y}, S_{2y}$, substitutes double-commutators containing these operators or their superpositions into the integral in Eq.(\ref{master}), performs the ensemble average, and then represents each of the resulting terms as one of the above operators multiplied by an integral  of an equilibrium correlation function with respect to $\tau = t^\prime - t$. Each such an integral then gives a relevant relaxation rate.

With the above procedure in mind, the diagonal relaxation rates entering Eq.~(\ref{C17}) can be obtained as follows:

\noindent (i) rate $\mu_{12}^d$ given by Eq.(\ref{mu12}) can be read from the second line of Eq.(\ref{S1zS2z});

\noindent (ii) rate $\mu_{1}^d$ given by Eq.(\ref{mun_dd}) can be read from the second, third and fourth lines of Eq.(\ref{S1z}); rate
$\mu_{2}^d$ can then be obtained by switching the indices ``1" and ``2" in the expression for $\mu_{1}^d$ (resulting in the same expression);

\noindent (iii) rate $\sigma_{12}$ given by Eq.(\ref{sigma12}) can be read from the second and the fourth lines of Eq.(\ref{S1z});

\noindent (iv) rate $\mu_{12}^{\alpha}$ given by Eq.(\ref{mu12_alpha}) can be read from the third line  of Eq.(\ref{S1zS2z});

\noindent (v) rates $\delta_1$ and $\delta_2$ given by Eq.(\ref{delta_n}) can be read, respectively, for the fourth and the fifth lines of Eq.(\ref{S1zS2z}) [$\delta_1$ can also be obtained from the fifth line of Eq.(\ref{S1z})].

The rates of the off-diagonal relaxation described by Eq.(\ref{T2}) can be extracted from the double-commutators (\ref{singlet/triplet}) and (\ref{plus/minus}) by first noting that $2 S_{1x} S_{2x} + 2 S_{1y} S_{2y} = S_{1+} S_{2-} + S_{1-} S_{2+}$, while $2 S_{1x} S_{2x} - 2 S_{1y} S_{2y} = S_{1+} S_{2+} + S_{1-} S_{2-}$. The structure of the double-commutator (\ref{singlet/triplet})  is such, that the master equation (\ref{master}) only couples $\langle 2 S_{1x} S_{2x} + 2 S_{1y} S_{2y} \rangle$ to itself, meaning that $\langle 2 S_{1x} S_{2x} + 2 S_{1y} S_{2y} \rangle(t)$ is a relaxation eigenmode for our system.  For the same reason, 
the double-commutator (\ref{plus/minus}) implies that  $\langle 2 S_{1x} S_{2x} - 2 S_{1y} S_{2y} \rangle$ is another relaxation eigenmode.

In terms of individual rates:

\noindent (vi) rate $\lambda_\text{\tiny ZQ}^d$ given by Eq.(\ref{lambda_plus_dd}) can be read from the second line of Eq.(\ref{singlet/triplet});

\noindent (vii) rate $\lambda_\text{\tiny DQ}^d$ given by Eq.(\ref{lambda_minus_dd}) can be read from the second line of Eq.(\ref{plus/minus});

\noindent (viii) rate $\lambda_\text{\tiny ZQ}^{\alpha}$ given by Eq.(\ref{lambda_plus_alpha}) can be read from the third line of Eq.(\ref{singlet/triplet});

\noindent (ix) rate $\lambda_\text{\tiny DQ}^{\alpha}$ given by Eq.(\ref{lambda_minus_alpha}) can be read from the third line of Eq.(\ref{plus/minus}).

\begin{figure}[t]
    \setlength{\unitlength}{0.1cm}
      \begin{picture}(100,141)
       \put(0,0){\includegraphics[width=1.0\columnwidth]{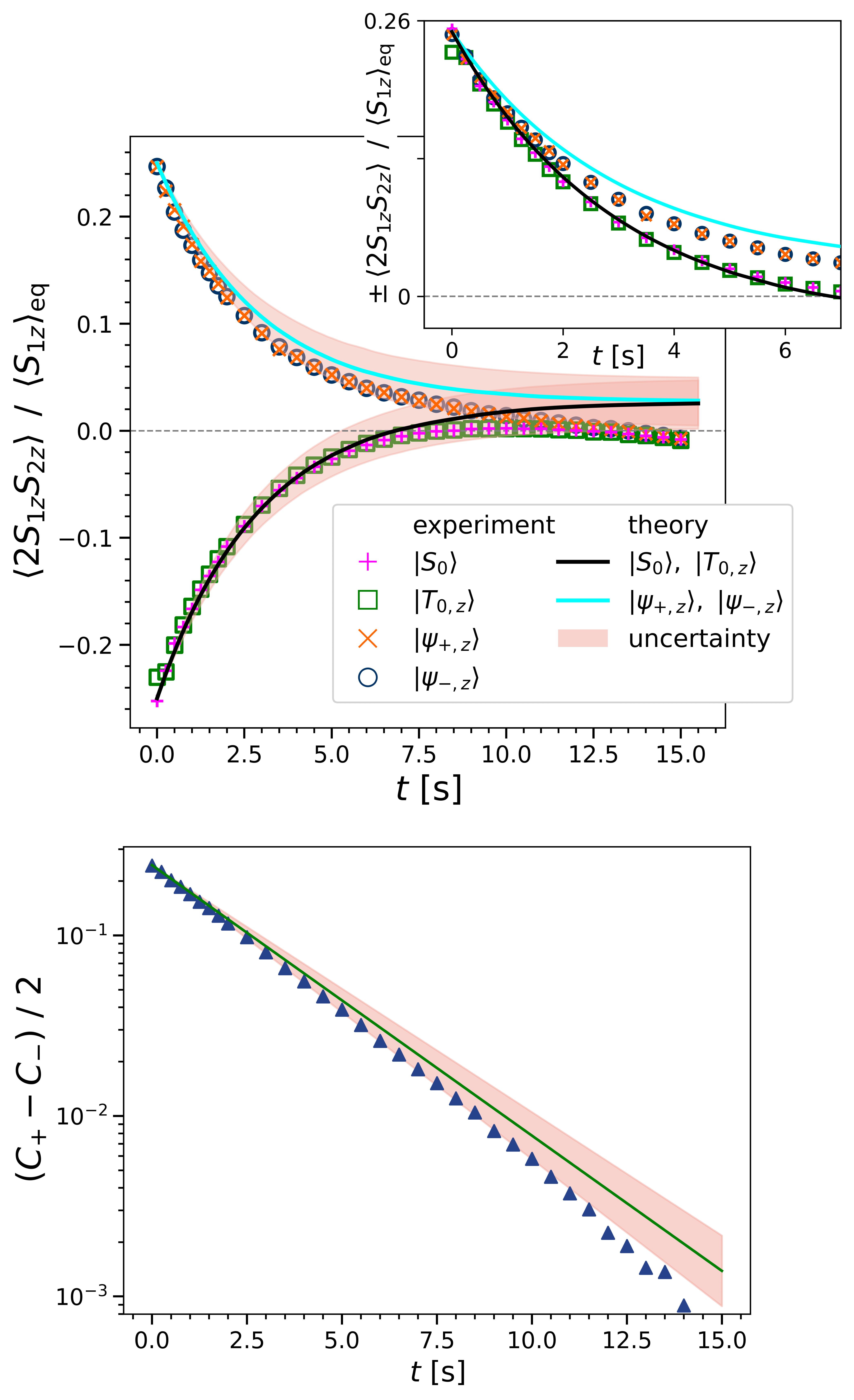}}

       \put(0,128){{\Large(a)}}
       \put(0,56){{\Large(b)}}

       \put(23,123){{\Large $^1$H}}

       \put(74.5,94.5){\colorbox{white}{\ }}
       \put(74.5,92.5){\colorbox{white}{\ }}
       {\color{white}
        \put(70,90.8){\rule{5mm}{5mm}}}
       \put(42.5,92.3){$|\text{Bell PPS} \rangle \  \text{--} \ t \text{ --}   \  (\pi/2)_{1y} \ \text{-- A1} $}
       \put(58,95){\tiny CPMG}

       \put(39.8,129){\colorbox{white}{\tiny }}
       \put(34,129){\colorbox{white}{\tiny\ }}

       \put(74.5,106.5){\colorbox{white}{\ }}
       \put(74.5,104.5){\colorbox{white}{\ }}
       \put(74.5,102.5){\colorbox{white}{\ }}
       \put(74.5,100.5){\colorbox{white}{\ }}
       \put(74.5,98.5){\colorbox{white}{\ }}
       \put(74.5,96.5){\colorbox{white}{\ }}
       
      \end{picture}
    \caption{(a) Relaxation of $\langle S_{1z}S_{2z} \rangle$ starting from four Bell PPSs  and measured as the antisymmetric component of $^1$H FID spectrum following the rf-pulse sequence given in the plot. Unlike the pulse sequence used in Fig.~\ref{fig:muZQ-muDQ}, the present one involves CPMG echo trains $(\pi)_{1y}, (\pi)_{2y} \ \text{---} \ \tau \ \text{---} \ (-\pi)_{1y}, (-\pi)_{2y} \ \text{---} \ \tau \ \text{---} ...$ with \mbox{$\tau = 1$~ms}. Black and cyan solid lines are the theoretical relaxation curves for the ZQ and DQ states obtained from Eq.(\ref{diag-relax-CPMG}). 
    Inset in (a): initial relaxation from the main plot with the sign of $\langle S_{1z}S_{2z} \rangle$ inverted for  $|S_0\rangle$ and $|T_0\rangle$ to visually compare the initial relaxation rates; these rates are consistent with those in the inset of Fig.~\ref{fig:muZQ-muDQ}(a). (b) Semilogarithmic plots of $(C_+ - C_-)/2$, where $C_+$  is the average of $\langle S_{1z}S_{2z} \rangle(t)$ for the two DQ states, and $C_-$  for the two ZQ states. Solid green line is the monoexponential theoretical curve determined by the rate $\mu_{12}$ in Eq.(\ref{diag-relax-CPMG}). 
    }
    \label{fig:muZQ-muDQ-CPMG}
\end{figure}

Finally, we note that the isotropic averaging of the dipole coupling coefficients (\ref{F0}-\ref{F2}) gives:
\begin{eqnarray}
    \langle F_0^2 \rangle &=& \frac{4}{5} k^2
    \label{F0av}   
    \\[1mm]
    \langle |F_1|^2 \rangle &=& \frac{3}{10} k^2
    \label{F1av}
    \\[1mm]
    \langle |F_2|^2 \rangle &=& \frac{3}{10} k^2.
    \label{F2av}
\end{eqnarray}
Substituting the above averages into Eqs.(\ref{mun_dd}, \ref{sigma12}, \ref{mu12}, \ref{lambda_plus_dd}, \ref{lambda_minus_dd}) and taking the limit $\Omega_1, \Omega_2 \ll 1/\tau_c$, which means $\mathcal{J}_0 \equiv \mathcal{J}(0) \approx \mathcal{J}(\Omega_n) \approx \mathcal{J}(\Omega_1 + \Omega_2) \approx \mathcal{J}(\Omega_1 - \Omega_2) $, we then obtain:
\begin{eqnarray}
    \mu_n^d &=&  \frac{1}{2} k^2 \mathcal{J}_0
    \label{mun_dd_av}
    \\[1mm]
    \sigma_{12} &=&  \frac{1}{4}k^2 \mathcal{J}_0
    \label{sigma12_av}
    \\[1mm]
    \mu_{12}^d &=&  \frac{3}{10}k^2 \mathcal{J}_0
    \label{mu12_dd_av}
    \\[1mm]
    \lambda_{\text{\tiny ZQ}}^d &=&  \frac{1}{5}k^2 \mathcal{J}_0
    \label{lambda_plus_dd_av}
    \\[1mm]
    \lambda_{\text{\tiny DQ}}^d &=&  \frac{9}{20}k^2 \mathcal{J}_0.
    \label{lambda_minus_dd_av}
\end{eqnarray} 
Equations(\ref{mun_dd_av}-\ref{mu12_dd_av}) imply that
\begin{equation}
    \frac{\mu_1^d  \,  +  \, \mu_2^d  \,  - \, \mu_{12}^d}{\sigma_{12}} = 2.8 ,
    \label{mu_d-sigma_ratio}
\end{equation}
which, finally, leads to Eq.(\ref{mu-sigma_ratio}) after, in that equation, the contributions from $\mu_n^{\alpha}$ and $\mu_{12}^{\alpha}$ become canceled. At the same time, Eqs.(\ref{lambda_plus_dd_av}, \ref{lambda_minus_dd_av}) lead to Eq.(\ref{lambda_d_ratio}).

\section{Measurements of $\langle 2 S_{1z}S_{2z} \rangle(t)$ with the CPMG echo train on $^1$H and $^{13}$C. }
\label{SzSz_CPMG}

In this Appendix, we include the counterpart of the measurements of $\langle 2 S_{1z}S_{2z} \rangle(t)$ reported in Fig.~\ref{fig:muZQ-muDQ} but with the CPMG sequence, which decouples the $^1$H---$^{13}$C pair from $^{15}$N and from the inhomogeneity of the static magnetic field.  [The same CPMG sequence was used   for measuring $\langle 2 S_{1x}S_{2x} \rangle(t)$ in Fig.~\ref{fig:lambdaZQ-lambdaDQ}.] This is a control experiment, since the CPMG sequence as such is not required to measure the diagonal relaxation characterized by $\langle 2 S_{1z}S_{2z} \rangle(t)$,  $\langle  S_{1z} \rangle(t)$ and $\langle S_{2z} \rangle(t)$.

The CPMG sequence repeatedly flips the signs of $\langle  S_{1z} \rangle$ and $\langle S_{2z} \rangle$, which modifies the solutions of the relaxation equations (\ref{C17}).   For an ideal CPMG sequence with closely spaced $\pi$-pulses, the values of $\langle  S_{1z} \rangle(t)$ and $\langle S_{2z} \rangle(t)$ are supposed to average to zero, so that the third equation in the system (\ref{C17}) converts into
\begin{equation}
\frac{d}{dt} \langle 2 S_{1z}S_{2z} \rangle = - \delta_1 \langle S_{1z} \rangle_{\text{eq}} - \delta_2 \langle S_{2z} \rangle_{\text{eq}} - \mu_{12} \langle 2 S_{1z}S_{2z} \rangle,
\label{diag-relax-CPMG}
\end{equation}
where $\langle S_{nz} \rangle_{\text{eq}} = \varepsilon_n$, and the values of $\delta_1$, $\delta_2$ and $\mu_{12}$ are given in Table~\ref{tbl:rates}. Equations (\ref{C17}) and (\ref{diag-relax-CPMG}) predict the same  initial behavior of $\langle 2 S_{1z}S_{2z} \rangle(t)$ starting from a Bell PPS, but at longer times the two time evolutions become different. 

The measured relaxation of $\langle 2 S_{1z}S_{2z} \rangle(t)$  with the CPMG sequence for all four initial Bell PPSs is shown in Fig.~\ref{fig:muZQ-muDQ-CPMG}(a). The initial decays in this case are consistent with those obtained without the CPMG sequence and presented in Fig.~\ref{fig:muZQ-muDQ}(a) --- see, in particular, the insets in the two figures. The quantitative comparison between the two experiments is facilitated by the fact that  the theoretical curves superimposed on the experimental datasets were obtained from  Eqs.~(\ref{C17}) for Fig.~\ref{fig:muZQ-muDQ}(a) and from Eq.~(\ref{diag-relax-CPMG}) for Fig.~\ref{fig:muZQ-muDQ-CPMG}(a) using the same rates 
$\delta_1$, $\delta_2$ and $\mu_{12}$ and hence exhibiting the same initial behavior. At longer times, the measurements in the presence of the CPMG sequence depart from those without it, as expected. However, as one can see in Fig.~\ref{fig:muZQ-muDQ-CPMG}(a), the CPMG-based measurements at longer times also deviate from the solutions of Eq.~(\ref{diag-relax-CPMG}). We attribute this discrepancy to a baseline drift associated with non-ideal implementation of the CPMG sequence due to factors discussed in Section~\ref{relaxation-Bell}.

Assuming the same baseline drift for all four Bell PPSs (as done in Section~\ref{relaxation-Bell}), one can subtract the measured relaxation curves of $\langle 2 S_{1z}S_{2z} \rangle$ for the ZQ states from those for the DQ states, and then expect on the basis of Eq.~(\ref{diag-relax-CPMG}) that the difference exhibits monoexponential relaxation controlled by the rate $\mu_{12}$. The latter expectation is shown in Fig.~\ref{fig:muZQ-muDQ-CPMG} (b) to be in a satisfactory agreement with experiment.

\bibliography{biblio}

\end{document}